\documentclass[referee]{raa}            

\usepackage{graphicx,times}             
\usepackage{natbib}
\usepackage{amssymb,amsmath}
\graphicspath{{./}{figures/}}
\bibpunct{(}{)}{;}{a}{}{,}
\usepackage{listings}
\definecolor{codegreen}{rgb}{0,0.6,0}
\definecolor{codegray}{rgb}{0.5,0.5,0.5}
\definecolor{codepurple}{rgb}{0.58,0,0.82}
\definecolor{backcolour}{rgb}{0.95,0.95,0.92}
\lstdefinestyle{mystyle}{
    backgroundcolor=\color{backcolour},   
    commentstyle=\color{codegreen},
    keywordstyle=\color{black},
    numberstyle=\tiny\color{black},
    stringstyle=\color{codegreen},
    basicstyle=\ttfamily\footnotesize,
    breakatwhitespace=false,         
    breaklines=true,                 
    captionpos=b,                    
    keepspaces=true,                                   
    numbersep=5pt,                  
    showspaces=false,                
    showstringspaces=false,
    showtabs=false,                  
    tabsize=2,
    language=Python
}
\usepackage[pagebackref=true]{hyperref}
\usepackage{xcolor}

\begin{document}
 \title{LightCurve MoE: A Dynamic Sparse Routing Mixture-of-Experts Architecture for Efficient Stellar Light Curve Classification}

 \volnopage{ {\bf 20XX} Vol.\ {\bf X} No. {\bf XX}, 000--000}
   \setcounter{page}{1}
   \author{Cunshi Wang \inst{1,2} 
   \and Yu Bai \inst{2,3}
   \and Xinrui Song \inst{4}
   \and Jiacheng Xu \inst{4}
   \and Henggeng Han \inst{1}  
   \and Yuyang Li \inst{1,2}\footnote{\url{liyuyang22@mails.ucas.ac.cn}}
   \and Xinjie Hu \inst{2}
   \and Huiqin Yang \inst{2,3}
   \and Jifeng Liu\inst{1,2,3,5}
   }

   \institute{    
    College of Astronomy and Space Sciences, University of Chinese Academy of Sciences, Beijing 100049, China \label{UCAS}
    \and
   Key Laboratory of Optical Astronomy, National Astronomical Observatories, Chinese Academy of Sciences,
    20A Datun Road, Chaoyang District, Beijing 100101, People's Republic of China \label{NAOC}
    \and 
    Institute for Frontiers in Astronomy and Astrophysics, Beijing Normal University,  Beijing 102206, China \label{IFA}
    \and
    School of Mechanical, Electrical \& Information Engineering, Shandong (Weihai) University, Weihai, Shandong, China \label{SDU}  
    \and
    New Cornerstone Science Laboratory, National Astronomical Observatories, Chinese Academy of Sciences, Beijing 100101, People's Republic of China \label{NCS}
\vs \no
   {\small Received 20XX Month Day; accepted 20XX Month Day}
}

   \date{Received September 15, 1996; accepted March 16, 1997}

\abstract{The classification of stellar light curves has become a key task in modern time-domain astronomy, fueled by the rapid growth of data from large-scale surveys such as Kepler and TESS. Although deep learning models have achieved high accuracy in this area, their computational costs can limit scalability. To tackle this issue, we propose LightCurve MoE, a Mixture-of-Experts (MoE) architecture that combines dynamic sparse routing with a dual-gating mechanism to balance accuracy, efficiency, and robustness.
Our model includes five specialized experts, each using a different feature extraction method—such as wavelet transforms, Gramian angular fields, and recurrence plots—to capture unique patterns in the light curves. A dual-gating mechanism evaluates these expert outputs by analyzing both frequency and time-domain features, allowing the model to adaptively weigh each expert’s contribution.
During inference, only the top 3 out of 5 experts are activated per sample using a Top-$k$ routing strategy, reducing computational cost by 40\% compared to dense models while preserving strong accuracy ($\approx 96\%$). The model also includes entropy regularization and a technique to retain inactive experts during training, ensuring stable and effective learning.
By combining sparse computation with multi-modal feature fusion, LightCurve MoE offers a scalable solution for future large-scale photometric surveys like LSST and Global Open Transient Telescope Array (GOTTA), where processing efficiency is crucial due to the massive volume of daily data.
}

   \authorrunning{Cunshi Wang et al. }            
   \titlerunning{Transfer learning, MoE of LCs}  
   \maketitle


\section{Introduction} 
\label{intro}
Photometric variability is a common feature among a wide range of astronomical objects, and understanding its underlying mechanisms is a fundamental goal in astrophysics. Variable stars, in particular, have long been central to the development of time-domain astronomy, a field that has become increasingly vital in modern astrophysical research. Among them, Cepheid variables \citep{edd1917,pierce94,stetson96} are well-known pulsating stars that serve as critical standard candles for measuring distances across galaxies and even on cosmological scales.

Pulsating variable stars encompass several subclasses, including bright Asymptotic Giant Branch (AGB) stars, Classical Cepheids, multi-mode RR Lyrae (RR), $\delta$ Scuti stars ($\delta$ Sct) and $\gamma$ Doradus stars ($\gamma$ Dor) stars. These types differ in absolute magnitude, pulsation mode, and driving mechanism, leading to distinct period-luminosity relations \citep{sterken05,Understanding}. The accurate classification of these variable stars—especially pulsating ones—has significantly advanced our understanding of stellar formation and evolution, as well as the structure and dynamics of binary and n-body systems \citep{keb1,keb2,keb3,keb4,keb5,keb6,keb7,keb8}.

Modern astronomy has entered the era of big data. For instance, the Zwicky Transient Factory (ZTF, \citealt{ZTF14}) at Palomar Observatory uses a 1.2m telescope with a $47 deg^2$ field of view to scan the sky every two days, generating approximately 1TB of raw data nightly \citep{maha19ztf}. The upcoming Global Open Transient Telescope Array (GOTTA, \citealt{sitian21}) will deploy over one hundred 1-meter telescopes worldwide, each with a $25 deg^2$ field of view, monitoring the sky in the $g$, $r$, and $i$ bands. It is expected to observe more than $1,000 deg^2$ every 30 minutes, producing around 40GB of raw data per minute.

To handle such massive data volumes efficiently, machine learning techniques have become essential for automated analysis. Moreover, as the GOTTA plans to integrate Large Language Model (LLM)-based agents for intelligent control \citep{wang2024SWT}, it is crucial to limit the computational cost of machine learning models to ensure seamless integration with these AI-driven systems. Therefore, there is a pressing need for an effective and efficient pipeline for variable star identification and classification that supports low-cost computation and flexible input structures. In this work, we adopt transfer learning and a Mixture-of-Experts (MoE) approach to meet these requirements.

Modern astronomy has entered the big data era. For example, the Zwicky Transient Factory (ZTF, \citealt{ZTF14}), a time-domain survey at Palomar Observatory, uses a 1.2m telescope with a $47 deg^2$ field of view to scan the sky every two days. ZTF produces about 1TB of raw uncompressed data each night \citep{maha19ztf}. GOTTA \citep{sitian21}, will use an array of more than one hundred 1-meter telescopes with a $25 deg^2$ field of views, deployed worldwide to monitor the sky in $g$, $r$, and $i$ bands. It will monitor more than $1,000 deg^2$ of the sky every 30 minutes, and generate raw data in a count of 40GB per minute. To process the data faster, machine learning based methods are the solution to big data analytics. 

Considering the telescopes in GOTTA will be organized with the help of an embodied control center by using Large Language Model (LLM) based-agents \citep{wang2024SWT}, the calculation cost of machine learning methods should be restricted for the usage of LLM-based-agents. A more effective and efficient pipeline is in great demand to do variable identification and classification with low calculation cost to fit the construction of AI-empowered agents. We here choose the method of transfer learning and MoE for low calculation cost and different input structures.

Transfer learning provides an efficient and computationally economical approach for edge computing scenarios \citep{pan10}. By leveraging pre-trained models and adapting them to new tasks, this technique minimizes the need for large datasets and extensive training times. Specifically, by adjusting fully connected and classification layers, knowledge from image recognition can be effectively transferred to tasks such as light curve classification.

Previous studies have demonstrated the efficacy of transfer learning in astronomy. For example, \citet{maha19ztf} used machine learning to classify transient objects, while \citet{duev19ztf} showed that CNNs are 50 times faster than random forest algorithms in identifying fast-moving objects. Additionally, ZTF successfully integrated machine-learning-based classifiers into its pipeline for point-source classification, streak diagnostics, and stellar-galaxy differentiation \citep{ztfmas18}, highlighting the versatility and efficiency of these methods in astronomical applications. However, it is important to note that transfer learning requires input data to be in a consistent format.

The Mixture of Experts (MoE) is a machine learning architecture that simplifies complex tasks by dividing them into smaller, more manageable subtasks handled by specialized models, known as "experts" \citep{jacob91moe}. Central to the MoE framework is a gating network that dynamically selects or combines predictions from these experts based on the input data. This modular approach facilitates scalability and adaptability, making MoE particularly well-suited for handling heterogeneous data and large-scale problems.

Over the years, MoE has found applications across various fields, including natural language processing \citep{shazeer2017moe} and LLMs. Notably, in models like GPT-4 \citep{openai2024gpt4technicalreport} and DeepSeek \citep{deepseekai2025deepseekv3technicalreport}, MoE has demonstrated substantial improvements in both performance and efficiency compared to monolithic architectures. Using expert specialization and dynamic routing, MoE enables these advanced models to achieve superior results while maintaining computational efficiency.

In this study, we explore various deep learning architectures to classify variable stars using the Kepler dataset. The best-performing models will be evaluated on simulated light curves from the GOTTA Pathfinder telescope catalogs to gauge their effectiveness and provide insights.

Our analysis begins with light curves selected from Kepler and K2 observations (Section \ref{kepler}). We focus on different types of Cepheid variables, including RR Lyrae (Section \ref{RR}), $\delta$ Sct (Section \ref{DSCT}), and $\gamma$ Dor (Section \ref{GDOR}) stars. Additionally, we include eclipsing binaries (EB) from the Kepler archive (Section \ref{EB}) and hybrid variables (HYB, Section \ref{HYB}) that exhibit multiple classifications in the literature. The construction of our training catalog is detailed in Section \ref{otherlabel}, followed by preprocessing steps for the light curves (Section \ref{sec2.8}).

To prepare the light curves for classification, we transform them into image representations using three algorithms: Continuous Wavelet Transform (CWT, Section \ref{cwt}), Gramian Angular Fields (GAF, Section \ref{gaf}), and Recurrence Plot (RP, Section \ref{rp}). These transformations are designed to make the light curves suitable for CNN architectures.

In Section \ref{models}, we describe the different CNN architectures used for classification. To further enhance classification accuracy, we employ a MoE model with a dynamic routing mechanism (Section \ref{sec:moe}). The results of these experiments are presented in Section \ref{result}. Finally, in Section \ref{furturework}, we discuss future directions, including the use of larger datasets, additional experts in MoE, and advanced automatic data processing methods applicable to light curve classification.

\section{Variable types and Data reduction} \label{data}

The causes of stellar variability can be either intrinsic or extrinsic. Variability classification includes categories such as rotation (ROT), eclipsing binaries, pulsating variables, eruptive variables, and other types \citep{sterken05,Understanding}.

Our study focuses specifically on eclipsing binaries and pulsating variables, including RR, $\delta$ Sct, and $\gamma$ Dor stars. Stars that exhibit characteristics of more than one class are labeled as HYB. Classical Cepheid variables are excluded from our analysis due to their limited sample size, which is insufficient for effective classification.

\subsection{\textit{Kepler} and \textit{K2 mission}}
\label{kepler}

The Kepler spacecraft was designed primarily to search for Earth-like exoplanets \citep{kepler10}. It hosted an optical telescope with a 95 cm aperture and a field of view of $115.6 deg^2$, delivering high-precision light curves for over 200,000 targets. For stars with V-band magnitudes between 13–14, the photometric precision reached 100 ppm, while for brighter stars (V = 9–10), it achieved up to 10 parts per million (ppm) \citep{Han_2021}. After the failure of two reaction wheels, the original mission ended and was followed by the K2 mission. K2 continued observations using the remaining reaction wheels and thrusters, focusing on stars near the ecliptic plane with similar photometric precision, albeit in shorter observation campaigns limited to approximately 80 days each.

Two types of light curves are provided by the mission: Simple Aperture Photometry (SAP) and Presearch Data Conditioning (PDC) light curves \citep{2016ksci.rept....1V}. While SAP light curves represent raw aperture-summed fluxes, PDC light curves are corrected for systematic errors by the Kepler Operations Science Center. In this study, we use only the PDC-corrected data.

The data are available at two cadences: long-cadence (30-minute sampling) and short-cadence (1-minute sampling) \citep{Han_2021}. Due to the limited availability of short-cadence data, we focus exclusively on long-cadence observations. Using the Kepler Data Integration Platform developed by \cite{2019ApJS..241...29Y}, we remove inter-quarter variations and convert absolute flux values into relative flux. For each light curve segment, we apply polynomial fitting to mitigate the impact of outliers, computing the mean of the fitted values. The light curves are then normalized to ppm by subtracting and dividing by the mean value, a procedure consistent with that used in \cite{Han_2021}.

\subsection{Eclipsing Binary} \label{EB}

Eclipsing binaries feature nearly edge-on orbits where the components eclipse each other, leading to light curve variations closely tied to stellar radii, masses, and luminosity ratios \citep{Understanding}. 

These binaries are classified into EA, EB, or EW types based on their light curve shapes. EA-type binaries display minimal variation between eclipses, whereas EB-type binaries show changes due to the ellipsoidal shapes of their components. EW-type binaries, characterized by contracting ellipsoidal components, exhibit quasi-sinusoidal modulations in their light curves. In this study, we do not focus on these sub-type classifications.

\subsection{RR Lyrae } \label{RR}

RR Lyrae stars are pulsating variables located at the extension of the Cepheid instability strip in the Hertzsprung–Russell (H-R) diagram. These stars serve as important standard candles for measuring cosmic distances. They are typically A5 to F5-type stars with absolute magnitudes of around 0.5 mag, masses close to $\sim1\textup{M}_\odot$, and low metallicities ([Fe/H] $\sim$ 0.0001 - 0.001 dex). These stars lie on the horizontal branch and exhibit pulsations with periods ranging from approximately 0.1 to 1 day and photometric amplitudes of 0.3–1.5 mag in the V-band \citep{Understanding}.

RR Lyrae stars have completed core hydrogen burning and are undergoing core helium burning and shell hydrogen burning. Their pulsations arise from periodic expansions and contractions: expansion is driven by radiative blocking, while contraction is caused by gravity. RR Lyrae stars are classified into several subtypes—RRa, RRb, RRc, and RRd—based on their pulsation modes and light curve shapes. RRa and RRb types have longer periods, with RRa showing larger amplitudes and more asymmetric light curves. RRc stars have shorter periods and more sinusoidal light curves. Each subtype corresponds to different pulsation modes.

\subsection{$\delta$ Scuti} \label{DSCT}

$\delta$ Sct stars, also known as dwarf Cepheids, are the most common type of Cepheid variables. They lie at the lower end of the Cepheid instability strip in the H-R diagram, intersecting the main sequence \citep{Understanding,dup05}. These stars are typically A5 to F8-type giants or main-sequence stars, with pulsation periods ranging from approximately 0.03 to 0.3 days. Most $\delta$ Sct stars are still undergoing core hydrogen burning.

Subtypes of $\delta$ Sct stars are classified based on pulsation amplitudes and metallicity, such as High-Amplitude $\delta$ Sct (HADS) and stars exhibiting multiple low-amplitude modes (MULT). These variables exhibit various pulsation modes, including pressure (p) modes, gravity (g) modes, and non-radial modes. Cooler $\delta$ Sct stars tend to pulsate in the fundamental radial mode. In this study, we do not distinguish between these subtypes.

\subsection{$\gamma$ Dor} \label{GDOR}

$\gamma$ Dors stars are F0 to F2-type stars located at the lower part of the $\delta$ Sct instability strip in the Hertzsprung–Russell (H-R) diagram \citep{Understanding}. They have masses ranging from 1.4 to 2.5 $M_{\odot}$ and effective temperatures between 6500K and 8500K. Compared to 
$\delta$ Sct stars, $\gamma$ Dors variables exhibit longer pulsation periods (0.3 to 3 days) and lower temperatures. These stars pulsate in non-radial g modes, which are driven by the convective blocking mechanism \citep{dup04}. Their pulsation behavior differs from that of $\delta$ Sct stars, which typically show p modes or mixed-mode pulsations \citep{dup05}.

\subsection{Hybrid} \label{HYB}
The label HYB in the input catalog includes both $\delta$ Sct/$\gamma$ Dor hybrid stars and objects classified into multiple variability types. For instance, some binary systems contain components that are $\delta$ Sct, $\gamma$ Dor, or ROT \citep{AJ149.68, sty1511}. Hybrid stars lie in the overlapping region of the $\delta$ Sct and $\gamma$ Dor instability strips in the H-R diagram. These stars exhibit characteristics of both types, with pulsation modes encompassing both high-frequency p modes and low-frequency g modes \citep{xiong15, xiong16, sanchez17}.

\subsection{Input catalog} \label{otherlabel}

The input catalog is compiled from multiple sources based on Kepler archive data. A summary of the catalog construction is provided in Table~\ref{inputcata}.

The General Catalog of Variable Stars (GCVS\footnote{\url{https://heasarc.gsfc.nasa.gov/W3Browse/all/gcvsegvars.html}}, \citealt{gcvs88,gcvs94,gcvs95,gcvs21}), maintained by the Sternberg Astronomical Institute and the Institute of Astronomy, lists $\delta$ Sct stars, 21 hybrid variables, and one classical Cepheid.

\citet{AA620} present a catalog of 1,391 RR Lyrae stars cross-matched with Gaia DR2, including 42 from Kepler and 1,349 from K2. The catalog's purity is estimated at $92\% \sim 98\%$ using automated identification tools. We include 42 RR Lyrae stars from the Kepler sample \citep{AA620}. This catalog also contains one classical Cepheid and one Type II Cepheid (T2CEP).

The Kepler Eclipsing Binary Star Catalog (KEB, \citealt{keb1,keb2,keb3,keb4,keb5,keb6,keb7,keb8}) includes 2,908 EBs identified from Kepler and K2 data. These classifications are generated using Eclipsing Binaries via Artificial Intelligence, a neural network that infers EB parameters from light curves.

\citet{AJ149.68} presents 82 $\delta$ Sct, 204 $\gamma$ Dor and 32 hybrid stars from \textit{Kepler} data. The HYB includes $\delta$ Sct/EB, $\delta$ Sct/$\gamma$ Dor and $\delta$ Sct/ROT hybrids. 

Finally, \citet{sty1511} provide a catalog of 1,307 $\delta$ Sct, 738 $\gamma$ Dor and 429 $\delta$ Sct/$\gamma$ Dor hybrids based on Gaia Data Release 2. Their classification relies on pulsation frequencies and amplitudes to study low-frequency variability mechanisms.

\begin{table}
\bc
\begin{minipage}[]{100mm}
\caption[]{Input catalog \label{inputcata}}
\end{minipage}
\setlength{\tabcolsep}{1pt}
\small
 \begin{tabular}{lcccccc}
\hline\hline \noalign{\smallskip}
 & $\delta$ Sct & EB & $\gamma$ Dor & RR & HYB & CEP \\
\noalign{\smallskip}
  \hline
  \noalign{\smallskip}
  GCVS & 6 & 0 & 0 & 0 & 21 & 1 \\
\citet{AJ149.68} & 82 & 0 & 204 & 0 & 32 & 0 \\
\citet{AA620} & 0 & 0 & 0 & 1,391 & 0 & 2 \\
\citet{sty1511} & 1,307 & 0 & 738 & 0 & 429 & 0 \\
KEB & 0 & 2,908 & 0 & 0 & 0 & 0 \\
  \hline
  \end{tabular}
\ec
\tablecomments{0.86\textwidth}{The input catalog. EB stands for eclipsing binary, not the variable type semi-detached eclipsing binary. RR stands for RR Lyrae, and HYB for all hybrid types. CEP contains both classic and type II Cepheids.}
\end{table}

\subsection{Pre-processing algorithm} \label{sec2.8}
\begin{figure*}
    \centering
    \includegraphics[width=0.9\textwidth]{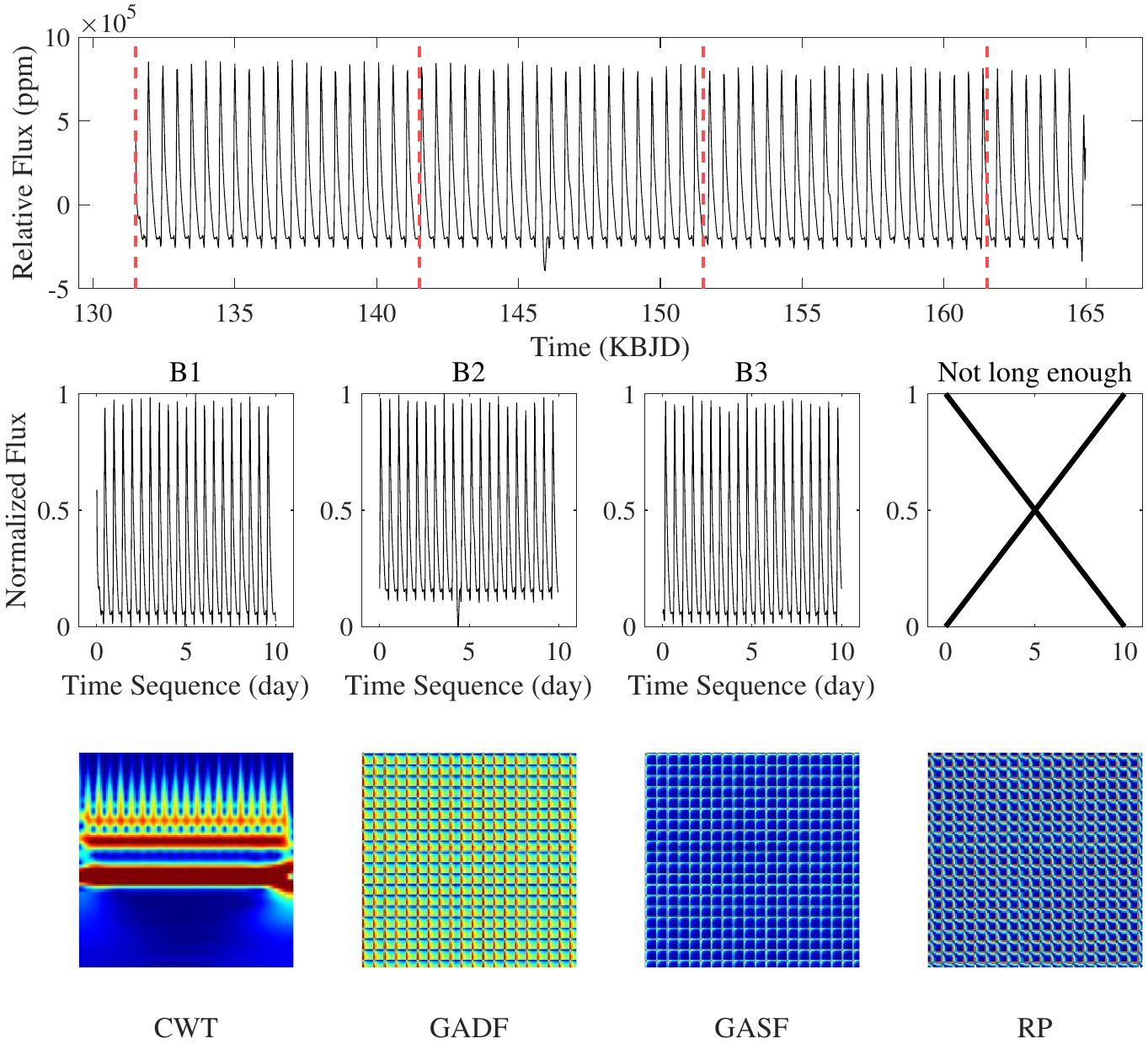}
    \caption{The transformation from light curves into images. The upper panel shows KIC 7176080 as an example. The middle panels are the partitions divided by the vertical red lines in the upper panel. The bottom panels show the output image of the four transformation methods. The CWT image is from B1. The GAF images are from B2, and the RP image is from B3.   \label{process1}}
\end{figure*}

A pre-processing step is constructed to manage and clean the light curves for our sample set (Figure~\ref{process1}, upper panel). This step helps to highlight features and increase the size of the training sample. We first remove one classical Cepheid (CEP) and one Type II Cepheid (T2CEP) to avoid an overly biased training set, as such bias can introduce significant uncertainty in the results \citep{shai14}. Stars labeled with multiple classes across different catalogs are treated as hybrids and assigned to the class HYB. This process enriches the HYB class with additional 'other' type objects.

To fully utilize each quarter of Kepler observations, we divide the data from each quarter into distinct segments, treating each segment as an individual sample in our training set. The continuous observational span within a single Kepler quarter, free of gaps, is approximately 20 days. We adopt a partition length of 10 days — half of this interval — resulting in $1 \sim 9$ partitions per quarter. These partitions correspond to the actual observation times within each quarter.

The Kepler observing cadence is approximately 30 minutes, although most quarters contain observation gaps between intervals. Light curves with gaps exceeding half of 0.25 days (about 6 hours) are excluded to avoid introducing interpolation errors, as the typical period of our target variables is around 0.5 days. The remaining partitions are then interpolated onto a uniform time grid with 0.02-day (about 30 min) intervals, similar to the GOTTA observation cadence. Afterward, each light curve is normalized to the range [0,1]. These pre-processed light curves serve as input for our transformation methods, which convert them into image representations (Figure~\ref{process1}, middle panels). Table~\ref{sampleset} summarizes the final sample sizes.

\begin{table}
\bc
\begin{minipage}[]{100mm}
\caption[]{Sample set \label{sampleset}}
\end{minipage}
\setlength{\tabcolsep}{1pt}
\small
\begin{tabular}{lcccc}
\hline\hline \noalign{\smallskip}
Label & Raw data & Sample object & Training sample  \\
\noalign{\smallskip}
  \hline
  \noalign{\smallskip}
$\delta$ Sct & 1,337 & 1,337  & 111,528 \\
EB & 2,821 & 2,821 & 226,937 \\
$\gamma$ Dor & 917 & 917  & 65,788 \\
HYB & 541 & 541 & 43,030 \\
RR & 1,391 & 1,373 & 9,306 \\
  \hline
  \end{tabular}
\ec
\tablecomments{0.86\textwidth}{The column raw data shows the number of stars inside the input catalog. The training sample is the final sample set after all pre-processing steps. The second column sample object shows the number of objects in the sample.}
\end{table}

\section{Transform light curves into images} \label{image}

Light curves are crucial for studying variable stars, as they encode information such as type, period, amplitude, and oscillation modes. However, extracting these parameters is computationally expensive, especially for large datasets. CNNs offer a powerful alternative by automatically learning robust deep features \citep{zhao17cnn, sadouk18cnn, tang20}. CNNs generally perform better on two-dimensional images than on one-dimensional time series. Therefore, we explore several transformation methods — including CWT, GAFs, and RP — to convert light curves into image formats. These images are then fed into pre-trained CNN models (Figure~\ref{process1}, bottom panels), after being resized to match the input dimensions.

\citet{szklenar22} used phase-folded light curve images for training, constructed by folding data from multiple epochs into a single period. While this approach may lose some period information, the authors incorporated the period as an additional input in a multi-input CNN framework. The resulting images may retain period-related patterns but are not guaranteed to preserve them explicitly.

\subsection{Continuous Wavelet transform} \label{cwt}
The Fourier transform (Equation \ref{fouriertrans}, where $i$ stands for the imaginary unit) is a powerful analytical tool widely used in numerical algorithms, analysis, differential equations, and time series analysis. In signal processing, it decomposes functions of space or time (the 'x' in Equation \ref{fouriertrans}) into functions of spatial or temporal frequency (the $\xi$ in Equation \ref{fouriertrans}). For variable star analysis, Fourier transform is a traditional method \citep{sterken05, Understanding}. Specifically, \citet{deb09cwt} applied Fourier decomposition to classify light curves of variable stars.

\begin{eqnarray}\label{fouriertrans} 
\hat{f}(\xi)= \int_{-\infty}^\infty f(x) e^{-2\pi ix\xi}dx 
\end{eqnarray}

However, the Fourier transform has its limitations. It cannot handle the signal whose frequency changes with time. It presents each frequency component rather than a time-frequency diagram. Fourier transform could not deal with a sudden change in frequency owing to its unattenuated basis on $\mathbb{R}$. 
\begin{eqnarray}\label{cwtformula}
\hat{f}(a,\tau)= \frac{1}{\sqrt{a}} \int_{-\infty}^\infty f(t) \Psi(\frac{t-\tau}{a})dt 
\end{eqnarray}

Wavelet transform, a Fourier-related transform, solves these problems by shifting and scaling of fast decaying wavelet basis (Equation \ref{cwtformula}). In the equation, $a\in \mathbb{R^+}$ is the scale (frequency), and $\tau \in \mathbb{R}$ is the translational value (time). The wavelet $\Psi$ belongs to both $L^1(\mathbb{R})$-space and $L^2(\mathbb{R})$-space. Further, it is identical in $L^2(\mathbb{R})$-space (Equation \ref{cwtformula2}). 
\begin{eqnarray} \label{cwtformula2}
\left \{
\begin{array}{l}
\int_{-\infty}^{\infty} |\Psi(t)|^2 dt =1 \\  
\int_{-\infty}^{\infty} |\Psi(t)| dt \textless  \infty \\
\int_{-\infty}^{\infty} \Psi(t) dt = 0
\end{array}
\right.
\end{eqnarray}

CWT shifts the translational value along the time of the signal. Shifting is done again when the scale enlarges to $2a$ (twice the scale). Then the equivalent frequency is reduced to a half, defined as an octave. The CWT also divides an octave into several voices, and the scale of n-voice per octave follows $a, 2^\frac{1}{n}a, 2^\frac{2}{n}a,..., 2^\frac{n-1}{n}a, 2a$ inside an octave. The more voice per octave, the finer the scale discretization. Wavelet transform can provide a scalogram, showing the relations among frequencies, times, and magnitudes. After the shifting and the enlargement of the scales, the CWT image is finally rescaled to a square matrix.

\begin{figure}
    \centering
    \includegraphics[width=0.5\textwidth]{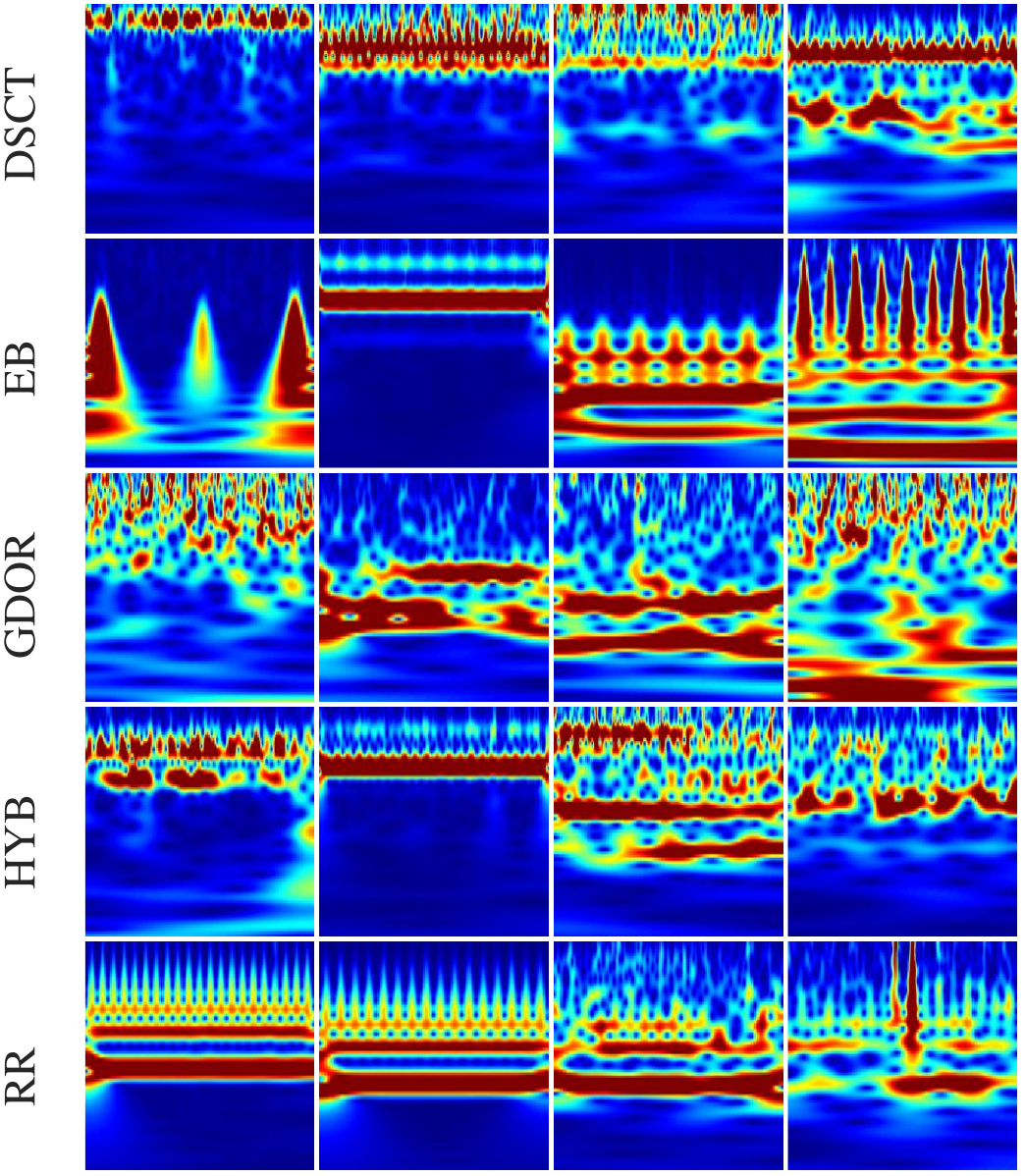}
    \caption{CWT images from different objects. See more images in Appendix \ref{applc}.}
    \label{figcwt}
\end{figure}

We choose Morse wavelet \footnote{\url{http://jmlilly.net/software}} \citep{olhede02cwt, lilly08cwt, lilly10cwt,lilly12cwt} with 12 voice per octave to transform our light curves into images, since Morse wavelet can be seen as a superfamily of analytic wavelets \citep{lilly12cwt}. The horizontal axis of the CWT scalogram shows the time scale, and the vertical axis shows the frequency. Some examples are shown in Figure \ref{figcwt} and Appendix \ref{applc}.

\subsection{Gramian Angular Field} \label{gaf}
We employed Gramian Angular Fields (GAFs) to transform light curves into image representations \citep{wang15gaf}, specifically using both Gramian Angular Difference Field (GADF) and Gramian Angular Summation Field (GASF). \citet{wang15gaf} compared GAFs with Markov Transition Fields (MTF) for time series representation in Tiled CNNs \citep{nigam11}, finding that MTF yielded lower classification accuracy across 20 datasets. This was later attributed by \citet{keo11} to the instability of the inverse transformation in MTF, a conclusion supported by \citet{ahmad21rp} through electrocardiogram classification experiments. Based on these findings, we adopted GAFs and excluded MTF from our analysis.

GAFs visualize time series by mapping them into polar coordinates and then forming an image matrix based on angular relationships between data points. The method involves first normalizing the time series using min-max scaling: $\tilde{x_i}=\frac{x_i-min(x)}{max(x)-min(x)}$,
where $x_i$ represents the i-th value in the equally spaced time series. The normalized values are then embedded onto the unit circle (Equation~\ref{eqgaf1}), enabling the calculation of pairwise cosine (for GASF) or sine (for GADF) relations across all time steps.

\begin{eqnarray} 
\label{eqgaf1}
\left \{
  \begin{array}{l}
     \phi_i = \arccos\tilde{x_i}  \\
     r_i = \frac{t_i}{n}
  \end{array}
  \right.
\end{eqnarray}

The $\phi$represents the angular coordinate, and $r_i$ denotes the corresponding radius. These are transformed based on either the sum of their cosine values (used in GASF) or the difference of their sine values (used in GADF). The mapping from the normalized time series ($\tilde{x_i} \in [0, 1]$) to the polar coordinate system is injective, ensuring that no information is lost during this transformation. The resulting GAFs are represented as $n\times n$ matrices (Equation~\ref{eqgaf2}), where each element encodes the angular relationship between two points in the time series.

\begin{eqnarray} \label{eqgaf2}
\begin{array}{l}
      GADF_{ij}= sin(\phi_i - \phi_j)\\
      GASF_{ij}= cos(\phi_i + \phi_j) \\
\end{array}
\end{eqnarray}
In these formulas, we show the (i, j) item in the GAF matrix. Back to the cartesian coordinates, they are equal to Equation \ref{eqgaf3}
\begin{eqnarray} \label{eqgaf3}
\begin{array}{l}
      GADF=\sqrt{I-\tilde{X}^2}^{\rm{T}} \cdot \tilde{X} - 
      \tilde{X}^{\rm{T}} \cdot \sqrt{I-\tilde{X}^2} \\
      GASF= \tilde{X}^{\rm{T}} \cdot \tilde{X} -
      \sqrt{I-\tilde{X}^2}^{\rm{T}} \cdot \sqrt{I-\tilde{X}^2} 
\end{array}
\end{eqnarray}
,where $\tilde{X}=(\tilde{x_1}, \tilde{x_2}, ..., \tilde{x_n})$ is a row vector, and the \textbf{T} stands for the transpose of the matrix.

Unlike methods based on CWT or Fourier transforms, GAFs emphasize the similarity between each pair of data points by expressing differences in sine or cosine formats. When viewed as a type of inner product, this transformation yields quasi-Gramian matrices. This explains why GASF and GADF use cosine and sine functions, respectively, to define their transformations \citep{wang15gaf}.

In analyzing these GAF matrices, each element $G_(i,j)$ reveals correlations through the superposition and disparity of their transformed values $|i-j|$, which can be interpreted as a form of temporal correlation. For light curves, periodic information is often manifested through repeated blocks within the GAF images, indicating primary maxima or sub-maxima. Variations in block colors may also highlight subtle light fluctuations across different periods.

\subsection{Recurrent Plot} \label{rp}

RP is a technique used for the nonlinear analysis and visualization of time series or dynamical systems. It has been applied across various domains, including the detection of chaotic attractors \citep{bradley02rp}. In the context of smart grid attack detection using CNN architectures, \citet{mostafa21gaf} compared RP with GAFs and found that the RP-CNN system achieved slightly better recall and $F_1$-score ($\sim 0.2\%$) than the GAF-CNN system. Both methods outperformed traditional classifiers such as Support Vector Machines, $k$-NN, and random forests. Additionally, \citet{ahmad21rp} conducted experiments on electrocardiogram classification using RP, GAF, and MTF with CNNs, finding similar performance between RP and GAF.

The RP is defined by:$$R_{i,j}= \Theta(\epsilon-\Vert x_i-x_j \Vert)$$
where $x_i$ and $x_j$ are states at different times, $\Vert \cdot \Vert$ denotes a norm, $\epsilon$ is a threshold, and $\Theta(x)$ is the Heaviside function, which equals 1 if $x \leq 0$ and 0 otherwise. We utilized an RP toolbox \citep{yang11rp, chen12rp} to transform light curves into images. This toolbox outputs $\epsilon$ as part of the recurrence plot to construct color images rather than monochrome ones.

RP evaluates recurrence in a non-linear time series by assessing the distance between each pair of states. The calculation formula allows us to determine this distance, along with trends, directly within the RP. This approach is more straightforward than those embedded within GAFs. Given its success in other fields, we evaluated the effectiveness of RP for light curve classification.

\section{Methodology} \label{models}
\subsection{Transfer learning}

\begin{figure*}
    \centering
    \includegraphics[width=0.8\textwidth]{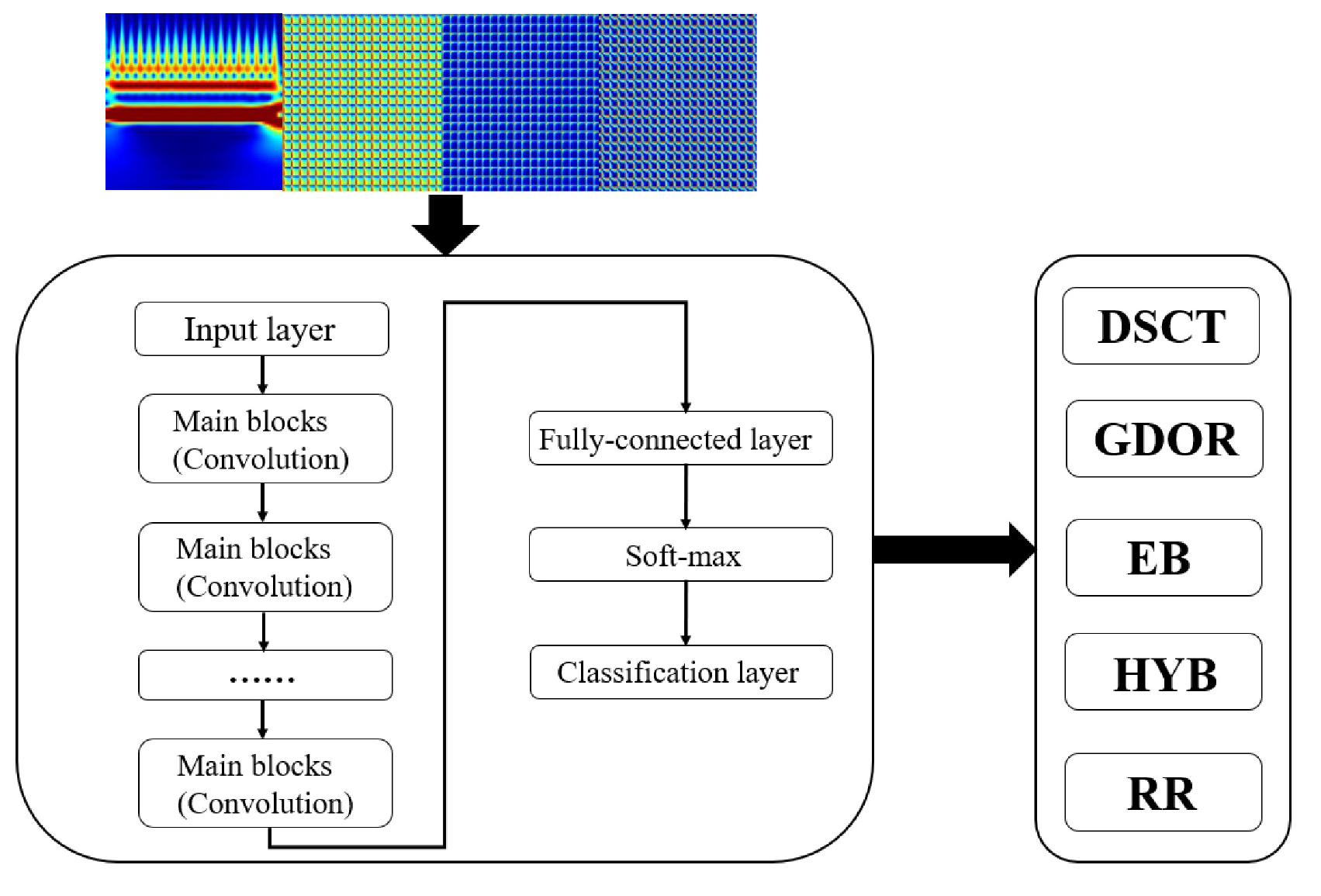}
    \caption{Summary flowchart of the transfer learning. The initial parameters at the left half panel in the main blocks are inherited from the origin models. The parameters in the last FC layer and the classification layer are new. The input images follow the order of CWT, GADF, GASF, and RP. \label{process2}}
\end{figure*}

We applied deep learning techniques for light curve classification using pre-trained CNN models. Initial parameters are sourced from ImageNet, while the FC and output layers are reinitialized to fit our specific task (Figure~\ref{process2}). Training is performed on GPUs, leveraging transfer learning to enhance feature detection. Each model underwent 30 epochs of training, with detailed configurations provided in Table~\ref{nets}.

\begin{figure*}
    \centering
    \includegraphics[width=0.8\textwidth]{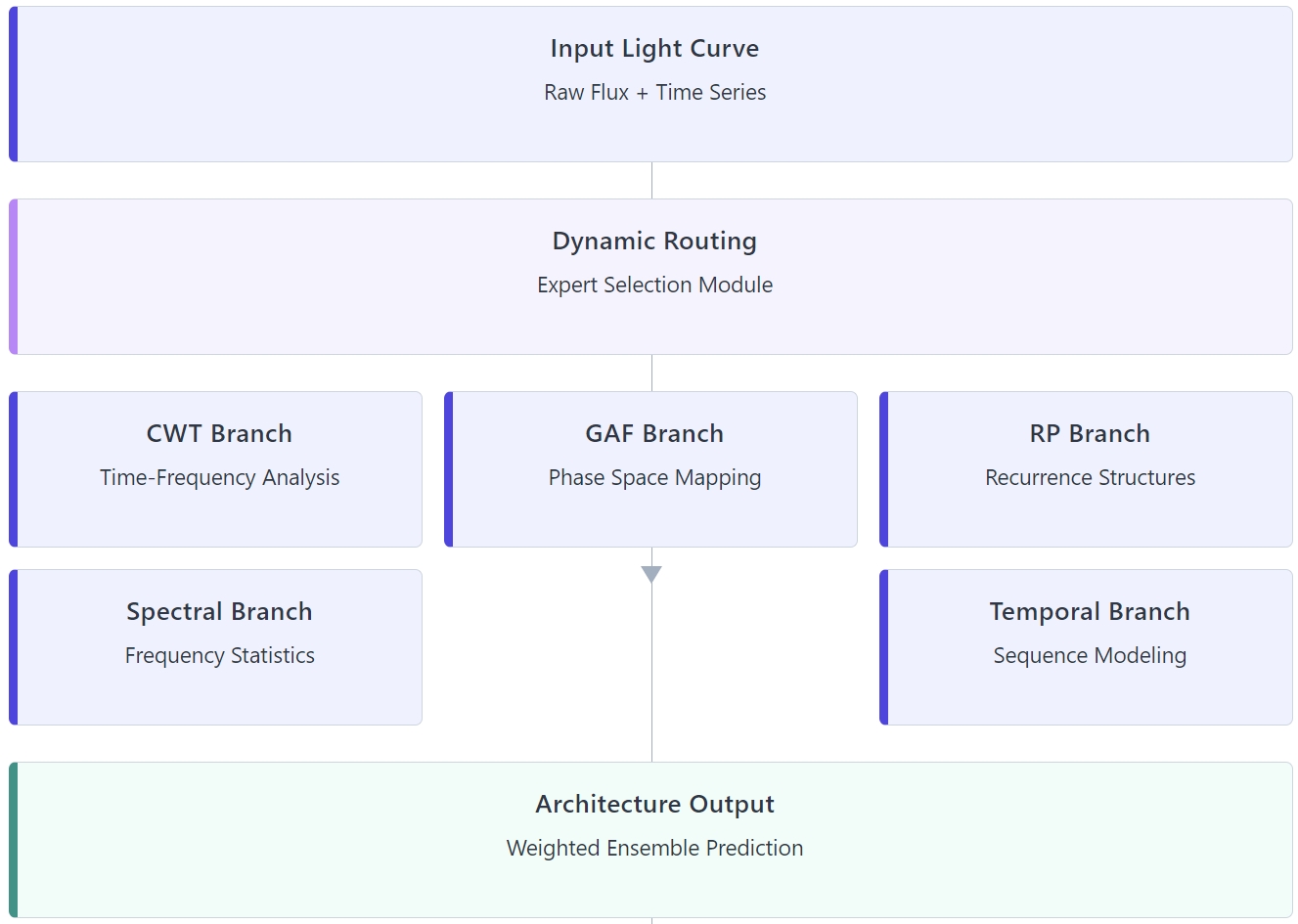}
    \caption{llustration of the LightCurve MOE, organized in a top-down pipeline: starting from raw light curve input, followed by a dynamic routing module that selects relevant experts. Five expert branches are CWT, GAF, RP, Spectral, and Temporal. The selected expert outputs are then fused to generate the final prediction. \label{process3}}
\end{figure*}

\begin{table}
\bc
\begin{minipage}[]{100mm}
\caption[]{Information of origin networks \label{nets}}
\end{minipage}
\setlength{\tabcolsep}{1pt}
\small
\begin{tabular}{lccccc}
\hline\hline \noalign{\smallskip}
Model & Stack & Depth & Size (MB) & Para & Input size  \\
\noalign{\smallskip}
  \hline
  \noalign{\smallskip}
VGG-19 & 16 & 19 & 535 & 144 & $224\times224$ \\
Inception-V3 & 11 & 48 & 89 & 23.9 & $299\times299$ \\
ResNet-101 & 33 & 101 & 167 & 44.6 & $224\times224$ \\
Xception & 11 & 71 & 85 & 22.9 & $299\times299$ \\
  \hline
  \end{tabular}
\ec
\tablecomments{0.86\textwidth}{The stack shows the number of stacks of the models' main idea, for example, Inception module for Inception-v3, separable convolution for Xception, residual block for ResNet and $3\times3$ convolution for VGG. The depth is defined as the largest number of convolution layers and FC layers of a path from input to output. The Para is the number of parameters, and counted in millions. }
\end{table}

\textbf{VGG-19}, developed by the Visual Geometry Group at Oxford, consists of 19 blocks with $3 \times 3$ convolutional layers and $2 \times 2$ max-pooling. Although effective, its three fully connected (FC) layers introduce redundancy. We adapted the final FC and output layers for our classification task and trained the model using stochastic gradient descent (SGD) with a momentum of 0.9 and an initial learning rate of $1 \times 10^{-4}$.

\textbf{Inception-v3} builds upon Inception-v1 by replacing large convolutional kernels with smaller ones and applying batch normalization to reduce computational cost while preserving accuracy. We modified the last two layers and used the same SGD optimizer settings.

\textbf{ResNet-101} addresses the degradation problem in deep networks through residual learning, enabling stable and efficient optimization. It uses residual and bottleneck blocks to mitigate issues such as vanishing gradients. We replaced the final two layers and applied SGD with the same hyperparameters.

\textbf{Xception} decouples feature mapping and combination in CNNs, achieving performance superior to Inception-v3 on ImageNet with a comparable number of parameters. The last two layers were reconfigured for our task, and the model was trained under identical conditions.

\subsection{LightCurve MoE} \label{sec:moe}

The MoE mechanism is originally proposed by \citet{jacob91moe}, aiming to enhance the overall model expressiveness and computational efficiency through introducing multiple sub-models (experts). Recently, as deep learning models have dramatically increased in scale, the MoE architecture has been widely adopted in large-scale pre-trained models to address the challenges of parameter explosion and excessive inference costs. By incorporating sparse activation mechanisms, MoE selectively activates only a subset of experts during each inference process, effectively reducing computational costs while maintaining high model capacity, which is the inspiration for our work.

As demonstrated in Switch Transformer research \citep{zoph2022stmoedesigningstabletransferable}, simplifying the routing mechanism to single-expert selection (Switch Routing) significantly reduces communication complexity and load balancing issues. On another front, \citet{riquelme2021moe} systematically analyzed design challenges of sparse MoE architectures in "Sparse Mixture of Experts Models" \citep{du2022glamefficientscalinglanguage}, including expert load imbalance and training instability. They proposed optimization strategies like gate mechanism improvement, expert regularization, and load balancing loss, which inspired our implementation of dynamic routing and dual-gating mechanisms.

To improve time complexity while reducing computational costs, we approach both temporal and computational dimensions: simplifying feature extraction steps temporally (\S~\ref{cwt}) and employing sparse MoE structures to reduce Floating Point Operations (FLOPs). This enhances robustness in identifying complex light curves. Building upon our previous CNN framework, we developed LightCurve MoE with five complementary experts. Through frequency-time dual-gating mechanisms, it dynamically selects Top-$k=3$ experts for forward propagation while skipping others, achieving superior trade-offs between accuracy and speed in the Large Language Model (LLM) realm.

\subsubsection{Physical Feature Expert Pool}
\begin{enumerate}
    \item \textbf{Time-Frequency Analysis Branch}:\\
    Apply Continuous Wavelet Transform to 500-length light curves, generating time-frequency maps that capture transient pulsation modes and frequency evolution patterns. Frozen backbone ensures temporal resolution preservation.

    \item \textbf{Phase Space Reconstruction Branch}:\\
    Construct Gramian Angular Fields (GASF/GADF) to encode periodic correlations and symmetry properties in phase space trajectories, effective for detecting orbital harmonics.

    \item \textbf{Dynamic Pattern Embedding Branch}:\\
    Generate recurrence plots to visualize high-dimensional dynamical system behaviors, revealing chaotic/stable oscillation patterns through diagonal structures.

    \item \textbf{Spectral Statistics Branch}:\\
    Calculate spectral density from CWT coefficients across scales, quantifying dominant frequency distributions and harmonic ratios in power spectra.

    \item \textbf{Temporal Robustness Branch}:\\
    Introduce 15\% random temporal masking during imaging to simulate observation gaps, enhancing robustness against irregular sampling and sudden flux anomalies.

\end{enumerate}

All branches inherit ImageNet-pretrained weights with fixed feature extractors to maintain training stability while focusing on physical characteristic learning.

\subsubsection{Frequency-Time Dual-Gating Mechanism}
We designed a Dual-Domain Gating mechanism evaluating expert outputs from spectral structure and temporal statistics perspectives, fused into final expert routing scores.

For each expert's embedding vector $\mathbf{v}_k$, a single-head self-attention module calculates the response score:
$$ s_k^{\text{(freq)}} = \text{Attention}(\mathbf{v}_k) $$
This captures whether spectral patterns exhibit typical structures, such as periodic edges or harmonic bands.

For original sequences $x$, compute the statistics ($\mu_k,\sigma_k,\mu_k^{(2)}$) and project them through a linear layer with the process:
$$ s_k^{\text{(time)}} = \mathbf{W}[\mu_k,\sigma_k,\mu_k^{(2)}] $$

Then combine the scores with learnable proportion $\alpha=\sigma(a)\in(0,1)$:
$$ s_k = \alpha s_k^{\text{(freq)}} + (1-\alpha)s_k^{\text{(time)}} $$
and normalize via a Softmax layer
$$ w_k = \frac{\exp(s_k)}{\sum_{i=1}^5 \exp(s_i)} $$

At last, the entropy regularization during training follows
$$ \mathcal{L}_{\text{total}} = \mathcal{L}_{\text{task}} + \lambda \cdot \text{Entropy}(w_k) $$.

\subsubsection{Top-$k$ Dynamic Sparse Routing}
Given normalized weights $w_k$, the model automatically select $k$ experts to reach the best reault with the function:
$$ \mathcal{S} = \arg\max_k \{w_1,w_2,...,w_5\} $$ 
Only experts in set $\mathcal{S}$ participate in computations, and these are called Top-$k$ experts. This sparse design activates 3 experts per sample, theoretically saving 40\% computation.

During training, non-active experts retain gate scores for softmax regularization. Entropy constraints prevent router concentration/division. Maintaining original objectives, this achieves remarkable efficiency improvements while preserving generalization capabilities - a key component of LightCurve MoE.

\section{Result} \label{result}

We applied the pre-trained models to classify the transformed images derived from light curves, and the resulting accuracies are summarized in Table~\ref{resultcata}. The MoE model was trained using Muxi, a GPU developed in China by MetaX Tech\footnote{\url{https://www.metax-tech.com/}}, while the other models were trained on NVIDIA RTX3090 GPUs. All transfer learning architectures were evaluated using a 5\% hold-out validation set.

\subsection{Imaging methods and architectures}
The average training accuracies for GADF, GASF, RP, and CWT are $90.5\%$, $90.03\%$, $91.0\%$, and $95.5\%$, respectively (Table~\ref{resultcata}). CWT shows clear advantages in the analysis of Kepler light curves due to its multi-resolution time-frequency analysis, which effectively captures transient events and multiscale features such as periodic modulations and noise separation. Other methods also perform well in specific cases: GAF encodes global periodic patterns through angular mappings, making it suitable for detecting symmetry and trend-related features, while RP reconstructs phase space to reveal deterministic or chaotic structures, showing sensitivity to complex nonlinear behaviors.

However, these methods have limitations when applied to Kepler data. For example, GAF may lose high-frequency details during dimensional compression, and RP suffers from low information density due to its binary image representation and sensitivity to parameter settings. Given that future surveys (e.g., LSST, GOTTA) will encompass a broader range of astronomical phenomena, a single feature extraction method is insufficient. Therefore, the expert combination architecture in LightCurve MoE becomes essential. By dynamically routing and fusing CWT-based time-frequency features, GAF-encoded periodicity, and RP-derived phase-space information, our model achieves multi-modal complementarity, enhancing generalization and robustness across diverse datasets.

Notably, when the Top-$k$ setting is set to 3, the MoE model stabilizes at approximately 96\% accuracy—substantially outperforming the average performance of individual models and significantly improving upon the VGG-19 baseline, while maintaining a smaller model size.

The integration of a dynamic routing mechanism is a key factor in improving the performance of the MoE model. This approach has been shown to be effective in various domains, as reported in prior studies such as those by \citet{kim16} and \citet{Christy22}, demonstrating its adaptability and robustness.

The average training accuracies highlight a clear trade-off between model scale and performance. VGG-19 achieves the highest accuracy (97\%) but at the expense of a large parameter count. In contrast, lightweight models such as BiLSTM + Attention (93\%) and Conv1D + Transformer (85\%) trade accuracy for compactness (See Table \ref{tab:model_comparison}, \citealt{Li_2025SW}). Notably, the proposed LightCurve MoE effectively bridges this gap, achieving 96\% accuracy while being up to 91\% smaller in size, with a minimal variant as small as 6 MB. This makes it significantly more efficient than conventional models while maintaining strong predictive performance.

\begin{table}
\bc
\begin{minipage}[]{100mm}
\caption[]{Model size and accuracy \label{resultcata}}
\end{minipage}
\setlength{\tabcolsep}{1pt}
\small
\begin{tabular}{lccccc}
\hline\hline \noalign{\smallskip}
Model & Size (MB) & CWT & GADF & GASF & RP   \\
\noalign{\smallskip}
  \hline
  \noalign{\smallskip}
VGG-19 & 495.0 & 97.2\% & 91.8\% & 91.4\% & 92.6\% \\
Inception-v3 & 77.6 & 94.5\% & 90.0\% & 89.4\% & 89.3\% \\
ResNet-101 & 151.0 & 95.7\% & 90.4\% & 90.3\% & 91.8\% \\
Xception & 74.2 & 94.5\% & 89.9\% & 89.0\% & 90.3\% \\
  \hline
  \end{tabular}
\ec
\tablecomments{0.86\textwidth}{Size and accuracy of all trained models. The column size shows the average model size of all four imaging methods, and the unit is in megabytes. The last four columns, 'CWT', 'GADF', 'GASF', and 'RP', show the accuracy of each imaging method.}
\end{table}

\begin{table}[htbp]
\centering
\caption{Model Performance and Scale Comparison}
\label{tab:model_comparison}
\setlength{\tabcolsep}{12pt} 
\renewcommand{\arraystretch}{1.2} 
\begin{tabular}{lrr} 
\hline
\textbf{Model}            & \textbf{Size (MB)}    & \textbf{Accuracy (\%)} \\
\hline
Conv1D + Transformer      & 2                     & 85                     \\
BiLSTM + Attention        & 5                     & 93                     \\
Inception-v3              & 78                    & 95                     \\
Xception                  & 75                    & 95                     \\
ResNet-101                & 151                   & 96                     \\
\textbf{LightCurve MoE}   & \textbf{6--44}        & \textbf{96}            \\
VGG-19                    & 495                   & 97                     \\
\hline
\end{tabular}
\end{table}

\subsection{Confusion matrix and classification analysis}

The confusion matrices are presented in Appendix~\ref{appcm}. True positive (TP) refers to instances correctly predicted as belonging to a given class; false positives (FP) denote stars incorrectly predicted to belong to that class, and false negatives (FN) refer to those wrongly excluded. Recall is defined as $\rm recall = \frac{TP}{TP+ FN}$, and precision as $\rm precision = \frac{TP}{TP+ FP}$. These metrics reflect the model’s classification capability.

Among all classes, the hybrid (HYB) type achieves the lowest recall, likely due to its mixed feature composition that overlaps with multiple variable star types. The FN errors are primarily misclassified as $\delta$ Sct, suggesting that $\delta$ Sct features dominate in some HYB cases. Some $\delta$ Sct stars are also misclassified as $\delta$ Sct/$\gamma$ Dor hybrids. According to \citet{sty1511}, nearly all $\delta$ Sct stars exhibit hybrid behavior, often showing low-frequency peaks in their periodograms—indicative of binarity or rotation. This overlapping structure affects the recall performance of both $\delta$ Sct and $\gamma$ Dor.

In contrast, RR Lyrae stars achieve the best classification performance, attributed to their high data purity and clear light curve characteristics.

\section{Discussion} \label{discuss}

\subsection{Imaging representation ability} \label{imagerepresent}

In our study, we evaluate four methods for transforming light curves into images. The CWT demonstrates superior representation ability compared to GAFs, RPs. This finding aligns with the established principle that machine learning performance is closely tied to feature selection \citep{shai14}.

CWT's advantage stems from its capability to represent time-dependent frequency features, which are not captured by GAFs or RPs. By leveraging Fourier transform principles, CWT effectively visualizes signals in the time-frequency domain, illustrating how light curves evolve over time \citep{Understanding, sterken05}. This makes CWT particularly adept at isolating different pulsation modes, which are crucial for accurate classification.

Conversely, GAFs and RPs convert time series data into symmetric matrices, leading to redundancy in half of the generated images. Moreover, these methods fail to extract periodic components from light curves, whereas CWT excels at distinguishing various pulsation modes. The ability to isolate such modes is essential for distinguishing between different types of variable stars.

In summary, while GAFs and RPs can be effective in certain scenarios—such as encoding global periodic patterns and reconstructing phase spaces—they lack the nuanced detail provided by CWT. Our results underscore the importance of selecting appropriate feature extraction methods tailored to the specific characteristics of astronomical datasets, particularly when dealing with complex and multi-scale phenomena like variable star light curves.

\subsection{Experiments}
\subsubsection{Simple 1-D CNN}
We also evaluate a 1-D CNN architecture that operates directly on preprocessed light curves, bypassing image conversion. The model, based on a ResNet structure with 15 residual blocks and a total of 36.5 million parameters (considerably fewer than ResNet-101), is trained for 150 epochs with an initial learning rate of 0.001, decaying by a factor of 0.2 every 1,000 iterations. Under 15\% holdout validation, the model achieves an accuracy of 94.8\%. However, training time is significantly longer than that required for image-based feature extraction and transfer learning.

While this 1-D CNN outperform GAF and RP methods, its accuracy remains lower than that obtain using CWT-based image representations. These results suggest that CWT may provide more effective feature encoding for CNN-based classification of light curves. In contrast, GAF and RP appear to be less suitable for capturing discriminative temporal patterns. Furthermore, treating light curves as dynamical systems through these methods may be less effective than Fourier-based approaches such as CWT.

\subsubsection{Single expert test}

\begin{figure*}
    \centering
    \includegraphics[width=0.8\textwidth]{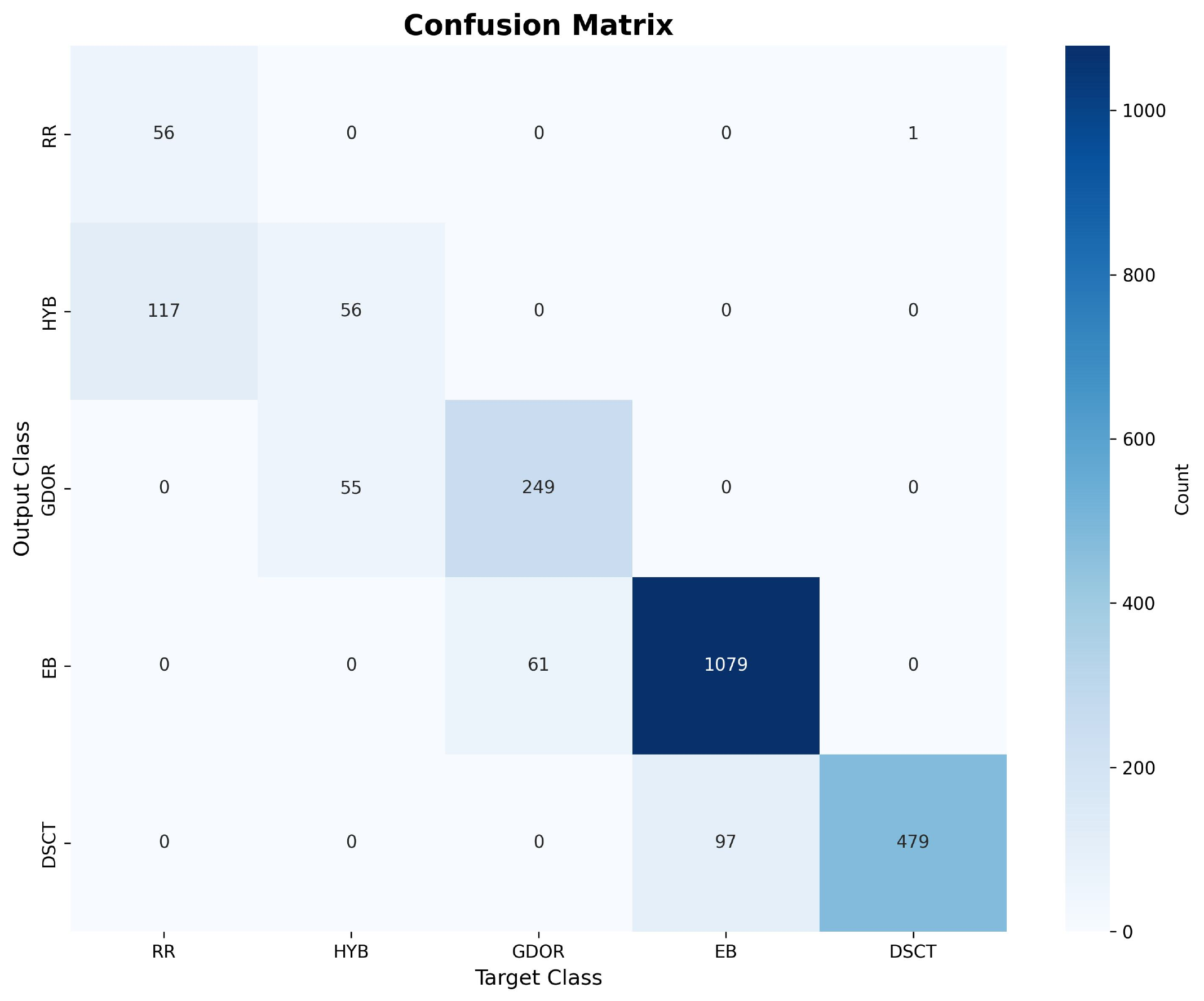}
    \caption{The confusion matrix of sending sequence data to CWT expert. \label{fig:MoE_CWT}}
\end{figure*}

In the MoE model, multiple experts process and make predictions on data that align with their respective input characteristics and capabilities. The final output is then aggregated by the MoE layer through Top-$k$ dynamic sparse routing. To evaluate its generalization ability, we test the model by feeding it with unfamiliar sequence data. Specifically, we route the input sequences to the CWT expert and achieved an accuracy of 85.29\%, as shown in the confusion matrix in Figure~\ref{fig:MoE_CWT}. Notably, the classification performance for the HYB class is poor, with many instances misclassified as RR. This shows its poor ability, given that the proportion of RR-hybrid samples in our dataset is relatively low, as most hybrid samples are of the $\delta$ SCT / $\gamma$ GDOR type.

This experiment demonstrates that the MoE architecture can achieve high accuracy across a diverse range of input types compared to a single model. When presented with data that deviates from an expert's specialty, the prediction accuracy naturally declines, making the expert a weak learner \citep{shai14}. However, the dynamic routing mechanism enables the MoE layer to effectively select and combine the most suitable expert predictions, thereby enhancing accuracy through boosting. As a result, the overall model maintains strong performance while being capable of handling a wider variety of input sequences compared to traditional methods.

\subsubsection{Comprison with \citet{Li_2025SW}}

As reported in \cite{Li_2025SW}, various models were evaluated on the \textit{Kepler} lightcurve prediction task using different types of input modalities, including time series, CWT images, textual descriptions, audio signals, and lightcurve images. The Lightcurve MoE achieves accuracy comparable to the Multimodal Large Language Model (MLLM)-based approach, slightly underperforms compared to models utilizing CWT image inputs (e.g., SwinTransformer and EfficientNet), but outperforms all other architectures, including 1D Convolution + Transformer, LightGBM, BiLSTM + Attention, 1D Convolution + BiGRU, 1D Convolution + BiLSTM, LLM-based models, and large audio language models. Looking ahead, as the development of LLMs and specialized expert architectures continues to advance, the potential of MoE models as end-to-end lightcurve classification systems becomes increasingly promising.

\subsubsection{Data augmentation}
We conduct experiments on augmented data to evaluate the impact of training with biased versus balanced datasets. In this test, the underrepresented classes RR, HYB, and $\gamma$ Dor are augmented to match the number of $\delta$ Sct samples in the training set. The augmentation pipeline first determines the required number of synthetic samples for each class and then randomly selects existing lightcurves uniformly as seeds for generation. Each seed is processed using one of two augmentation methods, selected with an 8:2 ratio: (1) adding zero-mean Gaussian noise with a standard deviation ranging from 0.05 to 0.2, or (2) smoothing the lightcurve using a Gaussian filter with $\sigma$ values between 0.3 and 0.8. All 'random' in this paragraph refers to a uniform distribution.

\begin{figure*}
    \centering
    \includegraphics[width=0.8\textwidth]{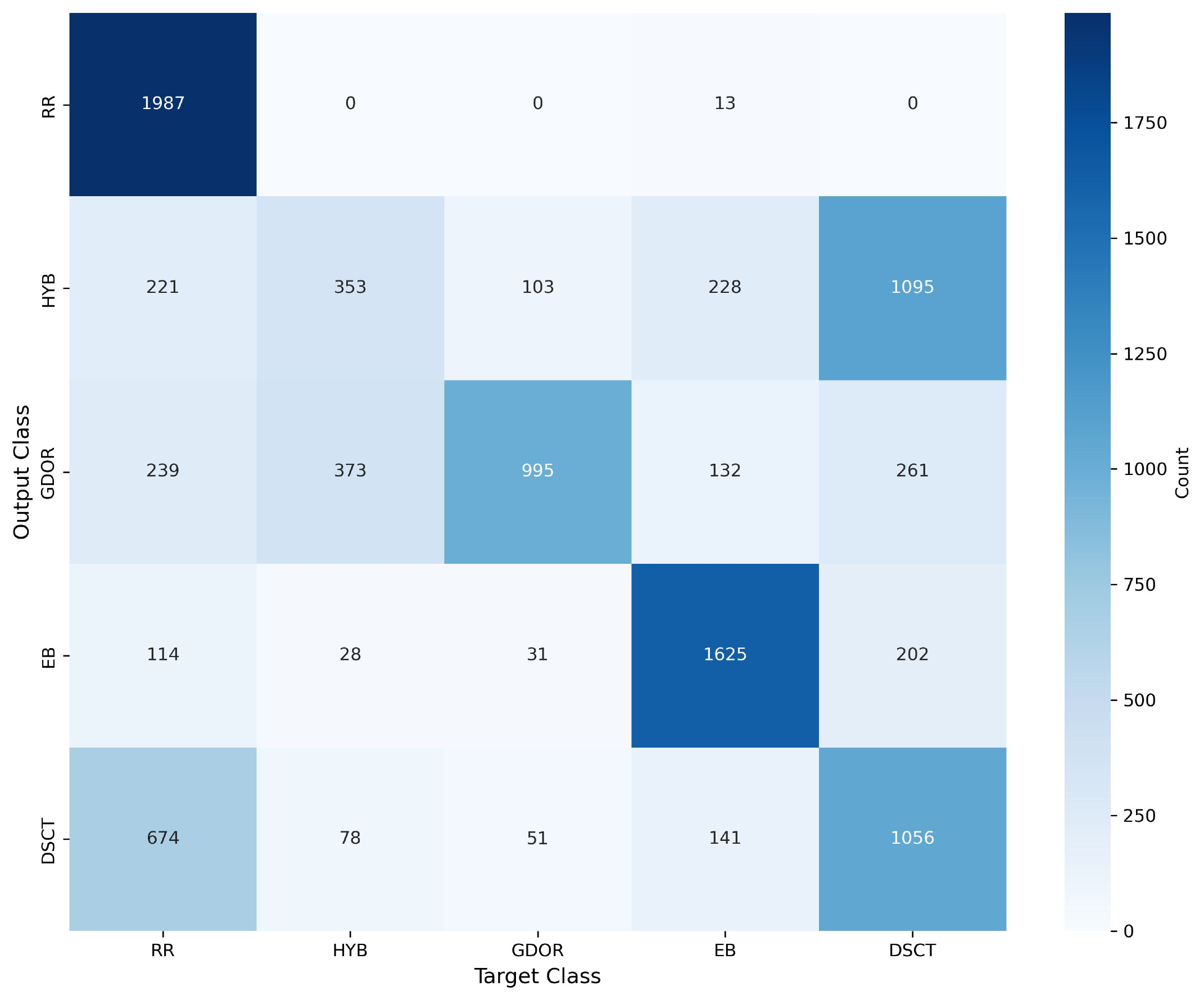}
    \caption{The confusion matrix in the data augmentation test.}
    \label{fig:augmoe}
\end{figure*}

However, the MoE model trained on the augmented dataset achieves only about 60\% accuracy (Figure \ref{fig:augmoe}), significantly lower than the model trained on the original, imbalanced data. This suggests that adding noise or Gaussian smoothing as the data augmentation techniques may not be suitable for lightcurve classification. Lightcurves encode precise physical features — such as pulsation modes, periodicity, and phase coherence. Adding noise or applying smoothing distorts these critical structures, especially in classes like $\delta$ Sct, $\gamma$ Dor, and the hybrids, where subtle temporal patterns are key to accurate classification. As a result, augmentation disrupts feature correspondence and degrades model performance.

Furthermore, several commonly used data augmentation techniques are not considered due to the specific nature of astronomical time series. First, sequence-level transformations like padding or interpolation would alter the fixed 10-day input length we have chosen for consistency, potentially disrupting periodic patterns crucial for classification. Second, downsampling or modifying the observational cadence was avoided, as it could amplify irregularities between observations and introduce instability into the temporal structure. Given that even mild noise augmentation led to performance degradation, we did not explore additional manipulations that might further compromise signal integrity. Finally, amplitude scaling or shifting is deemed ineffective due to prior normalization of flux values across all lightcurves, rendering such transformations neutralized before model input.

As shown in \cite{Li_2025SW}, our dataset has been previously used with promising validation results, further indicating that the intrinsic quality and representativeness of the data may be more critical than balancing class distributions through augmentation.

\subsection{Physical Interpretability and Computational Timecost}
\label{sec:physical_interpretability}

The LightCurve MoE architecture integrates domain-specific physical modeling with computational optimization through its five expert modules:

\begin{enumerate}
    \item \textbf{CWT Branch (Time-Frequency Analysis)}  
    Designed to resolve frequency-domain degeneracy between high-frequency $\delta$ Sct stars ($n=111,\!000$) and low-frequency $\gamma$ Dor variables ($n=65,\!000$). By isolating p-modes ($>20\ d^{-1}$) from g-modes ($<5\ d^{-1}$), this expert eliminates manual Lomb-Scargle harmonic thresholding while addressing the 13:1 sample imbalance in training data. This capability directly enables the feature engineering decentralization described in \citet{lilly12cwt}.

    \item \textbf{GAF Branch (Phase Space Mapping)}  
    Utilizes Gramian Angular Field representations for capturing nonlinear dynamics, particularly effective for RR Lyrae classification ($n=9,\!000$). Phase-space trajectories encode period-luminosity relationships that enable robust detection despite underrepresentation in training samples.

    \item \textbf{RP Branch (Recurrence Structures)}  
    Optimized for eclipsing binaries ($n=227,\!000$) through orbital harmonic reconstruction. This module quantifies eccentricity-dependent light curve asymmetry without traditional period folding pipelines, directly informing telescope scheduling algorithms through transient chaos detection.

    \item \textbf{Spectral Branch (Frequency Statistics)}  
    Implements gradient-preserving mechanisms to handle hybrid variable classification ($n=43,\!000$) where overlapping signals from competing processes (acoustic waves vs convection blocking) create observational degeneracies. Continuous routing weights transform discrete HYB labels into mixed-mode metrics (e.g., CWT:GAF:RP = 0.4:0.3:0.3).

    \item \textbf{Temporal Branch (Sequence Modeling)}  
    Specializes in transient phenomena detection (stellar flares, irregular variables) using recurrent architectures, complementing spectral/phase-based analyses.
\end{enumerate}

\textbf{Dynamic Routing Mechanism}:  
Our dual-gating system adaptively prioritizes experts based on input characteristics:
    
- $\delta$ Sct activation emphasizes CWT branch (high-frequency emphasis)
    
- EB systems trigger RP dominance (orbital harmonic detection)
    
- RR Lyrae engage GAF/Spectral synergy (periodicity + statistics)
    
- HYB variables receive balanced multi-expert activation

This architecture achieves both physical interpretability and computational efficiency through entropy-regularized routing that prevents over-reliance on any single feature space while maintaining robustness for rare classes like HYB. The Top-$k$ sparse routing strategy ($k=3$) reduces computation costs by 40\% compared to dense ensembles (Table~\ref{resultcata}) while maintaining 96\% accuracy - critical for real-time processing of GOTTA's 40 GB/min data stream on edge computing nodes.

The framework's adaptive resource allocation further enables operational flexibility: While $k=3$ provides optimal accuracy-efficiency trade-off, reducing $k=2$ maintains $>$90\% accuracy for simpler classifications. This design outperforms conventional CNNs like Xception (74.2 MB, 8$\times$ memory overhead) and ResNet-101 (151 MB, 1.5$\times$ cost for 1\% accuracy gain), demonstrating superior scaling behavior.

Comparative experiments validate this approach: VGG-19 and ResNet-50 failed to converge within 30 epochs when trained from scratch, while our architecture achieves SqueezeNet-level efficiency (94\% validation accuracy in 50 epochs) while embedding domain-specific physics into its computational core. This contrasts sharply with \citet{szklenar22} folded image CNN approach requiring 3.75s per 10,000 objects despite simplified inputs. 

The prediction speed of a single expert in the MoE model is approximately 20 objects per second, while using all experts together in Lightcurve MoE results in a throughput of about 5 objects per second. In comparison, other LLM-based lightcurve classification models \citep{Li_2025SW} achieve speeds ranging from 2 to 5 objects per second. Although this inference speed is significantly slower than that of CNN-based architectures, Lightcurve MoE offers an end-to-end solution that eliminates the need for lightcurve preprocessing — a major advantage in real-world deployment scenarios.

\begin{table}
\bc
\begin{minipage}[]{100mm}
\caption[]{Prediction time cost \label{timecost}}
\end{minipage}
\setlength{\tabcolsep}{1pt}
\small
\begin{tabular}{lccccc}
\hline\hline \noalign{\smallskip}
Model &  CWT & GADF & GASF & RP   \\
\noalign{\smallskip}
  \hline
  \noalign{\smallskip}
VGG-19 & 25.3 & 26.0 & 25.3 & 25.3 \\
Inception-v3 & 19.9 & 19.9 & 20.5 & 20.3 \\
ResNet-101 & 11.5 & 12.4 & 11.5 & 11.6\\
Xception & 40.9 & 40.9 & 41.2 & 41.4 \\
  \hline
  \end{tabular}
\ec
\tablecomments{0.86\textwidth}{The prediction time cost of each model. The unit is seconds per 10,000 light curves.}
\end{table}

\subsection{Future work} \label{furturework}

\subsubsection{More data}
The application potential of the MoE model and its dynamic routing mechanisms in large-scale astronomical datasets remains to be fully explored, especially given the vast amounts of data generated by projects like the LSST and the GOTTA. Future work will leverage these extensive data resources to further validate the classification accuracy of deep learning models.

Our objectives extend beyond merely enhancing model accuracy. We aim to strike a balance between training time and inference efficiency. This involves optimizing both the computational cost during training and the speed and resource requirements during real-time predictions. By doing so, we seek to develop more scalable and practical solutions for the automated classification of variable stars and transient events in future astronomical surveys.

\subsubsection{Temporal Dynamics Modeling}

Building upon the current frequency-domain capabilities, future iterations will enhance temporal sensitivity through dedicated sequence modeling modules.

Proposed experts include:

- \textbf{Long-cadence dependency bench} to resolve evolutionary patterns in variable stars over multiple pulsation cycles or orbital periods.
    
- \textbf{Mid-range sequential correlation bench} for transient event detection, such as eclipsing binary ingress/egress timing diagnostics.
    
- \textbf{Adaptive multi-scale fusion bench} that dynamically balances short-term fluctuations with long-term trends based on individual light curve morphology.

These temporal modeling capabilities will form a natural complement to existing frequency-domain experts by encoding progression dynamics inaccessible through static spectral analysis alone. For example, while the CWT branch isolates p-mode/g-mode frequency separation, the proposed temporal experts could reveal amplitude modulation patterns indicative of mode coupling or non-linear resonance effects.

A phased integration strategy is planned to ensure compatibility with existing systems and expertise: initial validation will be conducted using synthetic datasets to benchmark temporal sensitivity across known variability classes, followed by a gradual incorporation into the routing mechanism once stability has been demonstrated in controlled environments.

\subsubsection{Automatic data process}

With the emergence of LLM-based automatic data processors, it is now possible to leverage LLMs to decompose user-defined processing requirements into multiple sub-tasks. Each sub-task is then executed sequentially. In this framework, each sub-task is further divided into three components: an LLM-driven code generation module, a code interpreter for execution, and an LLM-based evaluation unit responsible for result validation. Upon successful completion of a sub-task, the output is passed to the subsequent sub-task; if errors occur, the system revises the process iteratively.

The automated data processor can interact with the local environment, supporting operations such as code execution, data loading, and writing. It also benefits from computational efficiency on local servers since LLMs can be accessed via API keys from cloud services.

However, there is a practical limitation to the volume of data that the system can handle effectively—approximately 200MB. While larger volumes may technically be processed, the robustness of the system tends to degrade over time due to increasing complexities in maintaining coherence between the generated code, the local environment, and the LLM itself.

To evaluate performance, we conducted tests using a randomly selected small sample set containing all 9,306 RRstars, along with 20,000 samples each for $\delta$ Sct, $\gamma$ Dor, EB and HYB. The single Type2CEP is also included. The prompt used for this experiment is detailed in Prompt \ref{prompt:auto}.

\begin{lstlisting}[caption={},label={prompt:auto}]
The input file is a light curve located at '''{The csv path.} '''. The 'Label' column corresponds to the class, and the rest of the columns represent light curves sampled at 0.02 day intervals over a total span of 10 days, which can be considered as time series data. I need you to classify these light curves using either a CNN or an RNN algorithm. 

Please ensure the following are saved: 
(1) Training and testing datasets, including both the light curve inputs and their corresponding class labels. 
(2) Model files. 
(3) Confusion matrix. 
(4) The impact (importance) of each input feature. 
(5) A README file explaining the purpose of each file. 
(6) Any other content you deem important. 

Save location should be  '''{The save location} '''. 

When encountering issues running the task, if it involves a sub-task, please rewrite the code for the sub-task instead of reverting to re-plan the entire task. 
\end{lstlisting}

\begin{figure*}
    \centering
    \includegraphics[width=0.8\textwidth]{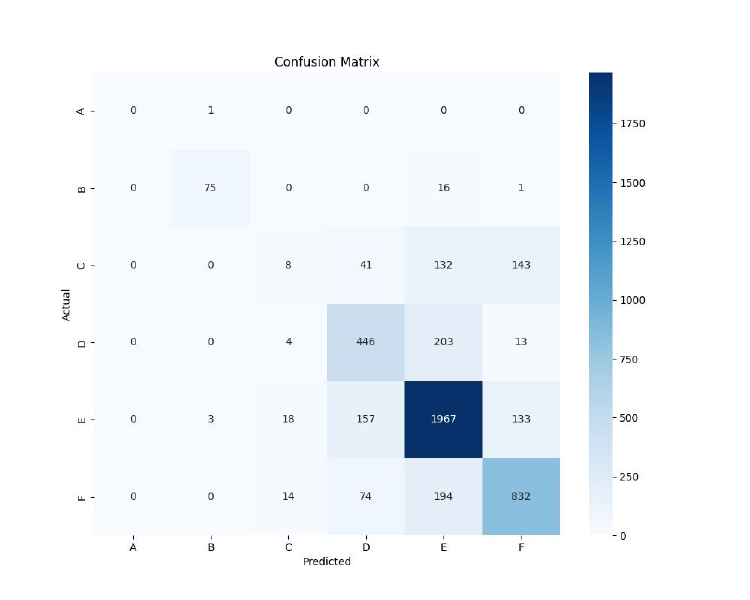}
    \caption{The confusion matrix given by the automatic data processing pipeline. The accuracy is 74.37\%. Remark that the A to F types are T2CEP, HYB, RR, GDOR, EB, and DSCT.}
    \label{fig:autocnn}
\end{figure*}

The data processing was carried out using the Qwen-Max model, integrated with the assistance of the ModelScope Agent framework\footnote{\url{https://github.com/modelscope/modelscope-agent}}. The overall classification accuracy achieved was 74.37\%. A confusion matrix illustrating the performance of the classifier is presented in Figure \ref{fig:autocnn}.

Furthermore, seven features were identified as having an importance greater than 99\%, with values of 5.6, 2.48, 9.6, 6.1, 1.32, 4.04, and 1.22, expressed in units of days and with the higher importance to the lower. These key features provide insight into the most significant temporal characteristics influencing the classification outcomes.

In future work, we aim to enhance the automated data processing pipeline by supporting additional data modalities and enabling more convenient and scalable input mechanisms. The goal is to achieve higher classification accuracy and produce more robust, reliable results for large-scale astronomical datasets.

\subsubsection{Data pipelines}
Looking ahead, we envision our MoE-based framework as a foundational component for next-generation time-domain astronomy pipelines. As an end-to-end model that directly consumes raw lightcurves, it is well-suited for integration into fully automated time-domain surveys such as the upcoming GOTTA.

In current survey pipelines, tools such as SExtractor and HotPants already demand substantial computational resources for image subtraction and photometry. Incorporating additional preprocessing steps required by transfer learning models can further increase the CPU load. Our MoE approach eliminates this burden by operating directly on raw lightcurves, thereby avoiding the need for manual feature extraction or normalization.

We also consider future hybrid architectures in which LLM-based agents dynamically schedule specialized models — such as transfer learning CNNs or other LLM-based models — as tools to handle specific tasks. While conceptually similar to MoE in its use of expert selection, this LLM-driven approach relies on intent recognition and semantic understanding rather than data-driven routing. It offers greater interpretability and flexibility for complex workflows, but at the cost of higher computational overhead and time cost.

In summary, while LLMs may enhance interpretability and task orchestration, lightweight, end-to-end models like MoE will remain essential for scalable and efficient real-time classification in time-domain astronomy.

\begin{acknowledgements}
This work was supported by the National Programs on Key Research and Development Project (grant No.2019YFA0405504 and 2019YFA0405000) and the National Natural Science Foundation of China (NSFC) through grants NSFC-11988101/11973054/11933004. Strategic Priority Program of the Chinese Academy of Sciences under grant number XDB41000000. We acknowledge the science research grants from the China Manned Space Project with NO.CMS-CSST-2021-B07. We also acknowledge the support from National Astronomical Observatories Chinese Academy of Sciences No. E4TG2001.
JFL acknowledges support from the New Cornerstone Science Foundation through the NewCornerstone Investigator Program and the XPLORER PRIZE

The paper includes data collected by the Kepler mission. Funding for the Kepler mission is provided by the NASA Science Mission Directorate. All of the data presented in this paper were obtained from the Mikulski Archive for Space Telescopes (MAST). STScI is operated by the Association of Universities for Research in Astronomy, Inc., under NASA contract NAS5-26555. Support for MAST for non-HST data is provided by the NASA Office of Space Science via grant NNX09AF08G and by other grants and contracts.
\end{acknowledgements}

\bibliographystyle{raa}
\bibliography{ms}

\appendix                  

\section{Light curves} 
\label{applc}

\begin{figure*}
    \centering
    \includegraphics[width=0.5\textwidth]{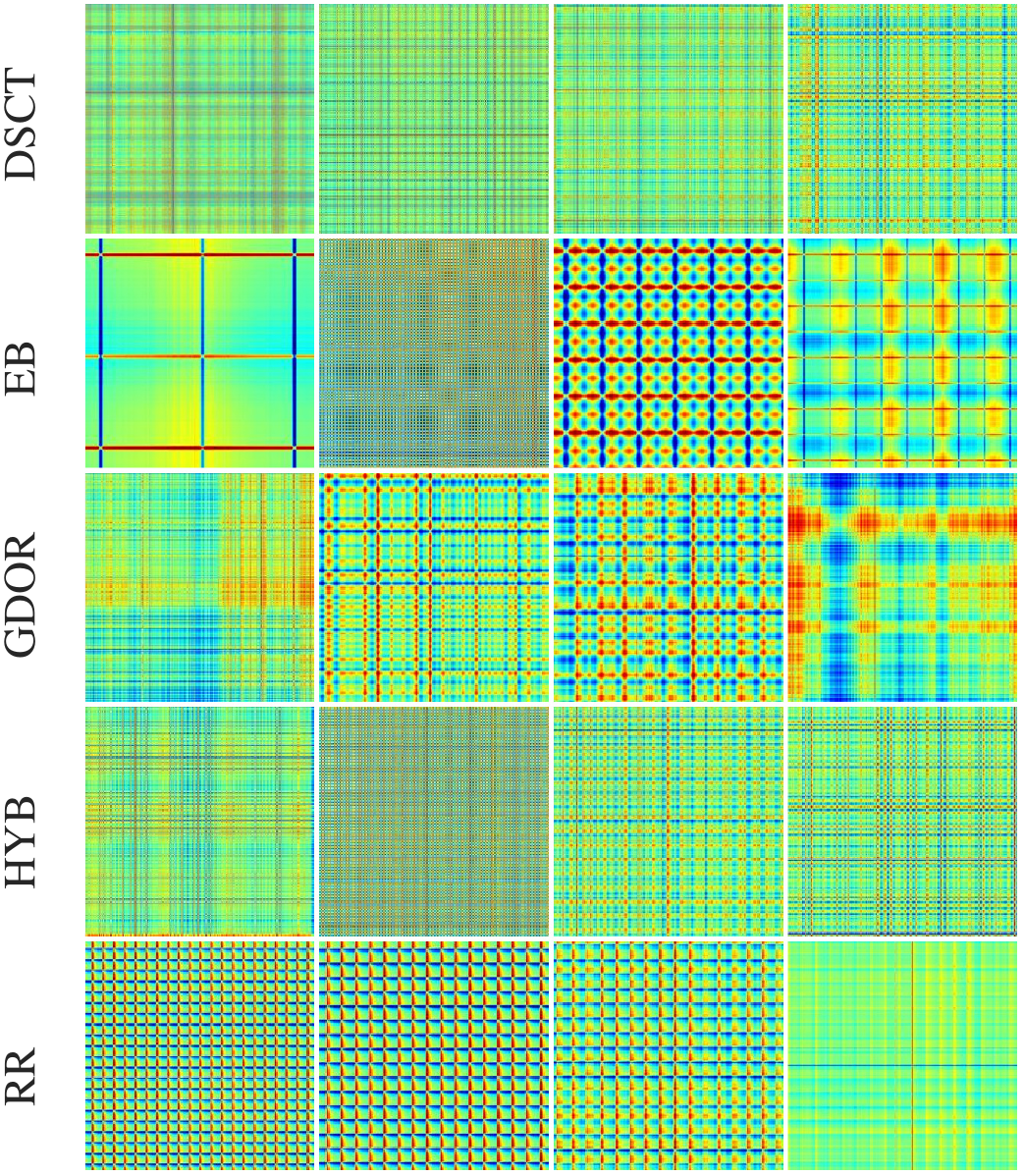}
    \caption{GADF images. Each object in this figure are the same as the object in Figure \ref{figcwt}.}
    \label{figgadf}
\end{figure*}

\begin{figure*}
    \centering
    \includegraphics[width=0.5\textwidth]{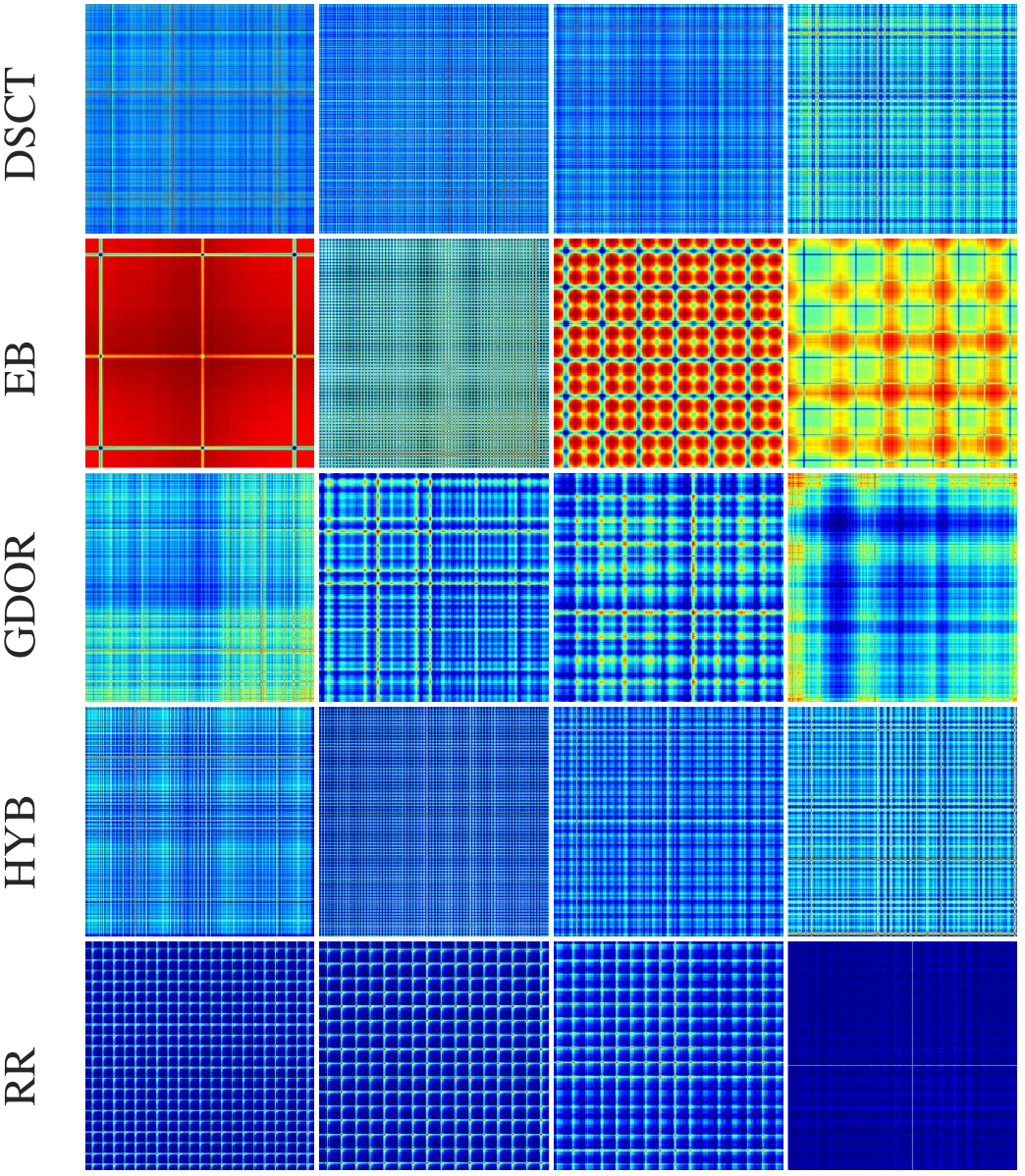}
    \caption{GASF images. Each object in this figure are the same as the object in Figure \ref{figcwt}.}
    \label{figgasf}
\end{figure*}

\begin{figure*}
    \centering
    \includegraphics[width=0.5\textwidth]{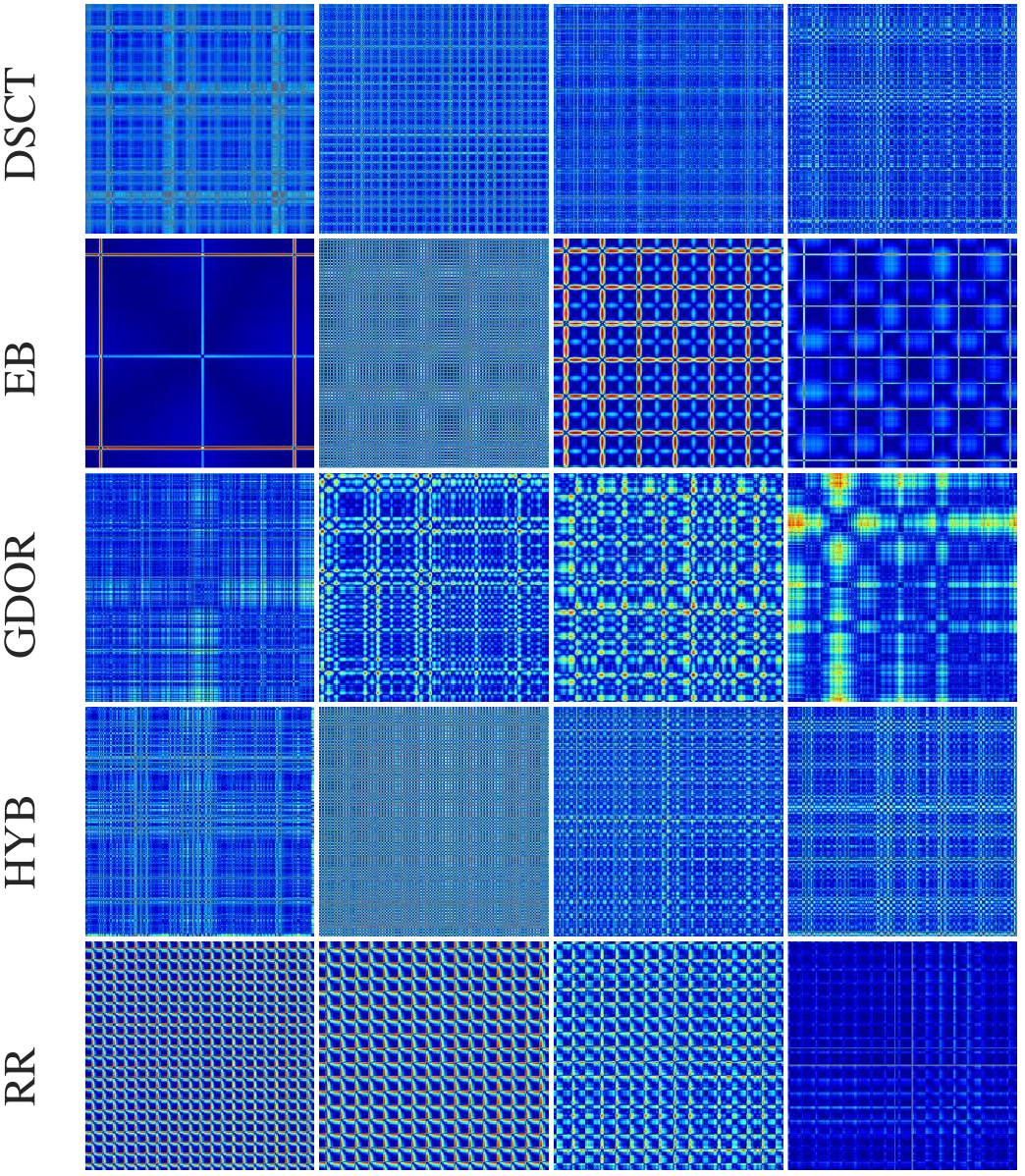}
    \caption{RP images. Each object in this figure are the same as the object in Figure \ref{figcwt}.}
    \label{figrp}
\end{figure*}

Figures \ref{figcwt}, \ref{figgadf}, \ref{figgasf}, and \ref{figrp} display the transformed images generated using the different methods described in Section~\ref{image}. All figures are arranged in the same order from top to bottom: $\delta$ Sct, EB, $\gamma$ Dor and HYB. Each column corresponds to the same object, with their associated KIC identifiers listed in Table~\ref{kicsample}.

\begin{figure*}
    \centering
    \includegraphics[width=0.4\textwidth]{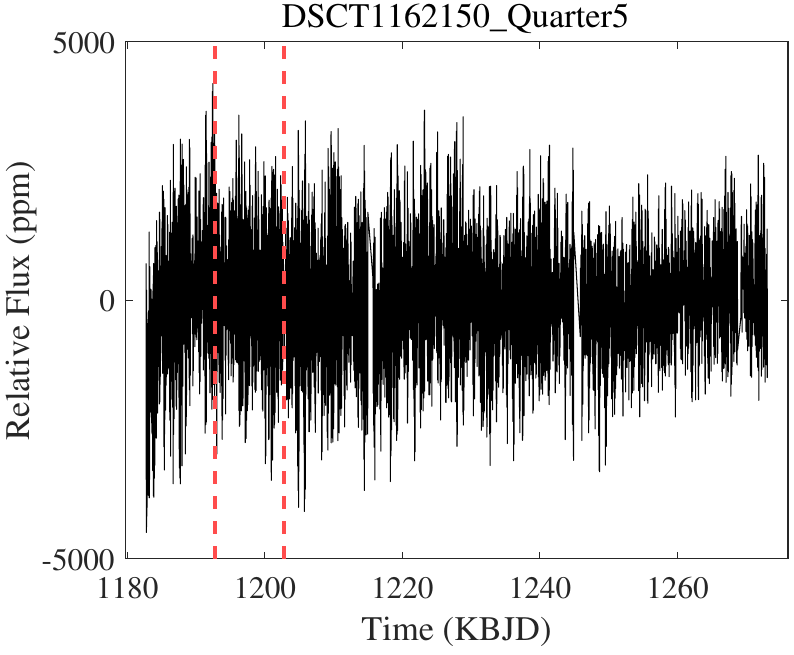}
    \includegraphics[width=0.4\textwidth]{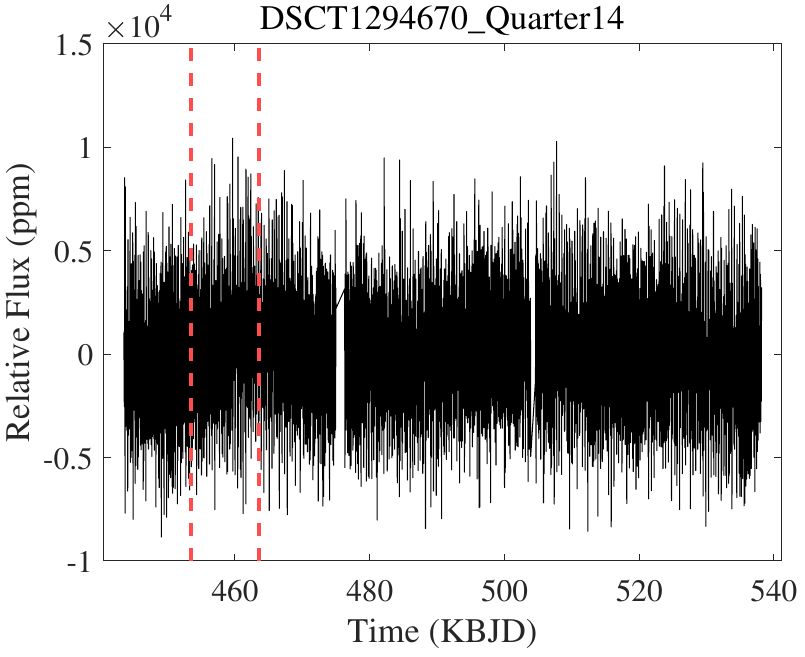}
    \includegraphics[width=0.4\textwidth]{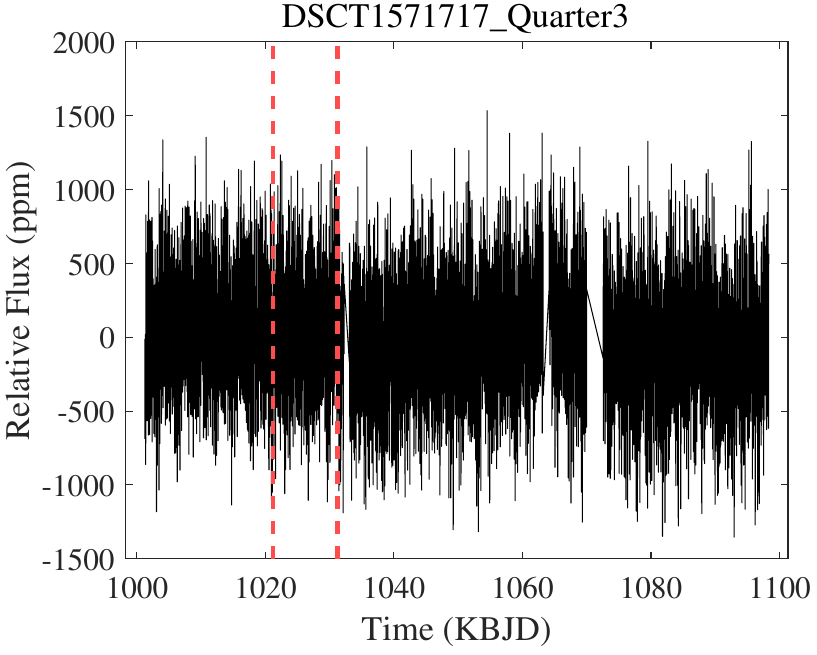}
    \includegraphics[width=0.4\textwidth]{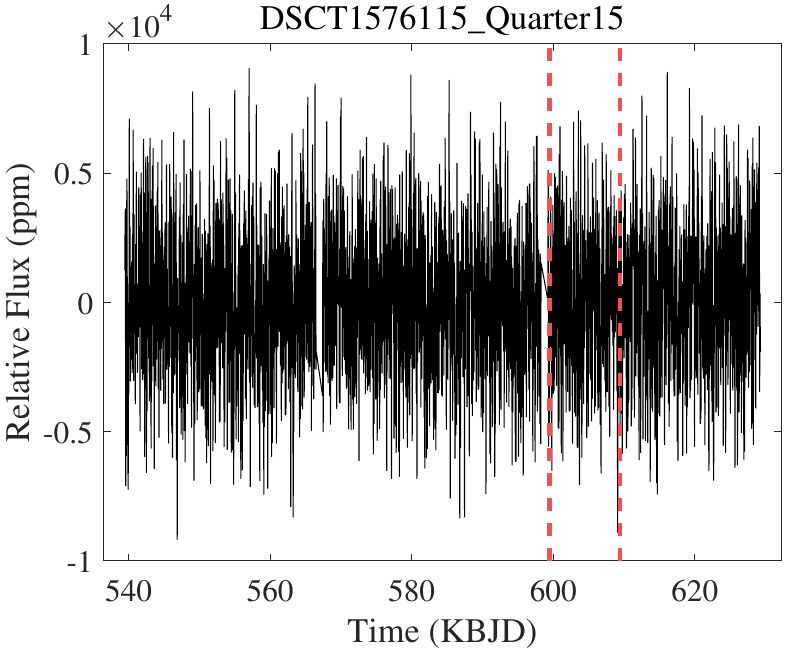}
    \caption{The DSCT light curves. The red line stands for the interval shown in Figure \ref{figcwt} to Figure \ref{figrp}.}
    \label{figdsct}
\end{figure*}

\begin{figure*}
    \centering
    \includegraphics[width=0.4\textwidth]{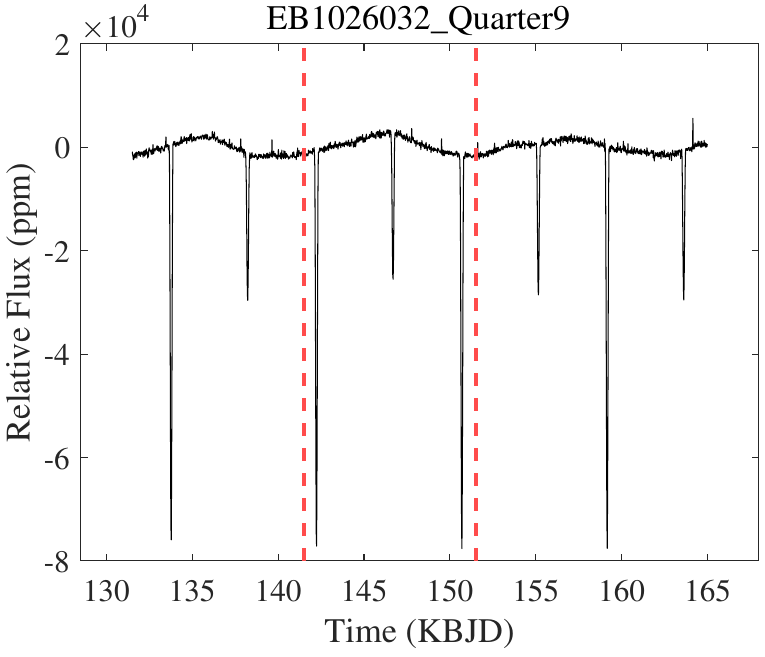}
    \includegraphics[width=0.4\textwidth]{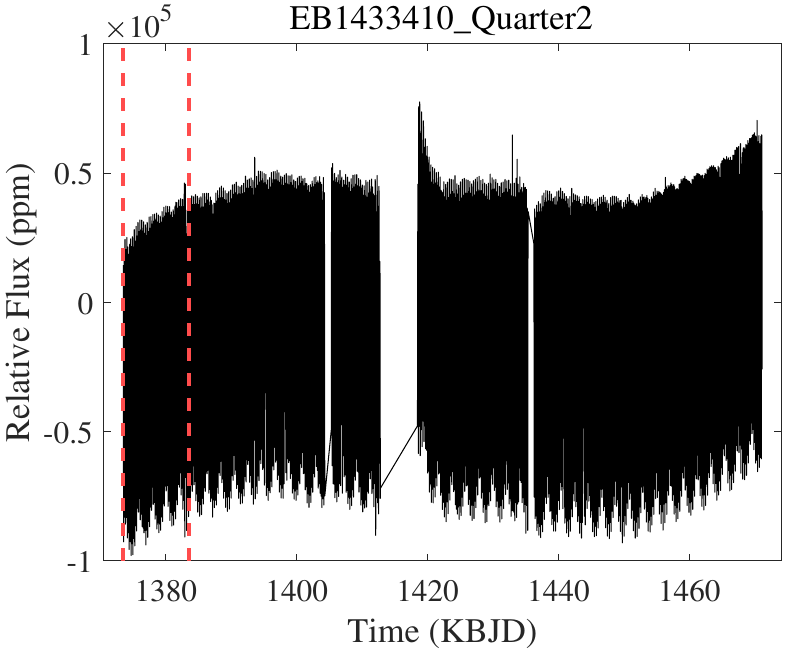}
    \includegraphics[width=0.4\textwidth]{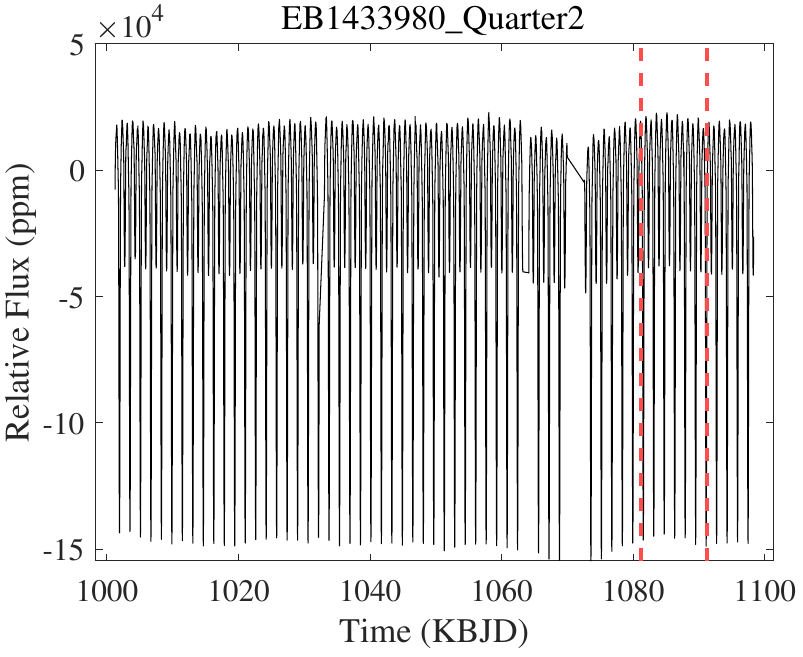}
    \includegraphics[width=0.4\textwidth]{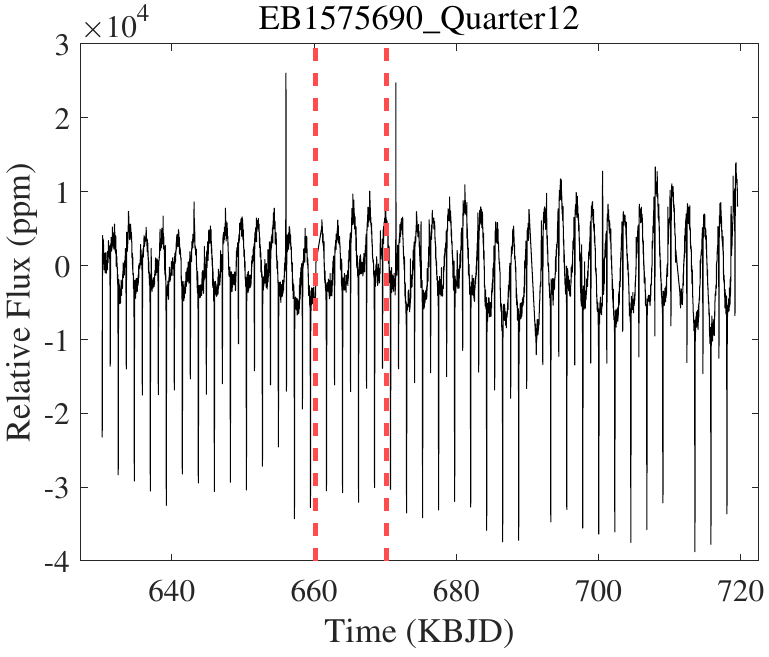}
    \caption{The EB light curves. The red line stands for the interval Figure \ref{figcwt} to Figure \ref{figrp}.}
    \label{figeb}
\end{figure*}

\begin{figure*}
    \centering
    \includegraphics[width=0.4\textwidth]{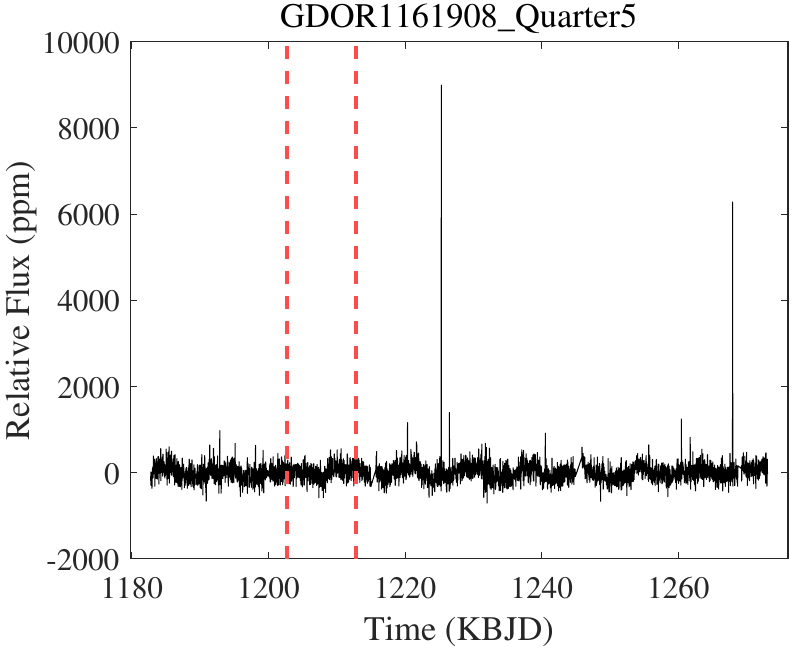}
    \includegraphics[width=0.4\textwidth]{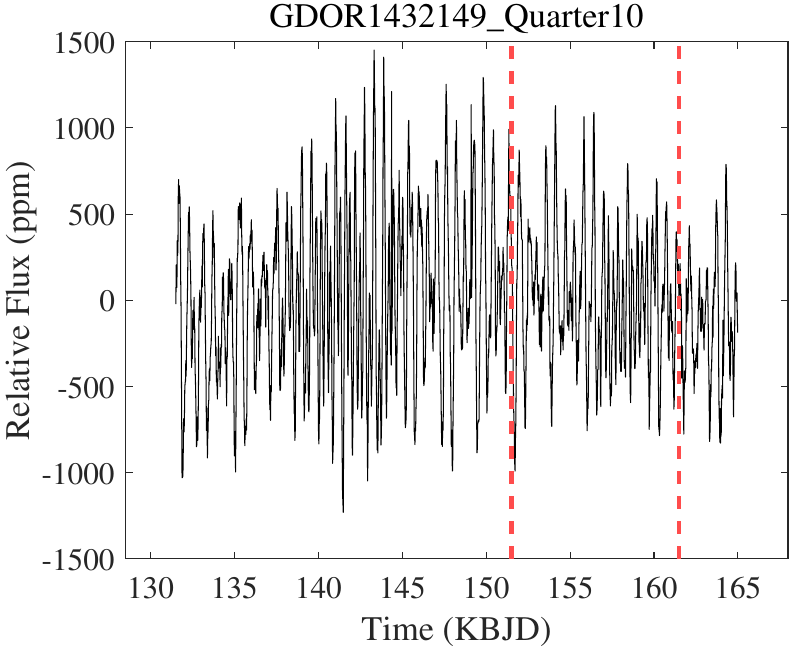}
    \includegraphics[width=0.4\textwidth]{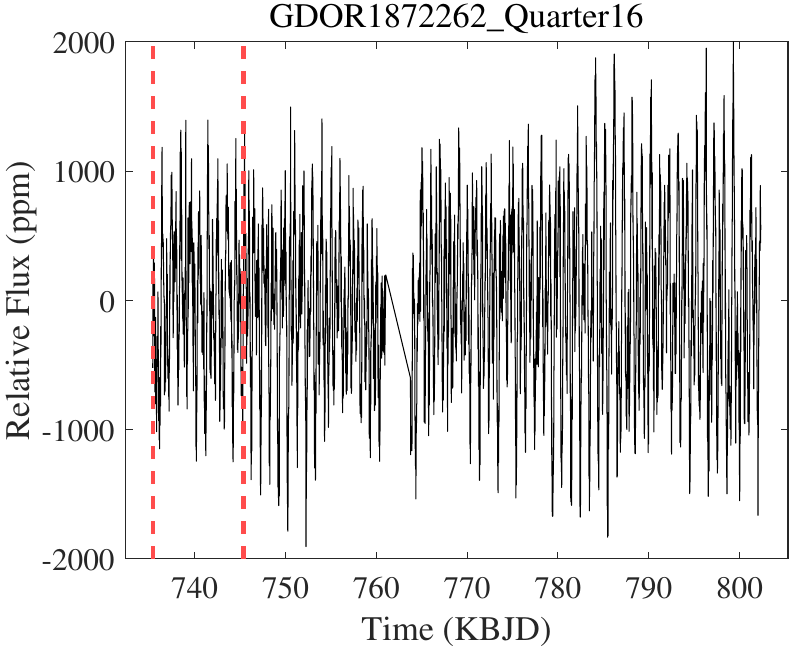}
    \includegraphics[width=0.4\textwidth]{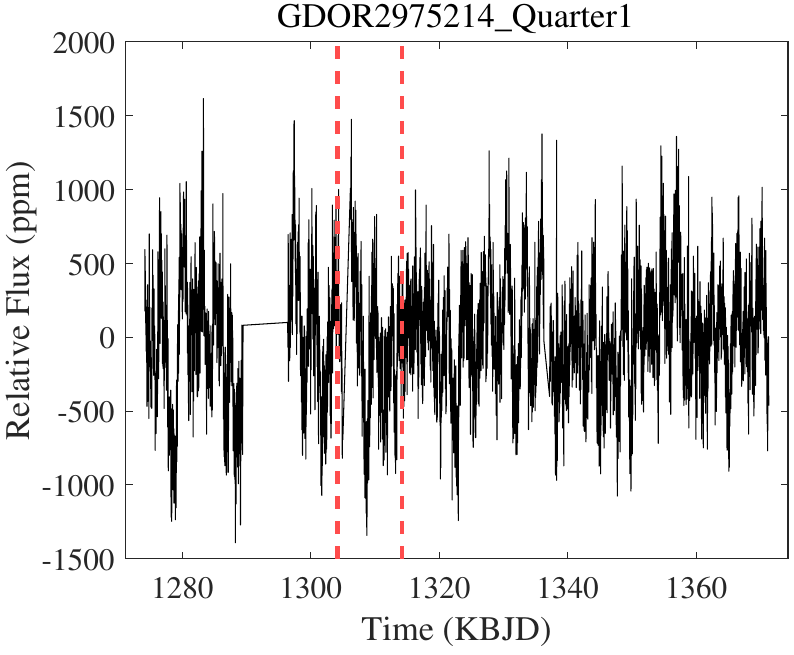}
    \caption{The GDor light curves.  The red line stands for the interval Figure \ref{figcwt} to Figure \ref{figrp}.}
    \label{figgdor}
\end{figure*}

\begin{figure*}
    \centering
    \includegraphics[width=0.4\textwidth]{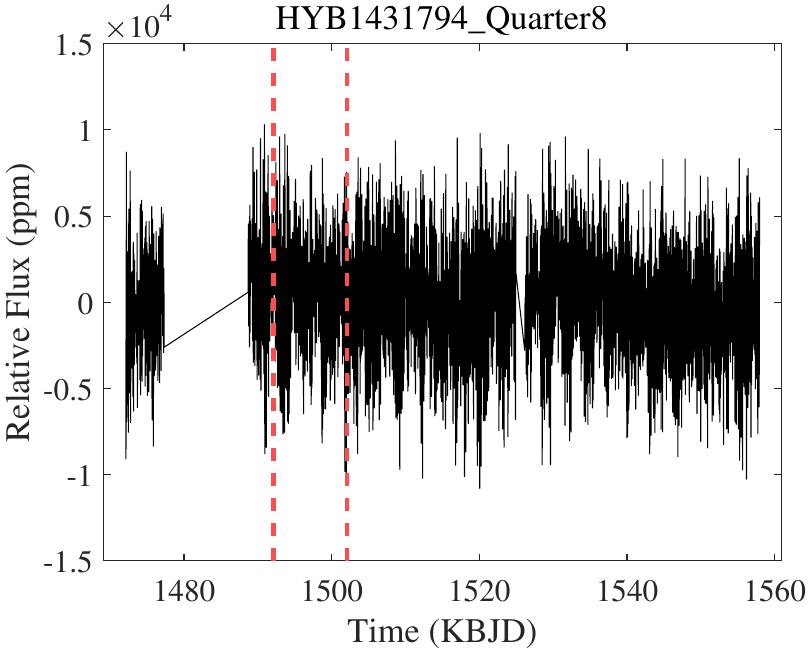}
    \includegraphics[width=0.4\textwidth]{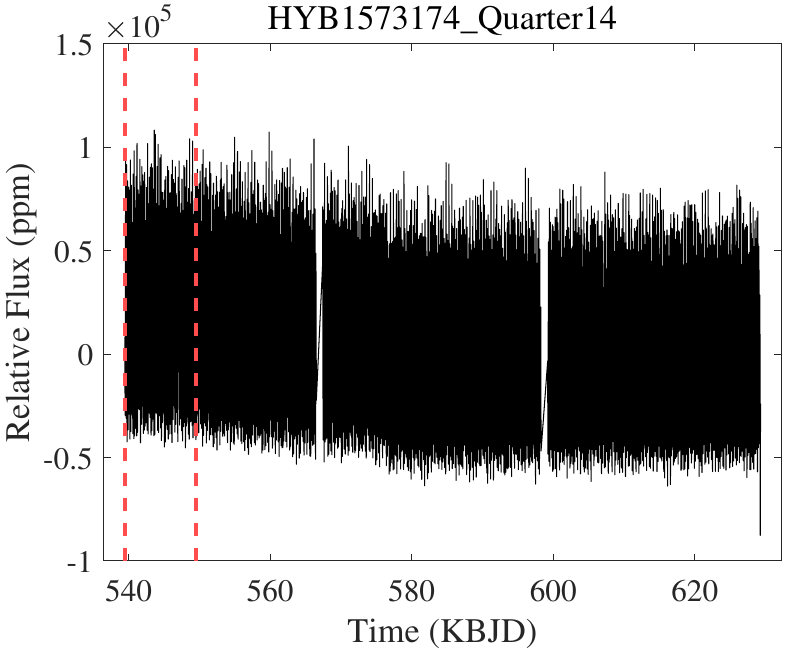}
    \includegraphics[width=0.4\textwidth]{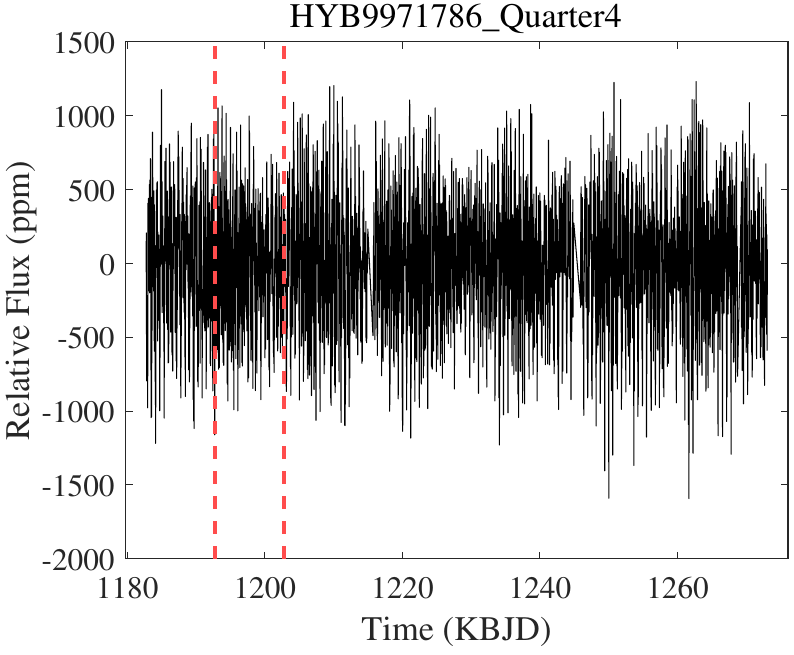}
    \includegraphics[width=0.4\textwidth]{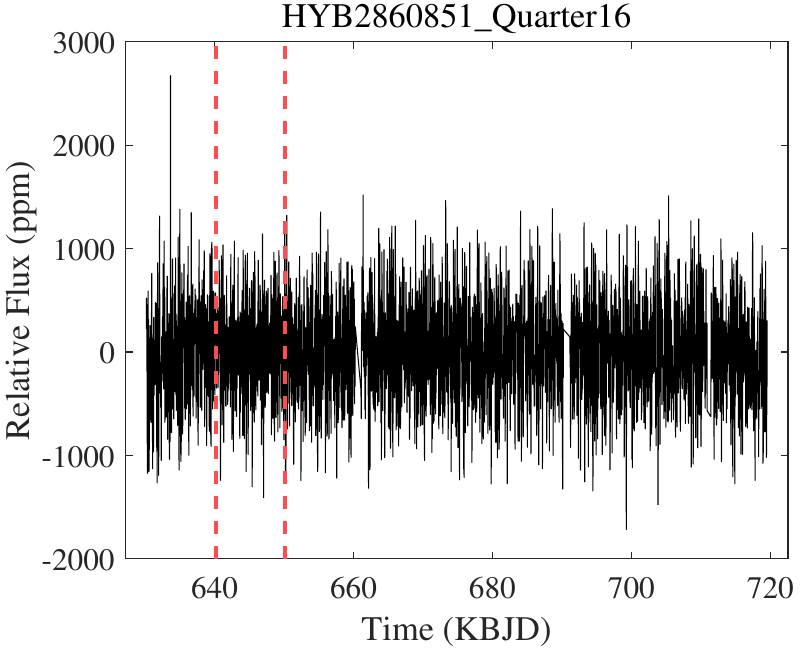}
    \caption{The HYB light curves. The red line stands for the interval Figure \ref{figcwt} to Figure \ref{figrp}.}
    \label{fighyb}
\end{figure*}

\begin{figure*}
    \centering
    \includegraphics[width=0.4\textwidth]{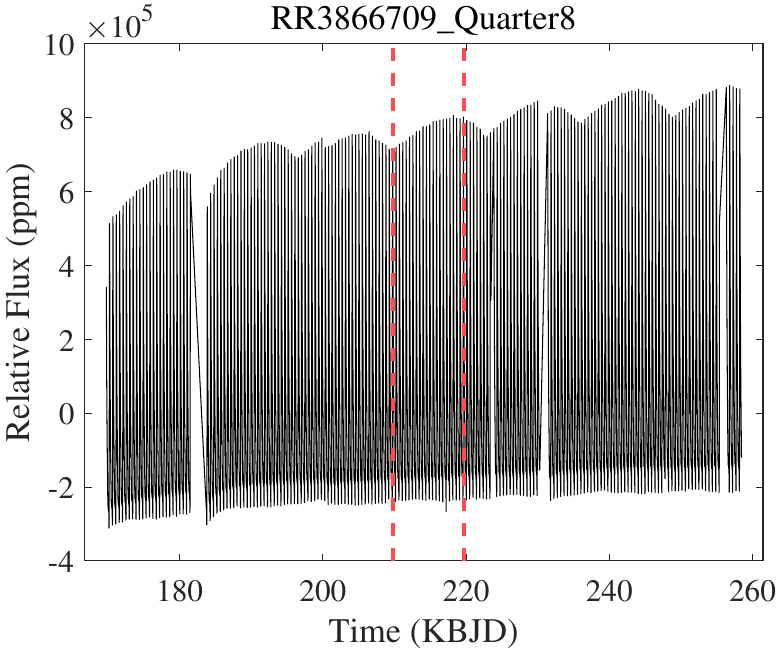}
    \includegraphics[width=0.4\textwidth]{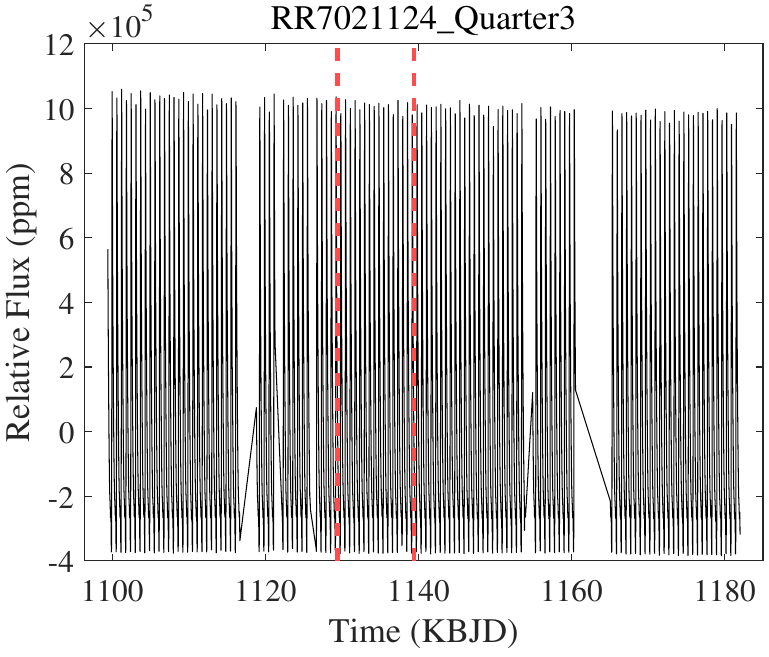}
    \includegraphics[width=0.4\textwidth]{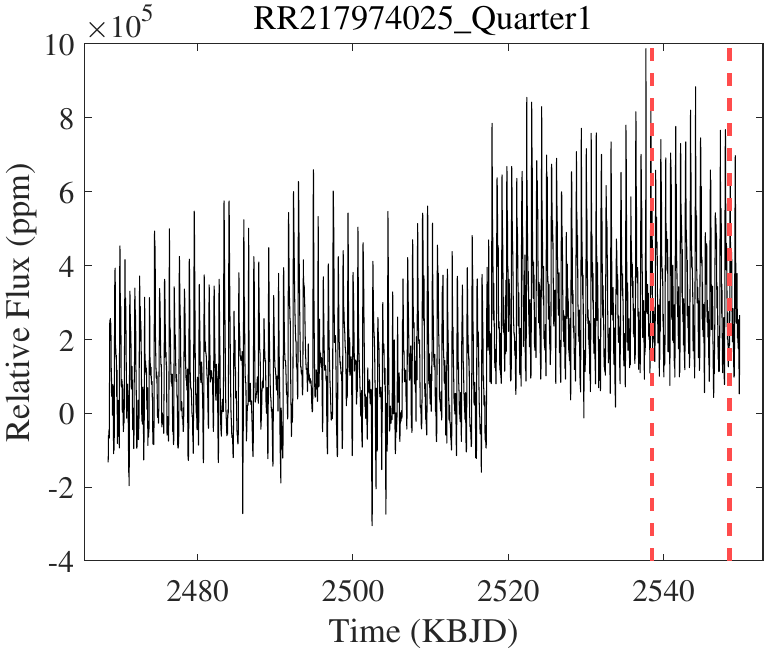}
    \includegraphics[width=0.4\textwidth]{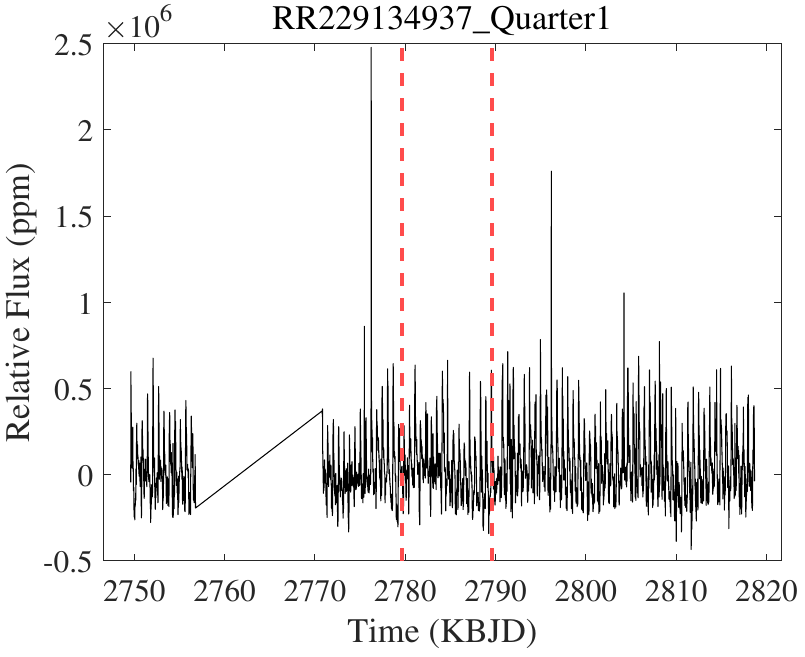}
    \caption{The RR light curves. The red line stands for the interval Figure \ref{figcwt} to Figure \ref{figrp}.}
    \label{figrr}
\end{figure*}

For reference, we also show the original light curves of these samples in Figures \ref{figdsct} to \ref{figrr}, with preprocessing details provided in Section~\ref{sec2.8}. Each light curve represents one quarter of \textit{Kepler} observations, and the region between the red lines indicates the portion used for image transformation, as shown in Figures \ref{figcwt}, \ref{figgadf}, \ref{figgasf}, and \ref{figrp}.

\begin{table}
\bc
\begin{minipage}[]{100mm}
\caption[]{Sample of light curves \label{kicsample}}
\end{minipage}
\setlength{\tabcolsep}{1pt}
\small
\begin{tabular}{lcccc}
\hline\hline \noalign{\smallskip}
Label & Column 1 & Column 2 & Column 3 & Column 4    \\
\noalign{\smallskip}
  \hline
  \noalign{\smallskip}
$\delta$ Sct & 1162150 & 1294670 & 1571717 & 1576115 \\
EB & 1026032 & 1433410 & 1433980 & 1575690 \\
$\gamma$ Dor & 1161908 & 1432149 & 1872262 & 2975214 \\
HYB & 1431794 & 1573174 & 9971786 & 2860851 \\
RR & 3866709 & 7021124 & 217974025 & 229134937 \\
  \hline
  \end{tabular}
\ec
\tablecomments{0.86\textwidth}{The KIC of light curves in Figure \ref{figcwt} to Figure \ref{figrp}. The last two RRs are from \textit{K2} catalog.}
\end{table}

\section{Confusion matrix}
\label{appcm}
We present all confusion matrices here. The different grayscale of the confusion matrix shows the number and fraction of each class. The bottom brightest blocks are the recall (fractions with larger font size) of each corresponding label, and the smaller percentage is ($1-recall$). The right brightest blocks are each class's precision (larger font size), and the smaller ones are ($1-precision$). The bottom right block shows the total accuracy (larger font size) and the error rate (the smaller one). 

\begin{figure*}
    \centering
    \includegraphics[width=0.4\textwidth]{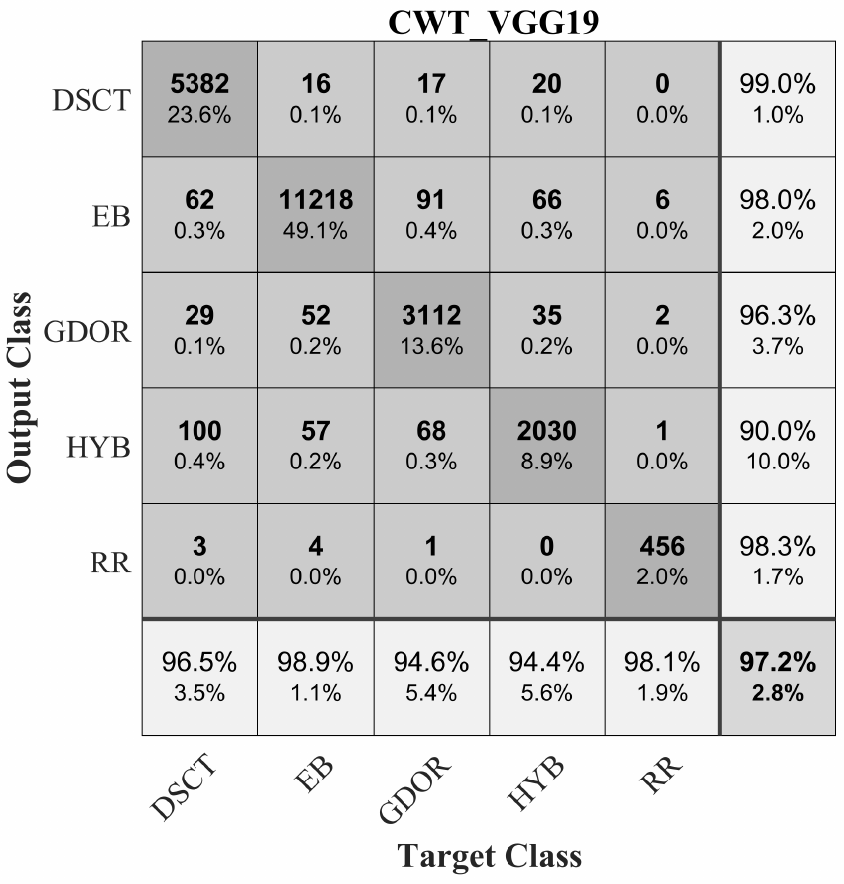}
    \includegraphics[width=0.4\textwidth]{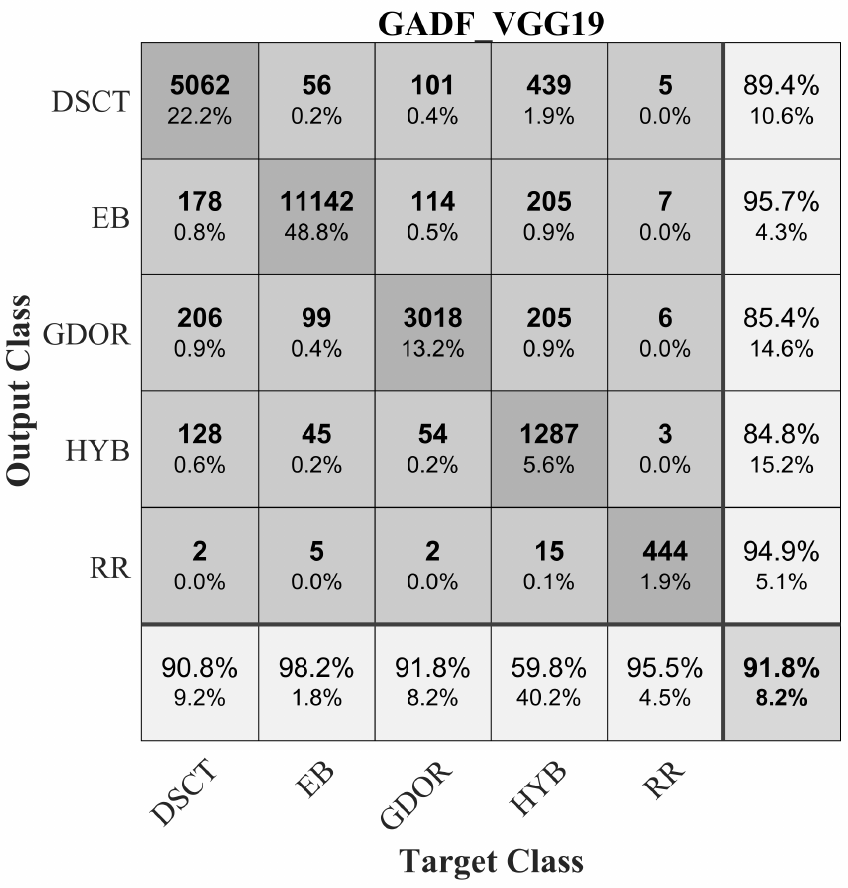}
    \includegraphics[width=0.4\textwidth]{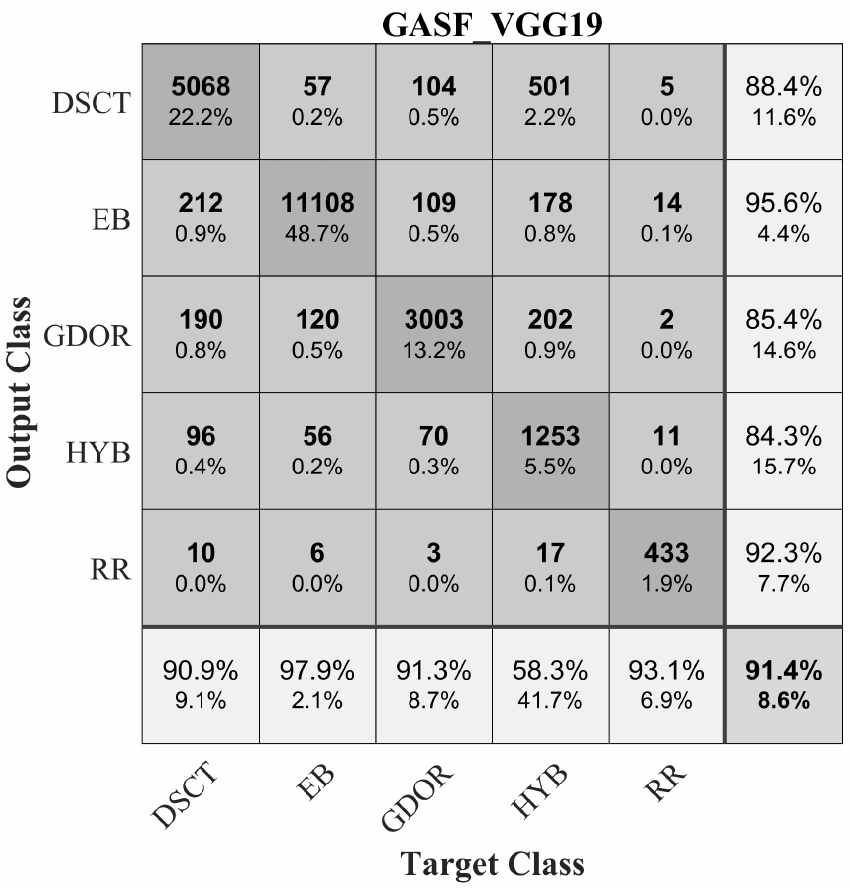}
    \includegraphics[width=0.4\textwidth]{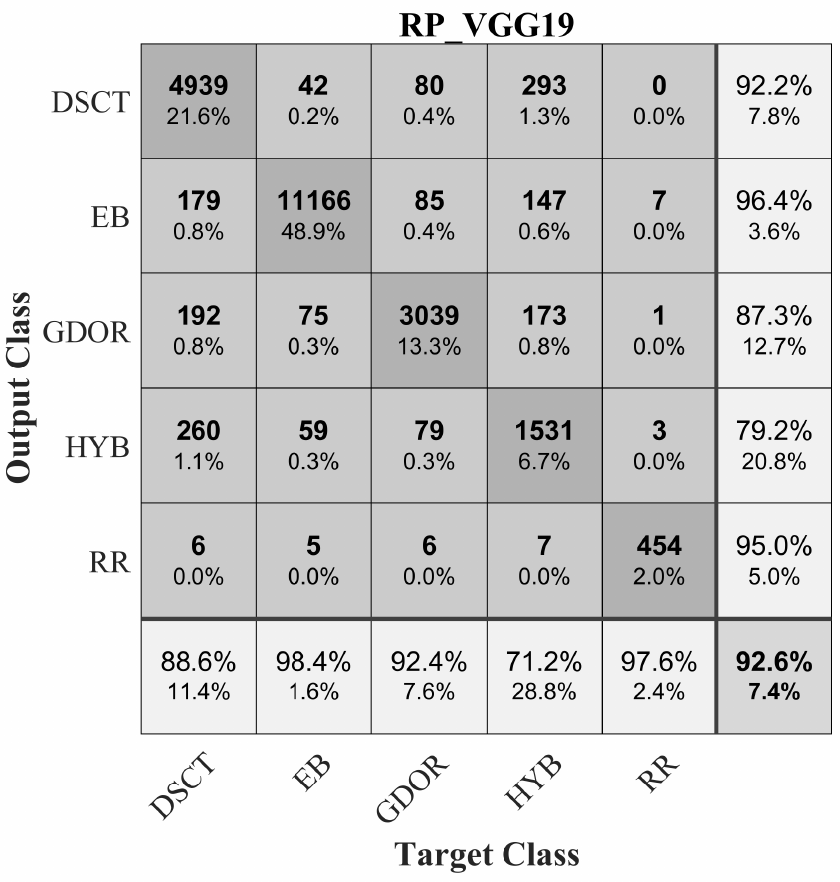}
    \caption{The confusion matrixes for VGG19 architecture.}
\end{figure*}

\begin{figure*}
    \centering
    \includegraphics[width=0.4\textwidth]{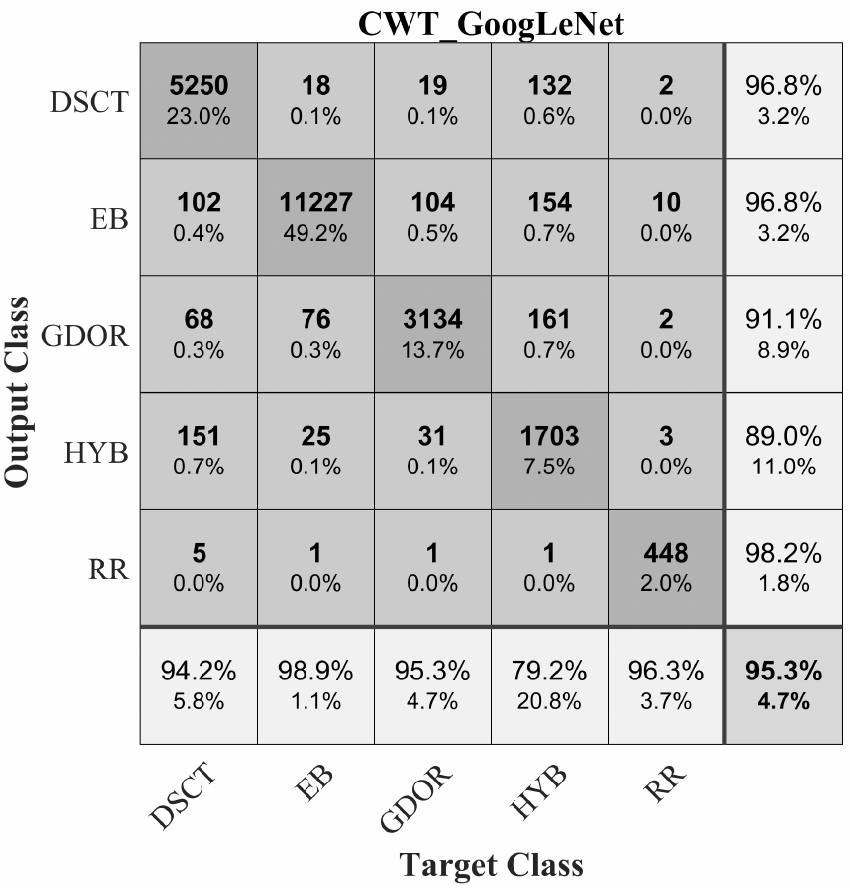}
    \includegraphics[width=0.4\textwidth]{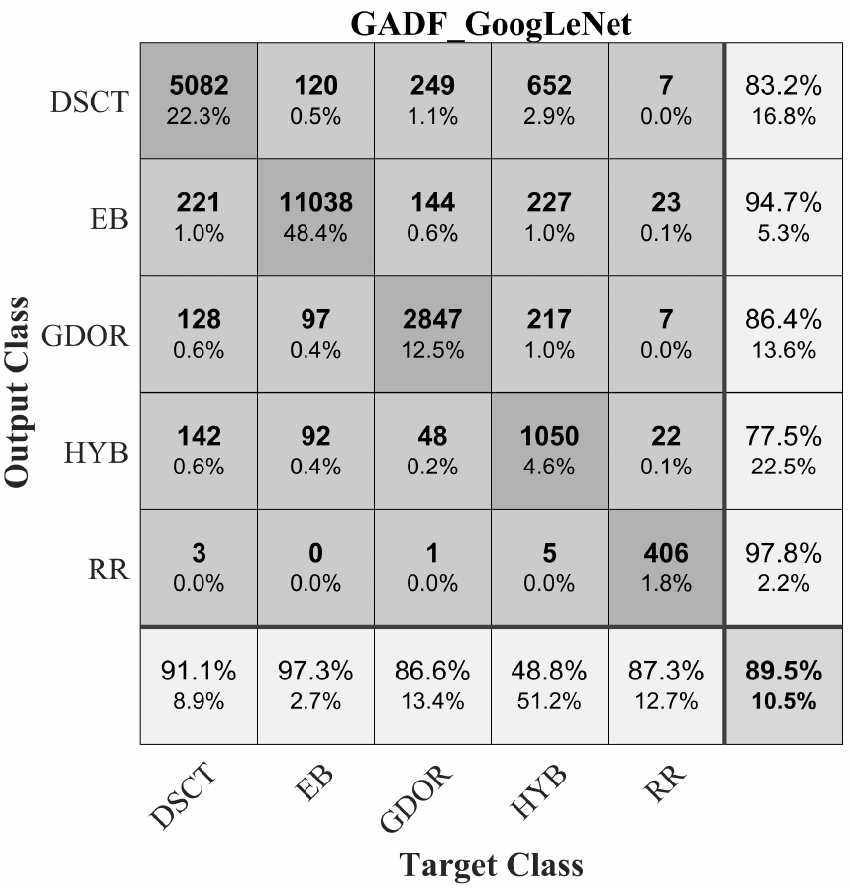}
    \includegraphics[width=0.4\textwidth]{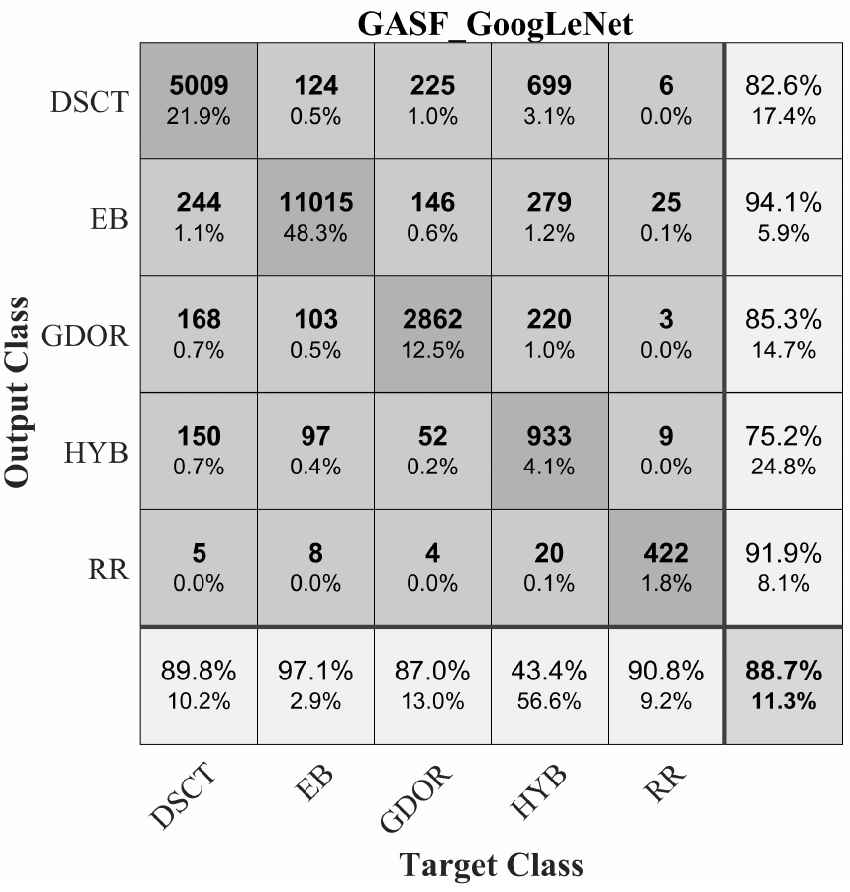}
    \includegraphics[width=0.4\textwidth]{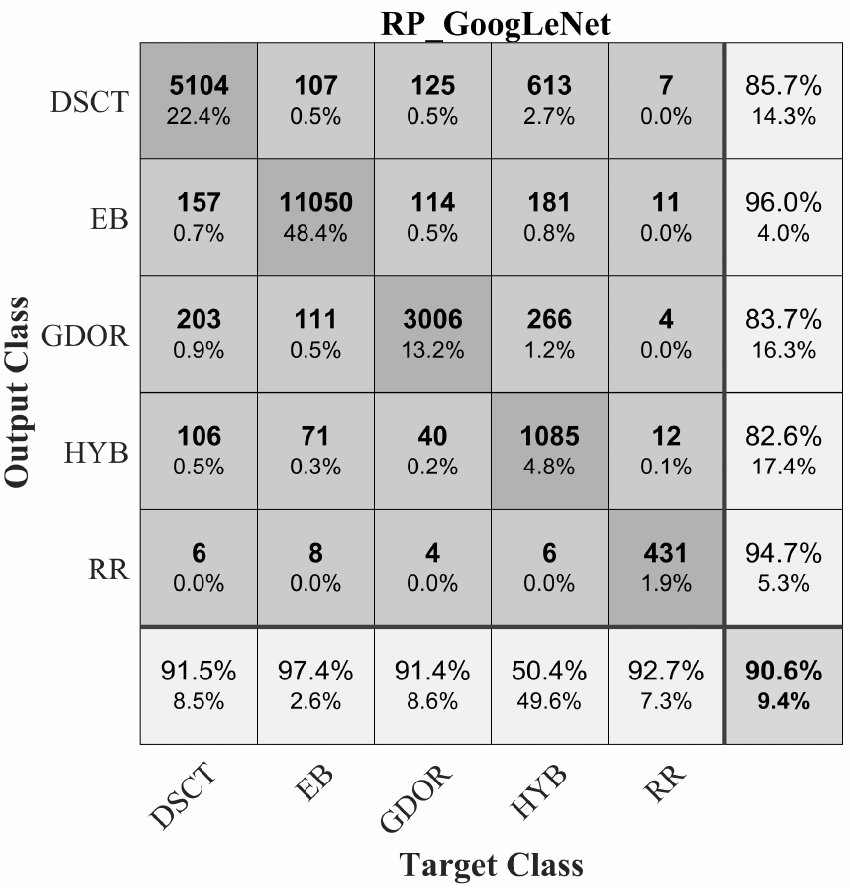}
    \caption{The confusion matrixes for GoogLeNet architecture.}
\end{figure*}

\begin{figure*}
    \centering
    \includegraphics[width=0.4\textwidth]{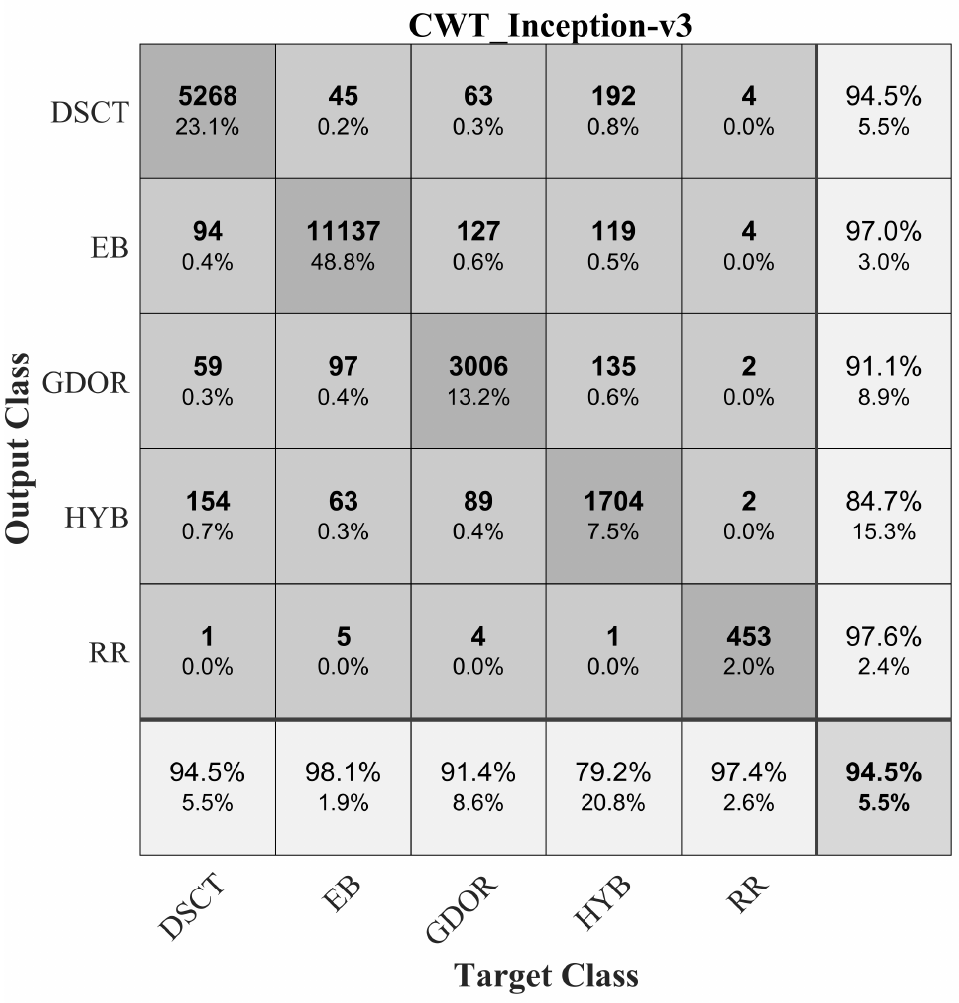}
    \includegraphics[width=0.4\textwidth]{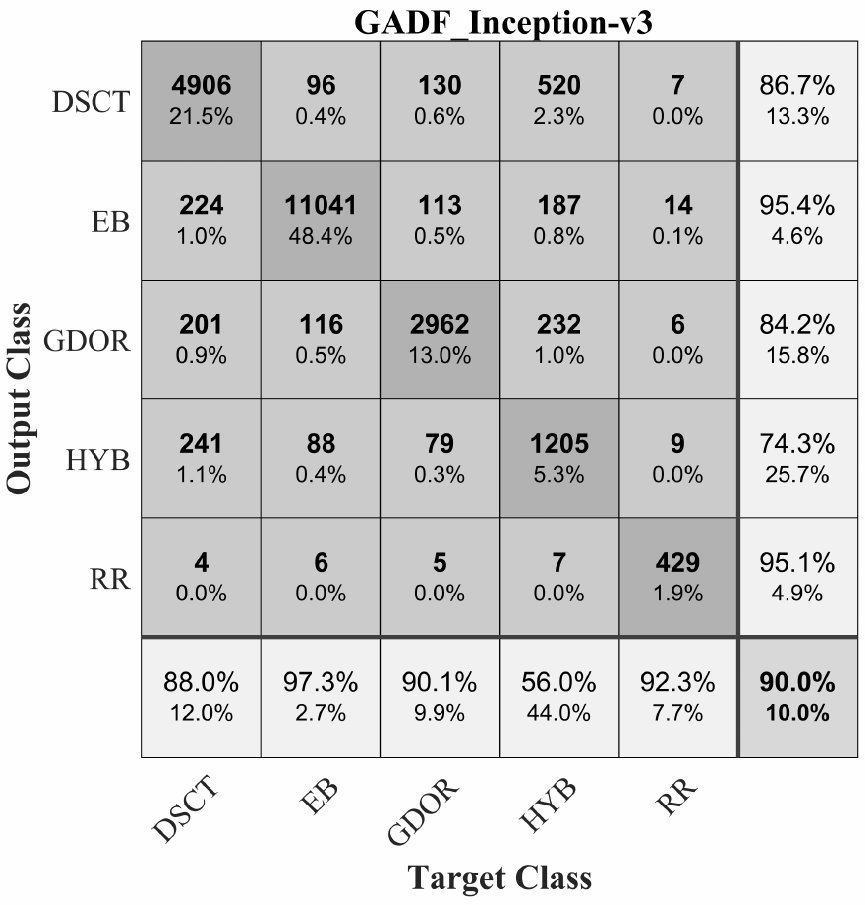}
    \includegraphics[width=0.4\textwidth]{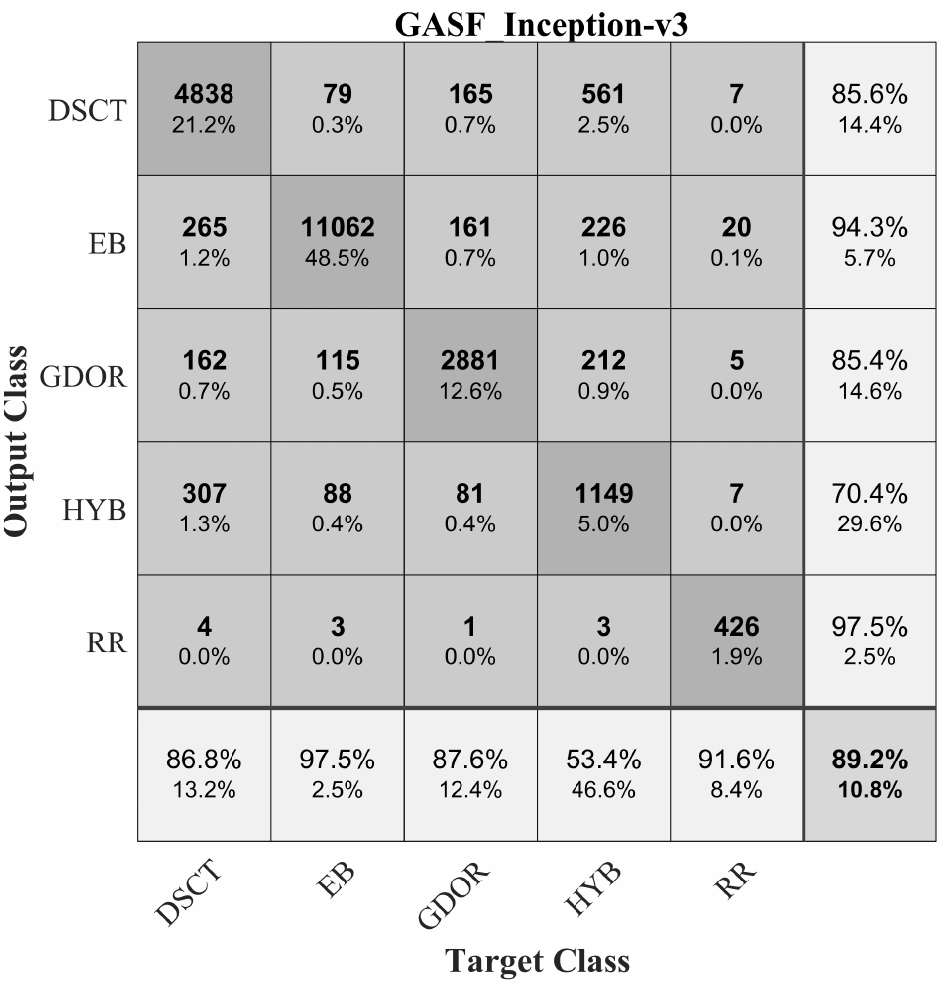}
    \includegraphics[width=0.4\textwidth]{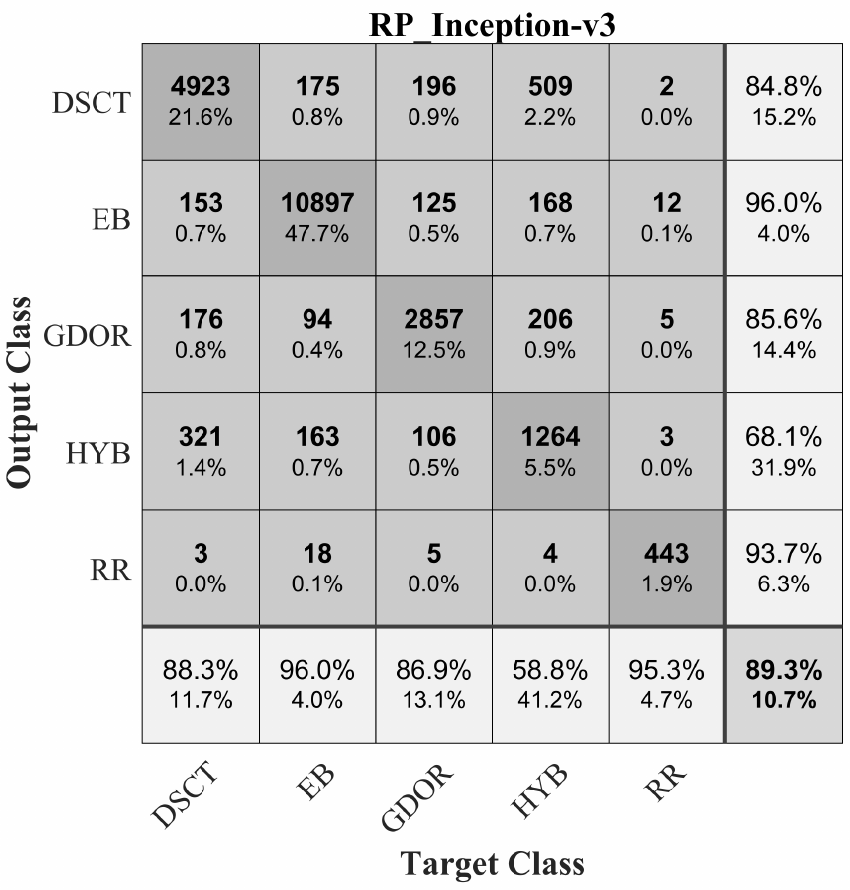}
    \caption{The confusion matrixes for Inception-v3 architecture.}
\end{figure*}

\begin{figure*}
    \centering
    \includegraphics[width=0.4\textwidth]{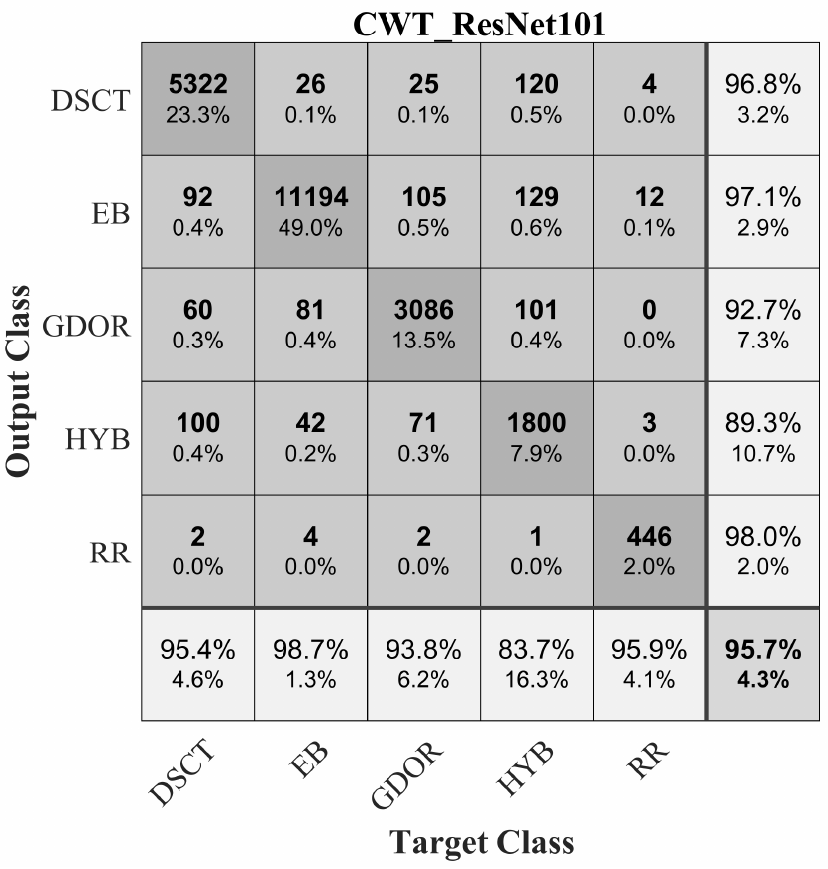}
    \includegraphics[width=0.4\textwidth]{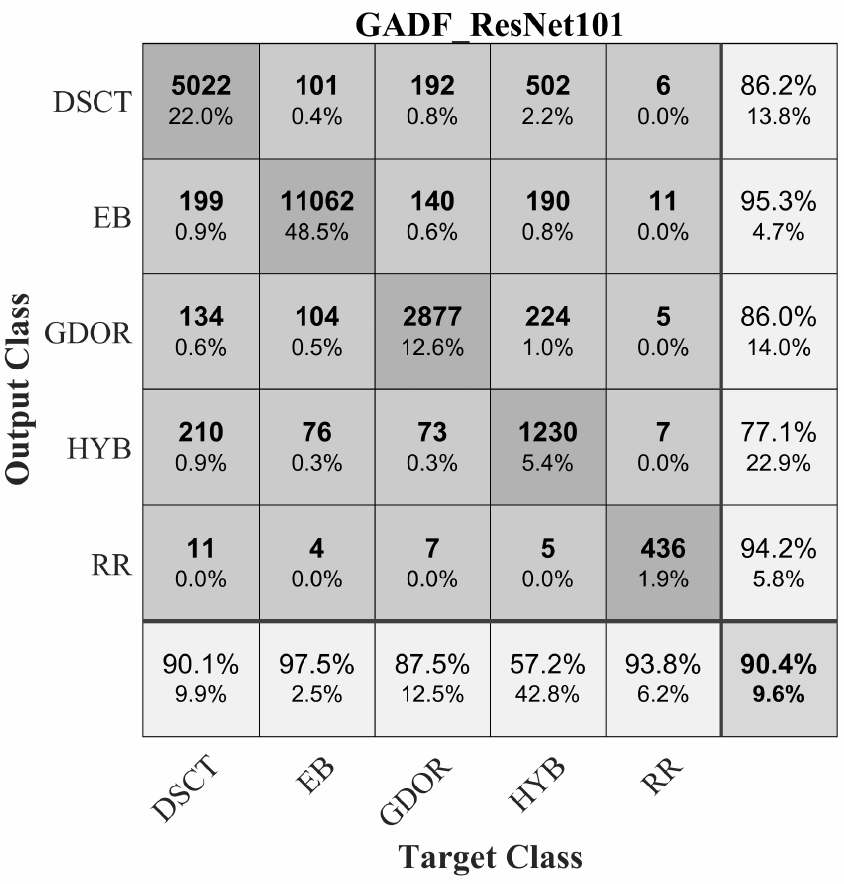}
    \includegraphics[width=0.4\textwidth]{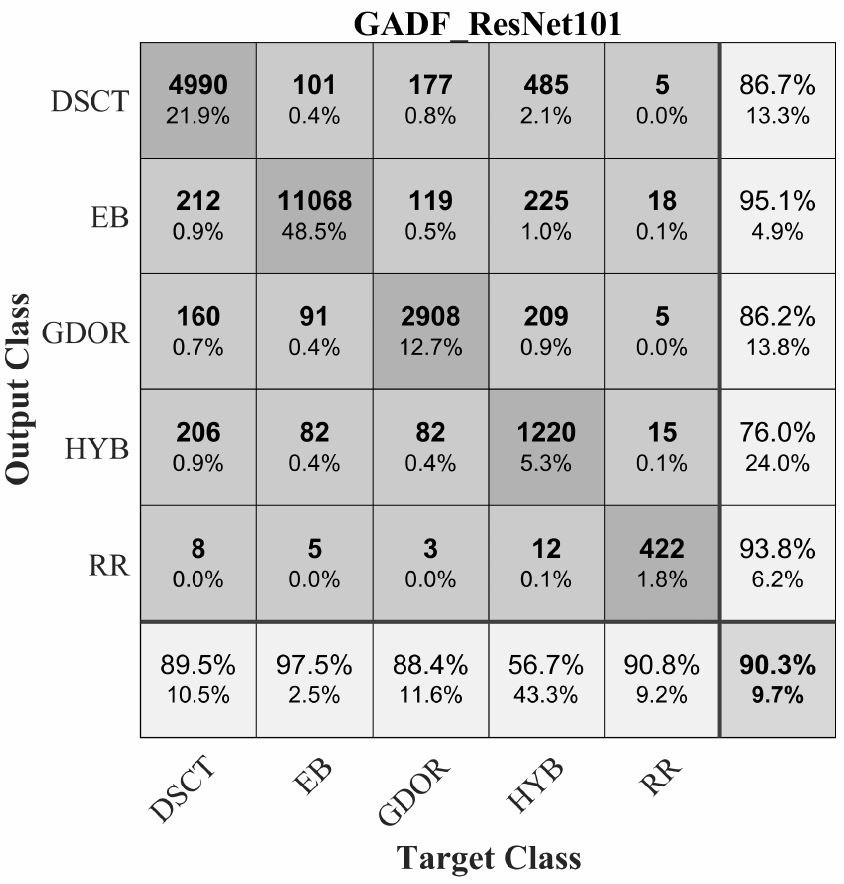}
    \includegraphics[width=0.4\textwidth]{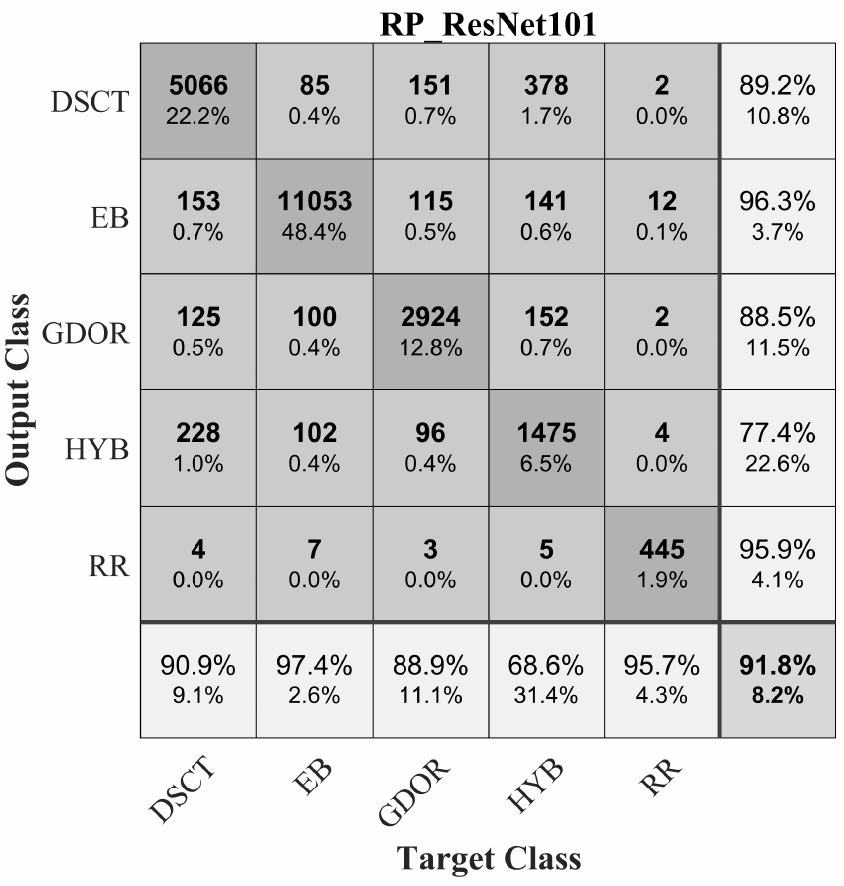}
    \caption{The confusion matrixes for ResNet101 architecture.}
\end{figure*}

\begin{figure*}
    \centering
    \includegraphics[width=0.4\textwidth]{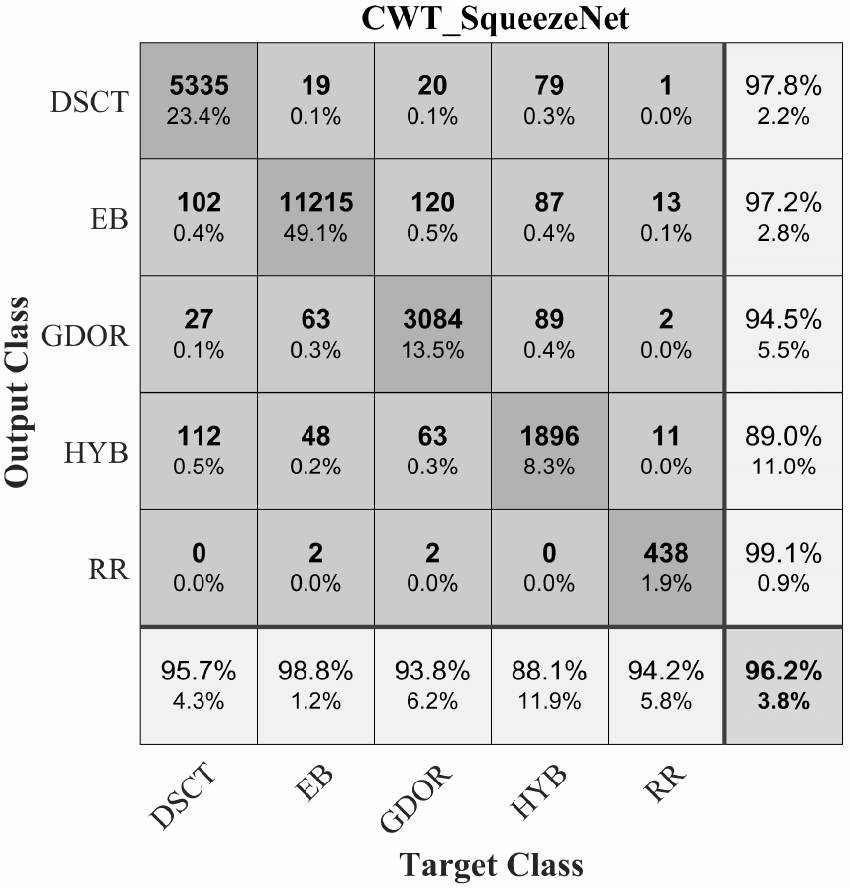}
    \includegraphics[width=0.4\textwidth]{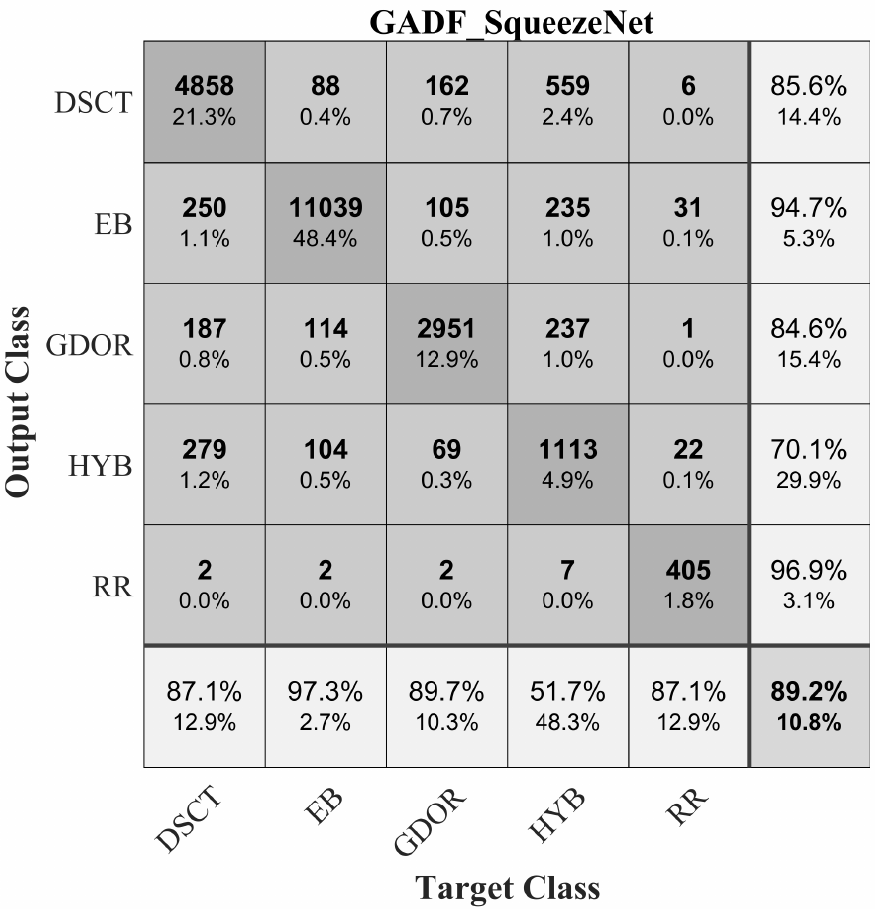}
    \includegraphics[width=0.4\textwidth]{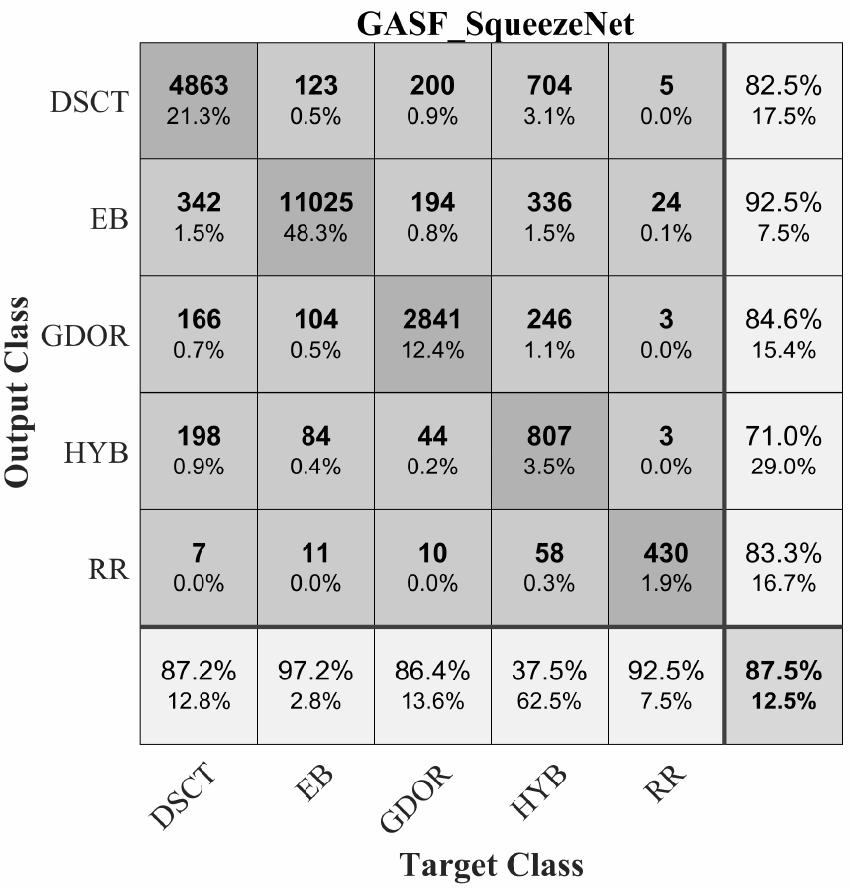}
    \includegraphics[width=0.4\textwidth]{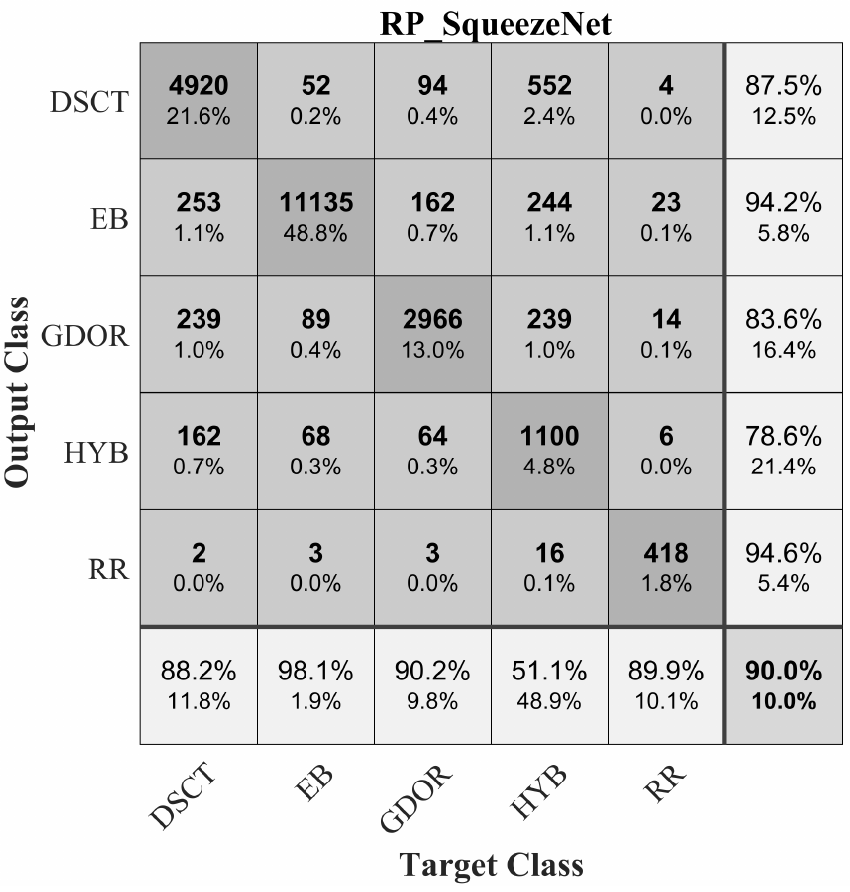}
    \caption{The confusion matrixes for SqueezeNet architecture.}
\end{figure*}

\begin{figure*}
    \centering
    \includegraphics[width=0.4\textwidth]{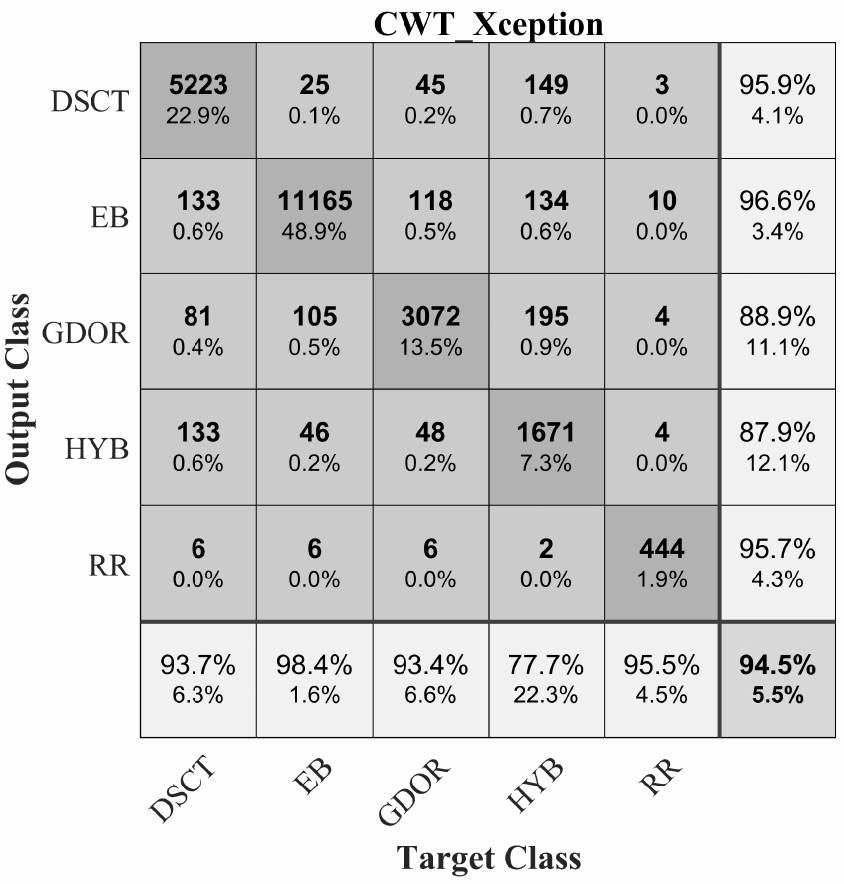}
    \includegraphics[width=0.4\textwidth]{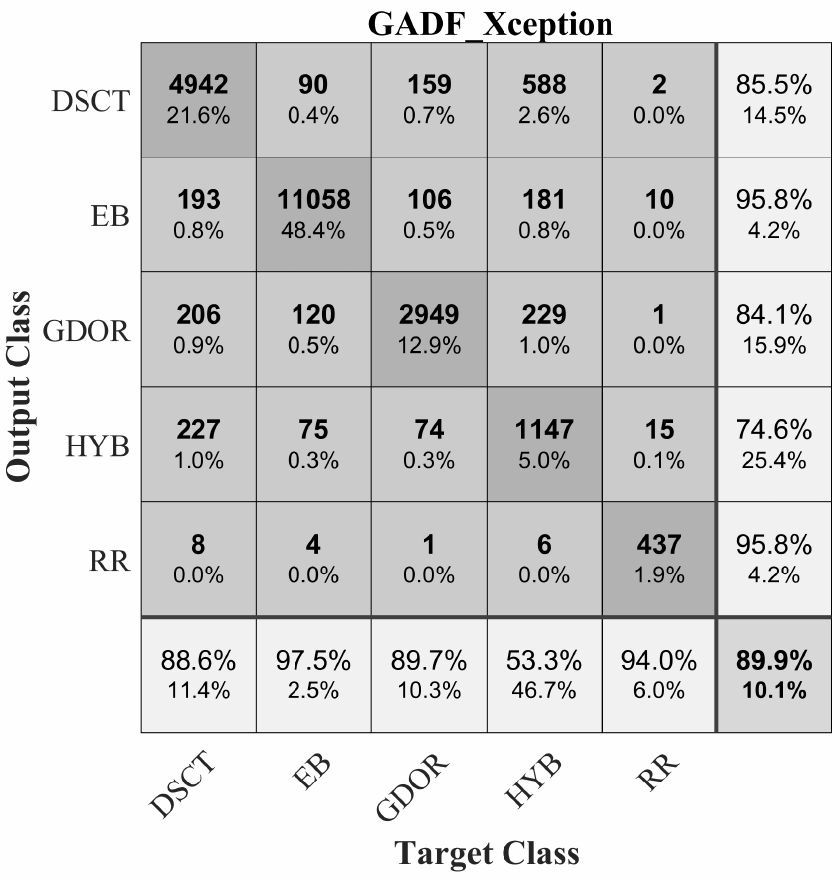}
    \includegraphics[width=0.4\textwidth]{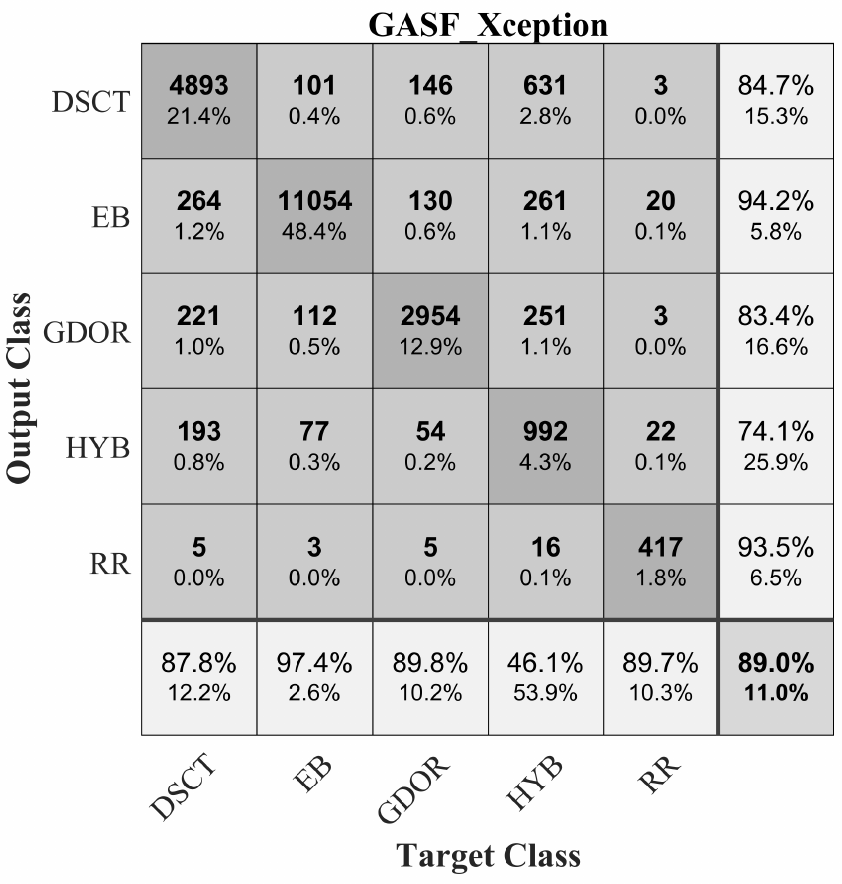}
    \includegraphics[width=0.4\textwidth]{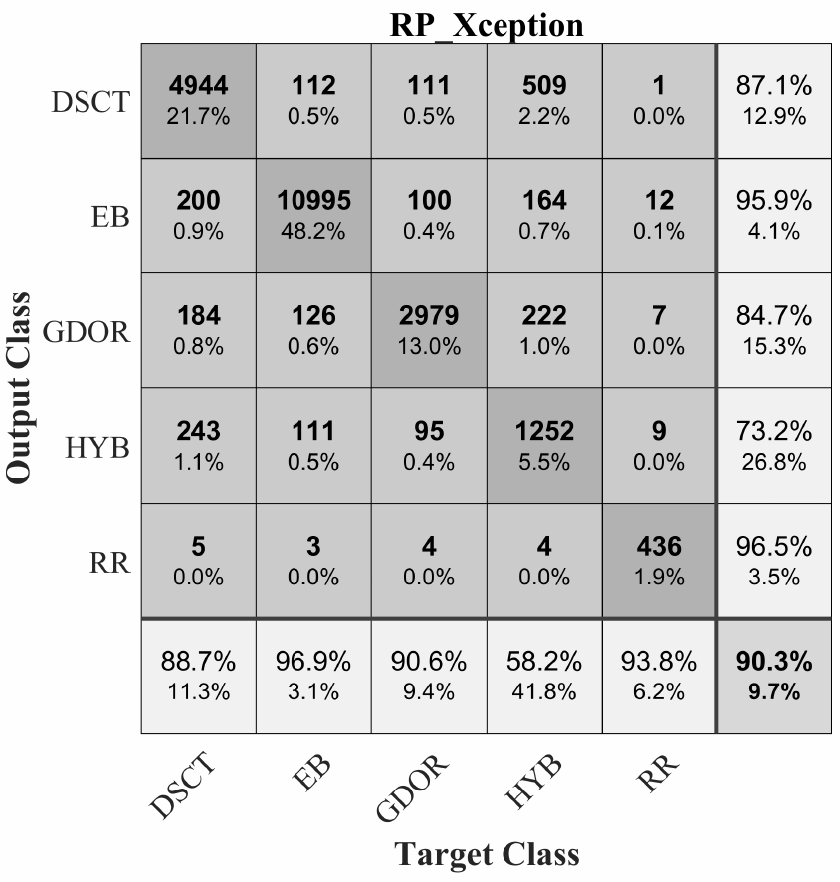}
    \caption{The confusion matrixes for Xception architecture.}
\end{figure*}

\section{Test with TESS}
\label{app:BT}

The MoE model is mainly based on the experts, and the transfer learning model have the highest accuracy under the CWT method. To evaluate the generalization capability of our models, we conducted a blind test using the CWT-based method on data from the Transiting Exoplanet Survey Satellite (TESS). TESS employs four 2$k$ $\times$ 2$k$ CCDs, each with a field of view of 24 $\times$ 24 deg$^2$ and a pixel scale of 21 arcseconds per pixel. The mission provides two observing cadences: 2 minutes and 30 minutes. Light curves derived from TESS’s wide-field imaging are often undersampled, which can introduce additional noise and potentially mimic astrophysical signals of interest, leading to false positives \citep{Oelkers18, Bramich08, Miller08}.

The light curves used in this study were obtained from \citealt{Huang_2020} and processed using the MIT Quick Look Pipeline (QLP). Low-frequency trends in the TESS light curves have been removed using high-pass filtering techniques.

The TESS blind test catalog was compiled from sources including \citet{Ant19}, \citet{Barcel20}, and \citet{Moln22}. Specifically, \citealt{Ant19} reports 57 $\delta$ Sct, 2 $\gamma$ Dor, 2 EB, and 10 HYB stars (including 5 $\delta$ Sct binaries); \citealt{Barcel20} includes 37 $\gamma$ Dor and 96 HYB candidates; and \citealt{Moln22} contributes 40 RR-type variables from its catalog.

Some stars in the blind test catalog exhibit non-standard pulsation behaviors, displaying unusual or rare characteristics in their light curves. For instance, TIC 30531417, classified as a $\gamma$ Dor star, exhibits an exceptionally strong pulsation amplitude with a regular frequency pattern, making its light curve distinctly different from those of typical $\gamma$ Dor variables \citep{Lares22}. In other cases, certain $\gamma$ Dor candidates show clear eclipsing features, leading to their manual reclassification as EBs. These EBs are easily identifiable due to the presence of prominent eclipse signatures in their light curves.

Prior to conducting the blind test, we performed a cleaning process on the TESS light curves to address contamination caused by cosmic rays that were not automatically removed during the initial data reduction. To achieve this, we employed a second-order difference analysis of the light curves. Specifically, data points affected by cosmic rays were identified and removed based on a $2\sigma$ criterion, which evaluates the difference between three consecutive data points. Here, $\sigma$ represents the standard deviation of the entire light curve. This method effectively removes outliers such as cosmic ray events without affecting other intrinsic features of the light curves.

In addition, we conduct a thorough visual inspection of the light curves and exclude observation sectors that exhibited artificial trends or persistent abnormal signals. These problematic light curves often display sunken patterns or abrupt, mountain-like distortions unrelated to intrinsic stellar variability. After applying these cleaning procedures, we obtain a final dataset consisting of 640 light curves from 194 distinct sources, including 51 $\delta$ Sct, 18 $\gamma$ Dor, 17 EB, 38 RR, and 70 HYB stars. All manual quality assessments were performed via simple visual scanning.

These data-cleaning strategies are not only applicable to TESS light curves but also generalize well to other time-series datasets of variable stars, making them a practical approach for improving model evaluation on real-world observational data.

Following the preprocessing technique described in Section~\ref{sec2.8}, we processed the TESS light curves and obtained a total of 590 samples for prediction. In the training set, the HYB class predominantly comprised hybrids of $\delta$ Sct and $\gamma$ Dor variables, along with some binary systems where one component exhibited rotational or pulsational variability.

However, it is important to note that our HYB training set may not fully represent the entire feature space of hybrid variables. As a result, prediction inaccuracies can occur when the signal in the test data differ from those observed in the training samples. This limitation primarily stems from the relatively small size of the HYB training dataset, rather than any deficiency in the deep learning models or the representational capacity of the applied transformation methods.

To mitigate potential bias, we have excluded the HYB class from the blind test results. It is evident that a more comprehensive characterization of hybrid variables will require additional labeled data and further subclassifications.

The accuracy results of the blind test are summarized in Appendix \ref{app:BT}. The models achieved the following accuracies: VGG19 at $79.4\%$, Inception-v3 at $78.2\%$, ResNet-101 at $74.2\%$, and Xception at $77.4\%$. Notably, VGG19 performed the best in this blind test scenario.

However, the observed accuracies below $80\%$ suggest that these models exhibit limited generalization ability when applied to a different catalog with distinct observation strategies, hardware configurations, and light curve characteristics. This highlights the importance of further model refinement and additional training data from diverse sources to improve robustness across varying datasets.

\begin{figure*}
    \centering
    \includegraphics[width=0.8\textwidth]{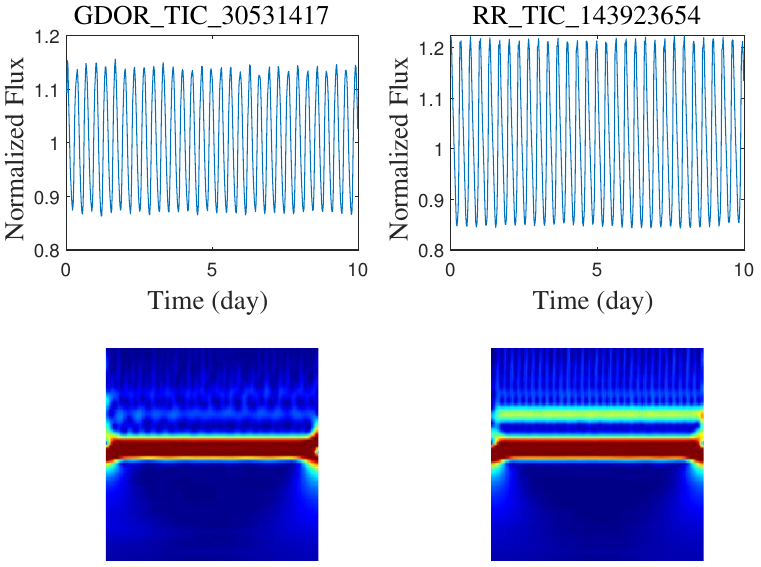}
    \caption{The light curve and CWT image sample for the EB leaking sources. The top left panel shows the light curve of the first piece of Sector 34 of object TIC 30531417, and the bottom left panel shows its CWT image. The right panel shows the first piece of Sector 28 of object TIC 143923654. \label{ebleak}}
\end{figure*}

\begin{figure*}
    \centering
    \includegraphics[width=0.4\textwidth]{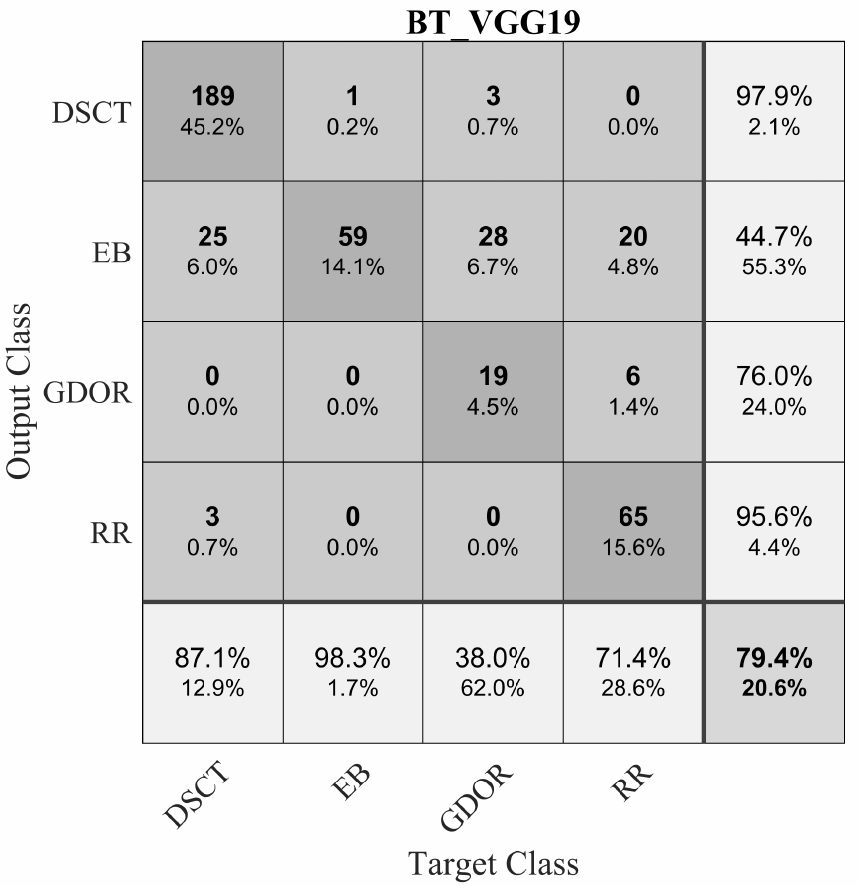}
    \includegraphics[width=0.4\textwidth]{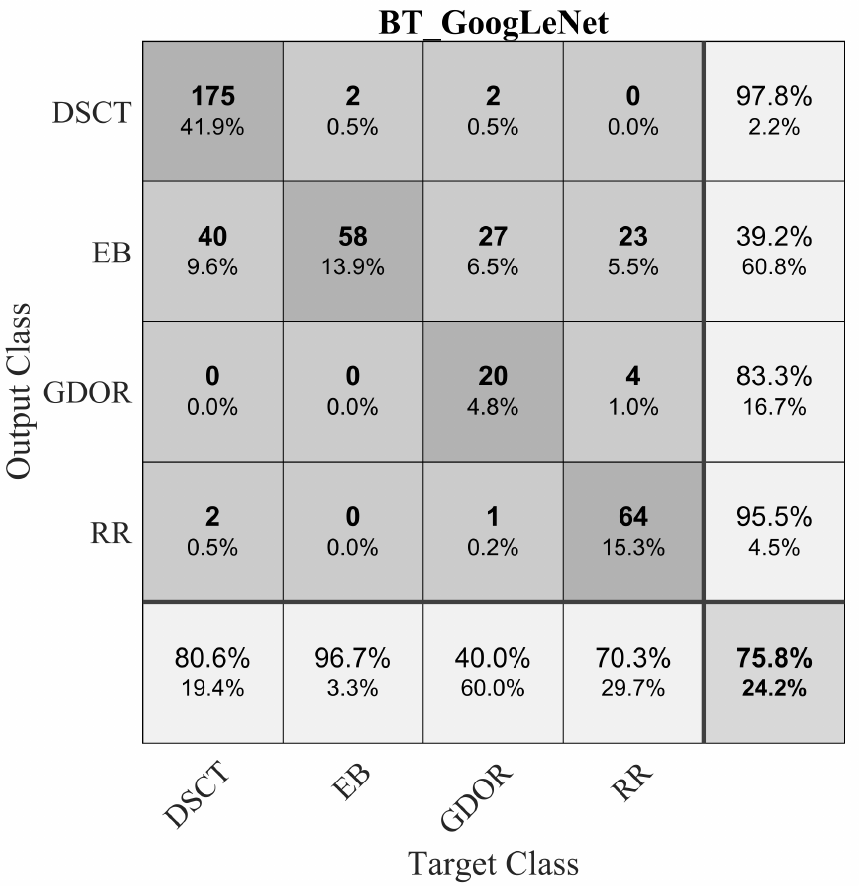}
    \includegraphics[width=0.4\textwidth]{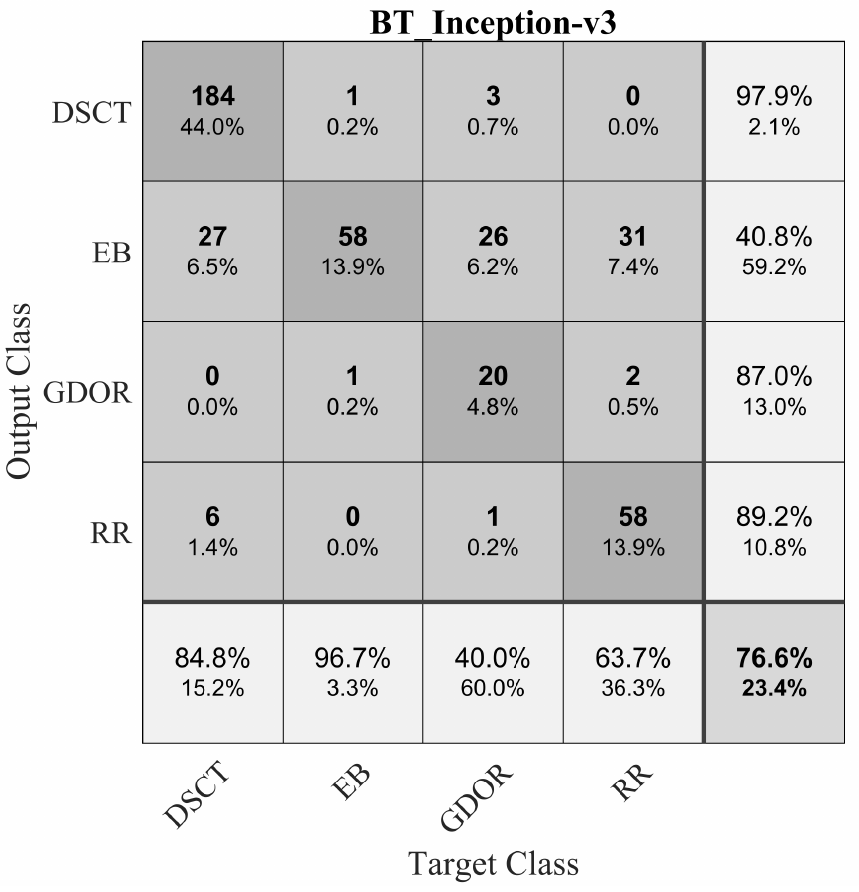}
    \includegraphics[width=0.4\textwidth]{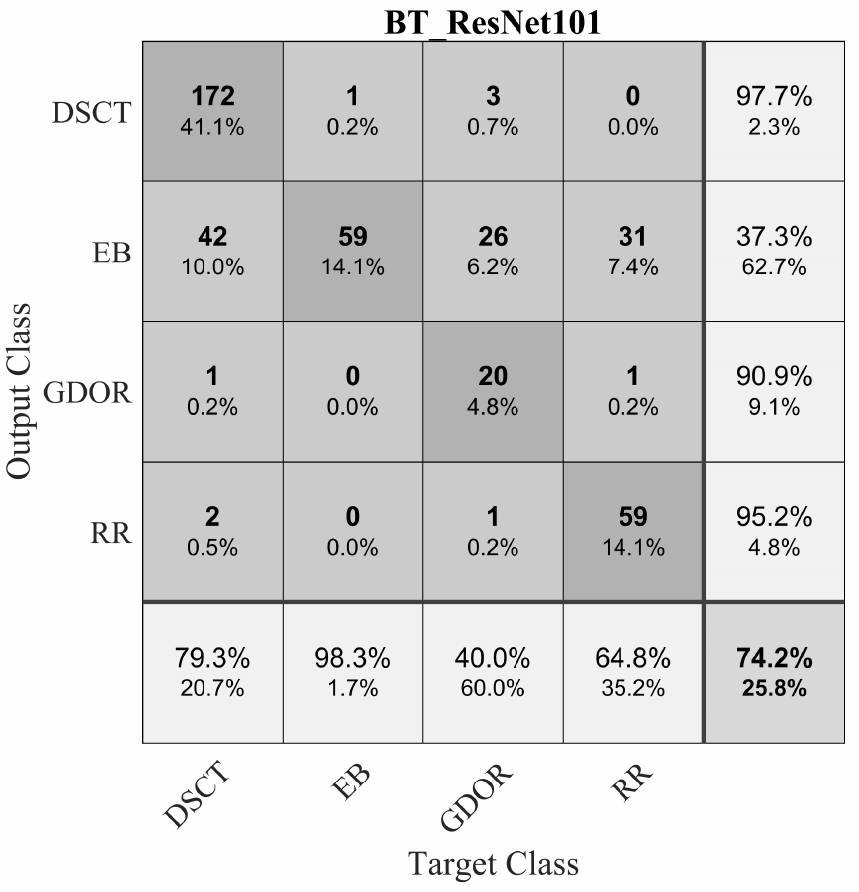}
    \includegraphics[width=0.4\textwidth]{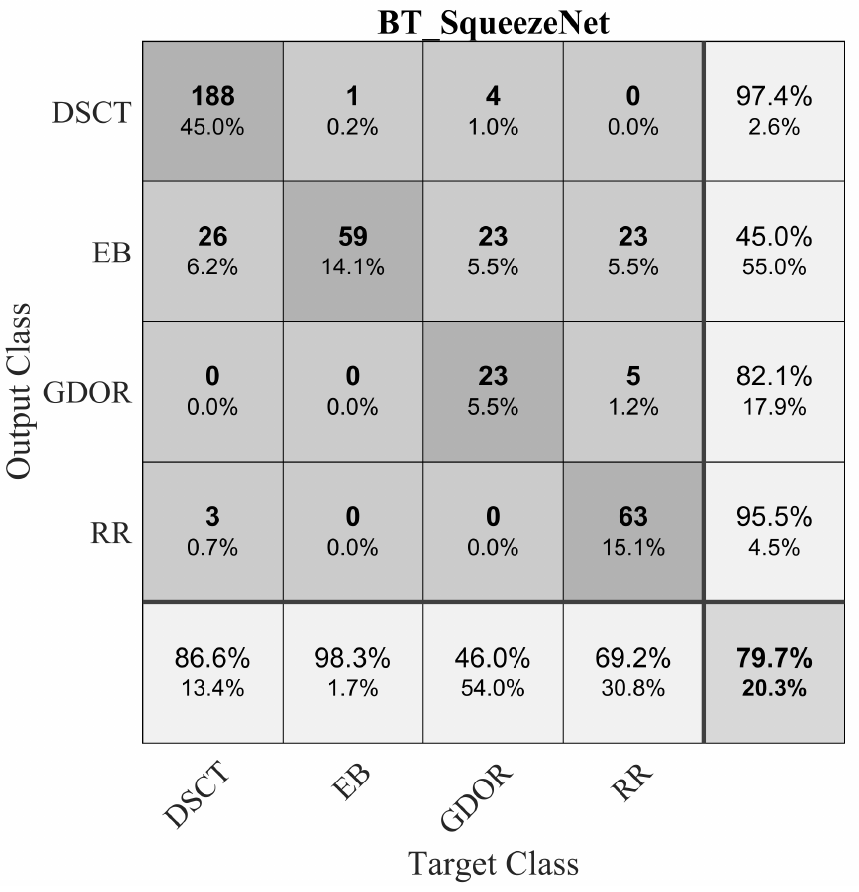}
    \includegraphics[width=0.4\textwidth]{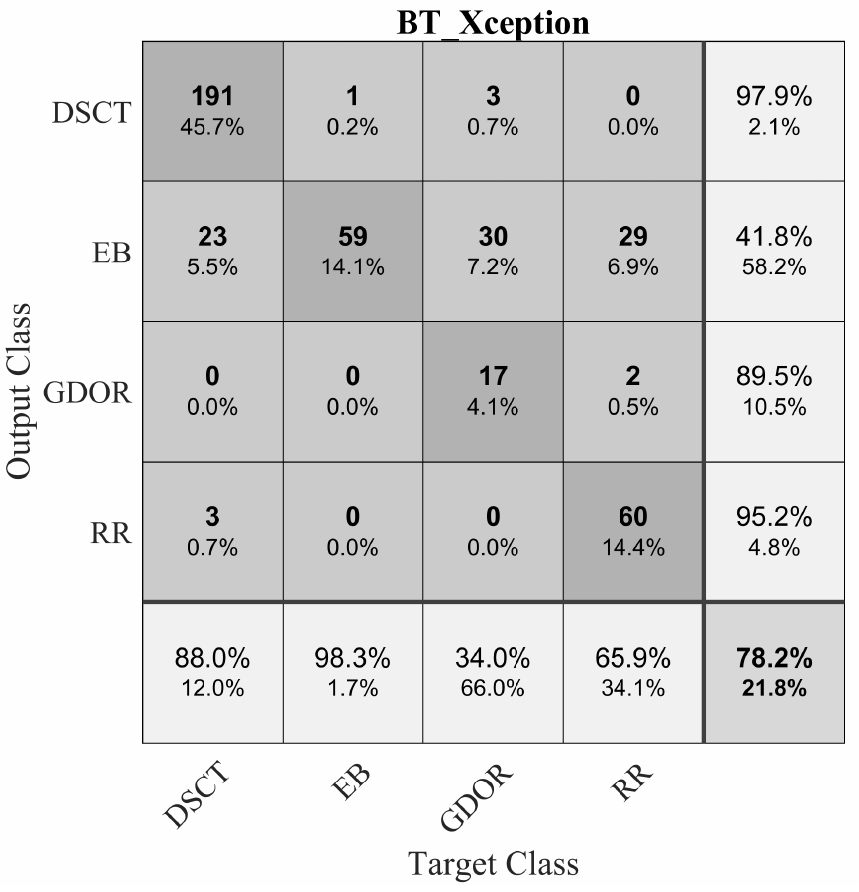}
    \caption{The confusion matrixes for blind tests.\label{fig:BT}}
\end{figure*}

The misclassification observed in our blind test can be attributed to the leakage of objects from different classes into the EB category (See Figure \ref{fig:BT}). This leakage occurs due to nearly constant frequencies and slightly varying amplitudes, which are also characteristic of certain $\gamma$ Dor subtypes, such as high-amplitude $\gamma$ Dor stars (see Figure \ref{ebleak}). The similarity in these features leads to confusion between $\gamma$ Dor and EB classes.

The limited generalization ability in the TESS blind test is primarily due to an insufficient sample size across varying surveys, minor on a small sample size. For example, \citealt{Jara20} employed 16,451 stars to construct a TESS sample set to detect exoplanets from light curves. \citealt{duev19ztf} used over 90,000 images to detect a transient object using CNN in ZTF. Similarly, \citealt{Cui22} analyzed more than 140,000 images for transient classification from light curves. Moreover, \citealt{Aguiree18} employed 246,474 stars from various surveys for variable classification. To improve the model's generalization ability, we require a larger sample set with data from different surveys with different observation strategies.

To mitigate this issue, a more effective approach is to augment the training set with additional samples. Combining light curves from both the Kepler and TESS missions could provide a richer and more diverse dataset, thereby improving the model's ability to distinguish between these closely related classes.

\end{document}